\documentclass[conference,compsoc]{IEEEtran}

\usepackage{00_preamble}
\usepackage{00_macros}

\begin{document}
\title{What Would Trojans Do? Exploiting Partial-Information Vulnerabilities in Autonomous Vehicle Sensing}
\date{}

\newif\ifAnonymize

\Anonymizefalse

\ifAnonymize

\else
    \author{
        \IEEEauthorblockN{R. Spencer Hallyburton}
        \IEEEauthorblockA{\textit{Duke University} \\
        spencer.hallyburton@duke.edu}
        \and
        \IEEEauthorblockN{Qingzhao Zhang}
        \IEEEauthorblockA{\textit{University of Michigan} \\
        qzzhang@umich.edu}
        \and
        \IEEEauthorblockN{Z. Morley Mao}
        \IEEEauthorblockA{\textit{University of Michigan} \\
        zmao@umich.edu}
        \linebreakand 
        \IEEEauthorblockN{Michael Reiter}
        \IEEEauthorblockA{\textit{Duke University} \\
        michael.reiter@duke.edu}
        \and
        \IEEEauthorblockN{Miroslav Pajic}
        \IEEEauthorblockA{\textit{Duke University} \\
        miroslav.pajic@duke.edu}
    }
\fi


\maketitle


\thispagestyle{plain}
\pagestyle{plain}

\begin{abstract}
Safety-critical sensors in autonomous vehicles (AVs) form an essential part of the vehicle's \emph{trusted computing base} (TCB), yet they are highly susceptible to attacks. Alarmingly, Tier 1 manufacturers have already exposed vulnerabilities to attacks introducing Trojans that can stealthily alter sensor outputs.
We analyze the feasible capability and safety-critical outcomes of an attack on sensing at a \emph{cyber level}. To further address these threats, we design realistic attacks in AV simulators and real-world datasets under two practical constraints: attackers (1) possess only \emph{partial} information and (2) are constrained by data structures that maintain sensor integrity.
Examining the role of camera and \lidar\ in multi-sensor AVs, we find that attacks targeting only the camera have minimal safety impact due to the sensor fusion system's strong reliance on 3D data from \lidar. This reliance makes \lidar-based attacks especially detrimental to safety.
To mitigate the vulnerabilities, we introduce security-aware sensor fusion incorporating (1) a probabilistic data-asymmetry monitor and (2) a scalable track-to-track fusion of 3D LiDAR and monocular detections (T2T-3DLM). We demonstrate that these methods significantly diminish attack success rate.
%
\end{abstract}
\section{Introduction} \label{sec:1-introduction}

Increasing inter-vehicle connectivity and complex supply chains have brought to light escalating vulnerabilities in both vehicle hardware and software, as noted by the National Highway Traffic Safety Administration (NHTSA)~\cite{nhtsa2016}. While connected and autonomous vehicles (AVs) offer the potential for safer and more efficient travel, they also introduce an unprecedented array of threats targeting cyber-physical systems (CPS)~\cite{zou2021cyber}. Attack vectors now span physical access to vehicle components, remote exploitation via wireless interfaces, compromised over-the-air (OTA) updates, and supply-chain attacks, underscoring the growing risks associated with these interconnected systems \cite{eiza2017driving}.

A particular concern is \emph{safety-critical} attacks where an adversary manipulates essential vehicle components to directly endanger occupants or nearby individuals. In this analysis, we identify components within the vehicle's \emph{trusted computing base} (TCB)—modules whose compromise can lead to severe safety incidents. Control systems that directly influence motion are an obvious part of this TCB. In AVs, control systems rely on accurate inputs from sensors and decision modules like perception, prediction, and planning that also make up the TCB.

Previous works assume attackers with significant control over vehicle inputs or even the capability to influence surrounding vehicles with attacker-driven commands~\cite{2022hally-frustum, thys2019fooling}. In reality, such extensive access is unrealistic. Our analysis adopts a practical threat model: an attacker with targeted access to sensor data by e.g., exploiting hardware Trojans, but no access to a vehicle's secured internal computation.

In particular, reports from Tier 1 manufacturers reveal critical vulnerabilities in AV sensing subsystems to attack vectors including physical access, over-the-air updates, and supply-chain infiltration~\cite{sikder2021survey}. Of particular concern to sensors are introduction of Trojans~\cite{beaumont2011hardware,subraman2019demonstrating}. Hardware Trojans involve subtle, malicious modifications to physical hardware to create triggers; exploiting these triggers can alter sensor outputs under specific conditions, misleading AV perception systems and causing errors in object detection, lane positioning, or obstacle recognition. Trojans are particularly dangerous due to their dormant nature, making them nearly invisible in conventional testing. For AVs, attacks targeting cameras and LiDAR are especially troubling, as they can disrupt core functions like object avoidance and navigation.

Prior works considered powerful attackers with e.g.,~full access to highly secured perception algorithms to design carefully crafted attacks~\cite{2019cao-spoofing,2020tuphysicalatt}. Instead, our threat model is an attacker with limited, partially-observable access to an AV sensor and its raw data packets. We denote this a ``cyber-level'' attack to distinguish it from attacks exploiting physical channels such as spoofing and the use of physical adversarial objects. Instead of requiring model access or iterative attack optimizations, the cyber-level threat model considers if attackers can cause devastating outcomes under limited yet easily-achievable sets of information in real time. In particular, a cyber-level attacker will not have access to systems other than the sensor; they may not even know where the sensor is placed on the vehicle.   


Our vulnerability analysis centers on a multi-sensor AV with monocular cameras and central LiDAR, as in industry-standard AVs including Baidu's Apollo~\cite{BaiduApollo}, GM's Cruise~\cite{cruiseblog}, and Google's Waymo~\cite{waymoblog}, and datasets including Waymo \cite{2020waymodataset}, nuScenes~\cite{2020nuscenesdataset}, and KITTI~\cite{2013kittidataset}. For attacks to induce safety-critical outcomes, they must: (1) impact the \emph{perception level}; (2) propagate through \emph{object tracking}; and (3) affect \emph{prediction} and \emph{control} leading to consequential changes in the AV's understanding or reaction to its surroundings.

Focusing on camera and LiDAR vulnerabilities, our findings show that while multi-sensor fusion mitigates the impact of camera-only attacks, LiDAR remains a high-risk component due to its essential role in 3D awareness. Through simulations on Baidu Apollo~\cite{BaiduApollo} and real-world datasets, we demonstrate how limited-access LiDAR attacks can lead to safety-critical events like emergency braking or collisions, even in multi-sensor AVs. This insight identifies LiDAR as an indispensable TCB member in AV safety.

Given these vulnerabilities, we propose novel defenses against LiDAR-based attacks by (1) introducing a sensor disagreement monitor to detect inconsistencies, and (2) leveraging decentralized track-to-track (T2T) fusion of 3D LiDAR and monocular (3DLM) detections to mitigate the discovered vulnerabilities such as the frustum vulnerability~\cite{2022hally-frustum} (T2T-3DLM). In tests, we demonstrate that the asymmetry monitor reduces the success of the false-positive attacks  to under 20\%, while T2T-3DLM lowers the rate of safety-critical incidents to under 5\% across all attacks. 

Our code as well as the videos of our experiments are publicly available at~\cite{AVcyberattack}.
The contributions of this work can be summarized as follows:
\begin{itemize}
    \item Define a realistic threat model reflecting achievable attacks on single sensors within multi-sensor AVs.
    \item Demonstrate limited effectiveness of camera-only attacks in multi-sensor settings.
    \item Establish LiDAR as a critical TCB component and conduct targeted attacks within a partial-information~model.
    \item Evaluate five LiDAR-based attacks using industry-grade simulators and longitudinal datasets.
    \item Introduce two novel defense strategies: a probabilistic sensor disagreement monitor and decentralized track-to-track fusion for robust, security-aware fusion.
\end{itemize}

\section{Background for Attacks on Sensing in AVs} \label{sec:2-background}

A growing body of literature has highlighted the vulnerability of 
AV sensors to a variety of attack vectors. This section first presents many recent analyses of attacks on AVs that used physical channels or adversarial objects to create safety-critical outcomes. We then discuss relevant attack vectors to motivate the need for our analysis of cyber-level~attacks. 

\subsection{Demonstrated Attacks Via Physical Channels}

Attacks on sensors via physical channels have been extensively studied in the literature. These include:

\vspace{4pt}
\noindent \textbf{Spoofing.} Spoofing attacks target GPS, LiDAR, and radar by introducing false signals to create misleading sensor data. GPS spoofing can manipulate an AV’s perceived location, causing navigation errors~\cite{quinonez2020savior}. LiDAR spoofing was presented in~\cite{2019cao-spoofing, 2020sun-spoofing} where fake points caused the AV to detect phantom objects. These spoofing attacks leverage AV reliance on spatial data, creating scenarios where the vehicle may brake unnecessarily or fail to respond to true obstacles.

\vspace{4pt}
\noindent \textbf{Jamming.} Jamming attacks overwhelm sensors with irrelevant or excessive signals, leading to disrupted sensor readings. For instance, Petit et al. explored LiDAR jamming by directing high-intensity signals that saturate the sensor and impair its ability to detect objects~\cite{2014cyberattacks}. These attacks force the AV to lose situational awareness, especially in dense or complex environments.

\vspace{4pt}
\noindent \textbf{Adversarial objects.} With the adoption of deep neural networks (DNNs) for perception, adversarial attacks have become a critical area of research. Perturbations to the environment including stickers or modified signs can mislead DNNs into misclassifying objects or detecting nonexistent lanes \cite{eykholt2018robust, thys2019fooling}. Moreover, adversarial objects can influence the outcomes of perception algorithms~\cite{2021robustfusion}. 

\subsection{Cyber-Level Attack Vectors}

While initial analysis of attacks on sensors exploited physical channels through spoofing or adversarial objects, vulnerabilities in the supply chain and over-the-air (OTA) updates highlight that sensors are susceptible to cyber-level attacks. Some feasible attack vectors are summarized below.

\vspace{4pt}
\noindent \textbf{Remote access.} With AVs increasingly reliant on Vehicle-to-Vehicle (V2V) and Vehicle-to-Infrastructure (V2I) communications, remote attacks have emerged as a prominent threat. Recent works illustrate how wireless interfaces such as Bluetooth, Wi-Fi, and cellular networks can be compromised to control AV sensor data remotely \cite{2010koschercheckoway, verma2015impact}. In one high-profile case, a remote attack on a Jeep vehicle gained control over multiple systems through its internet-connected systems \cite{remoteJeepHack}. These studies underscore the necessity of securing wireless interfaces to prevent adversaries from remotely accessing and modifying sensor data.

\vspace{4pt}
\noindent \textbf{Over-the-air (OTA) updates.} OTA updates allow manufacturers to patch AV software without physical intervention, but they also present a vector for attackers to insert malicious code~\cite{lopez2019security}. Supply chain vulnerabilities further complicate OTA and sensor security, as malicious actors may infiltrate at any point from manufacturing to deployment. Studies highlight the risks associated with third-party vendors, as a compromised update could introduce triggers, impacting AV safety without immediate detection.

\vspace{4pt}
\noindent \textbf{Supply-chain infiltration.} As AV sensors are often sourced from third-party vendors, attackers can introduce hardware Trojans or other malicious modifications during manufacturing or distribution. These alterations are usually dormant and only activate under specific conditions, making them hard to detect. Once triggered, they can modify sensor outputs, leading to incorrect object detection or navigation decisions, bypassing software-focused cybersecurity measures. \cite{lopez2019security}.

\vspace{4pt}
Cyber-level attacks that e.g., introduce Trojans via a compromised supply chain~\cite{beaumont2011hardware, subraman2019demonstrating} pose a severe security risk, as they are positioned inside conventional cryptographic security safeguards. Unfortunately, few AV security studies have analyzed the \emph{attacker capability and feasible impact} of these cyber-level threats. While existing research has extensively explored spoofing, jamming, and adversarial objects, there is a need for evaluations of partial-information cyber-level attacks in dynamic AV environments. 

\section{Cyber Attack Model} \label{sec:4-cyber-model}

Our model assumes the attacker does not possess privileged access to AV systems or knowledge of proprietary algorithms. Instead, the attacker operates at the cyber level, exploiting remote access, malicious OTA updates, or supply chain vulnerabilities to compromise a single sensor, such as a camera or LiDAR. This section focuses on the attacker’s goals, limitations, and capabilities in achieving safety-critical disruptions.

\subsection{Attacker’s Goals}
The primary objective is to induce safety-critical outcomes by altering perception data. Rather than assuming extensive system control, this model aligns with plausible tactics that real-world attackers might employ to infiltrate AV perception systems. Our analysis evaluates whether a sensor is in the TCB by testing whether its compromise can induce safety-critical outcomes at the following stages:

\vspace{2pt}
\noindent \textbf{Perception level.} By manipulating sensor outputs, the attacker can introduce errors in object detection and classification, such as injecting false positives.

\vspace{2pt}
\noindent \textbf{Object tracking level.} Ensuring that manipulated detections persist across frames, the attacker alters object tracks and causes distorted situational awareness.

\vspace{2pt}
\noindent \textbf{Prediction and control levels.} Using manipulated tracks to influence future object trajectory predictions, the attacker can trigger emergency responses, such as sudden braking or evasive maneuvers.

We now 
discuss partial-information attacks, emphasizing the attacker’s knowledge, constraints, and capabilities.

\subsection{Partial, Limited Information Context} \label{sec:cyber-model-partial-information}

\begin{figure}[!t]
    \centering
    \includegraphics[width=0.98\linewidth]{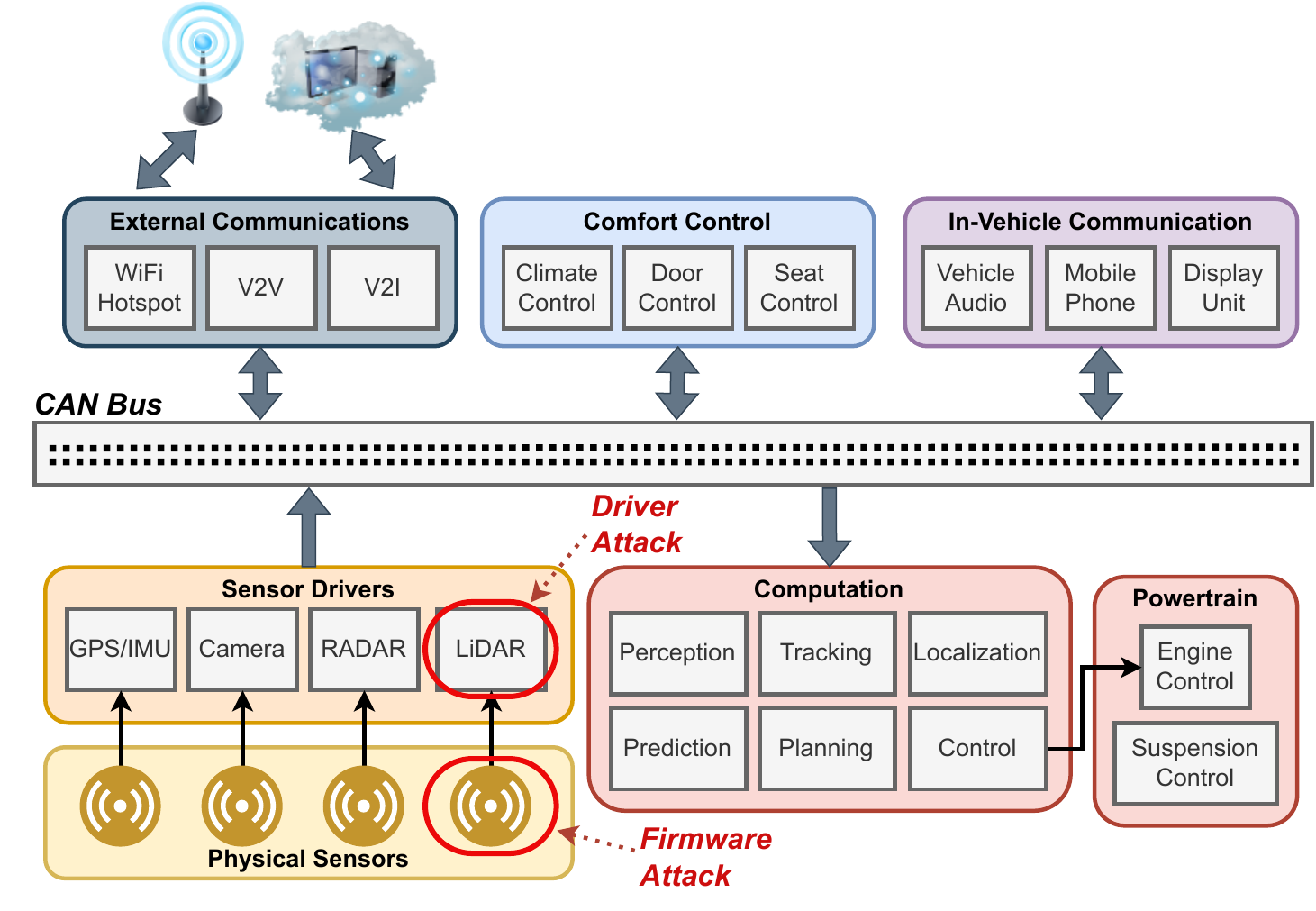}
    \caption{AVs connect diverse services mainly over a CAN bus. Attacks at the firmware/driver levels will only have access to isolated information at each subsystem.}
    \label{fig:av-network}
\end{figure}

Sensing data in AVs is transmitted to the computation subsystem (Fig.~\ref{fig:av-network}) as compact \emph{datagrams} containing essential information like timestamps and sensor readings in a header and payload structure. When an attacker compromises an AV’s sensing subsystem, they can access only these datagram contents, with influence limited to the sensing layer and no access to downstream data in the computation subsystem. Unlike ``white-box'' attackers with full model access (e.g., neural network weights) or ``black-box'' attackers with extensive situational control~\cite{2021invisiblecamlidar,2020attacktracking,2022hally-frustum,2019cao-spoofing}, our model restricts the attacker to limited access. In this context, the attacker operates under the following conditions:

\vspace{2pt} \noindent \textbf{Real-time attacks.} The attacker must manipulate sensor data in real-time, reflecting the rapid processing demands in AVs. This real-time requirement limits the attacker’s capacity to perform complex, iterative computations, necessitating precise yet swift decisions in altering perception data.

\vspace{2pt} \noindent \textbf{Capability limitations.} Data manipulation is restricted by the structured nature of sensor data formats and stringent timing protocols. These constraints prevent unrestricted data modification, as AV systems continuously monitor for timing integrity and anomalies, making any out-of-sequence or abnormal data easier to detect.

\subsection{Attacker Model}
We define the attacker’s knowledge and capabilities at a partial-information cyber level. First, we outline \emph{the information available in advance}, establishing a baseline for exploitation without full access. Next, we discuss \emph{insights derived from real-time monitoring} of sensor data transmitted in compact datagrams. We then focus on attacks targeting \emph{camera and LiDAR sensors}, examining the \emph{operational constraints} within this limited-access level. Finally, we establish \emph{the attacker’s potential capabilities}, given the challenges of restricted data manipulation and real-time processing.

\vspace{3.2pt}
\noindent\textbf{A-Priori information.}
With a cyber-level attack at the sensor level, the attacker will have little access to system or environment information. We only allow the following for a sensor attacker:
\begin{enumerate}[label=\textbf{K.\arabic*},topsep=0pt,itemsep=-1ex,partopsep=1ex,parsep=1ex]
    \item \label{know:structure} The attacker knows the size \& structure of datagrams but cannot manipulate them.
    \item \label{know:timestamp} The attacker knows the timestamps of datagrams and can manipulate them.    
    \item \label{know:config} The attacker has knowledge of sensor ``datasheet'' parameters such as resolution and field of view.
    \item \label{know:axes} The sensor is mounted with two ground-coplanar axes.
\end{enumerate}

\vspace{2pt}\noindent This information is standard given the sensor make and model. Other details (e.g.,~sensor height, presence of other sensors, runtime data, surrounding environment) will not be known in advance. This is in contrast to prior works (e.g., \cite{2022hally-frustum}) that leveraged accurate situational awareness a-priori.

\noindent\textbf{Induced information.}
The attacker achieves ``context-awareness'' through lightweight real-time monitoring algorithms. By accessing raw data, they can calculate the sensor rate from datagram timestamps and estimate higher-order details like the sensor's yaw angle or height above ground. However, they lack scene-specific details, such as object locations, which provide direct situational awareness.

\subsection{Datagram Structure for Integrity} \label{sec:datagram-integrity}
AV sensor data is transmitted in structured packets called \emph{datagrams}. The rigid structure and consistent delivery timing of these datagrams enable the AV to check data integrity. Attacks remain feasible and stealthy only if they comply with these structural and timing constraints.

\vspace{3.2pt}
\noindent\textbf{Notation} \label{sec:notation}
The point cloud $PC$ is a set of four-dimensional points $(X, Y, Z, \text{intensity})$ from a LiDAR rotation. The size of $PC$, stored as a matrix $[N \times 4]$, varies with each rotation. We define $D_k$ as a datagram at time $t_k$, and $P_{(i,j,k)}$ as a point in spherical coordinates $(\rho_{(i,j,k)}, \theta_i, \phi_j, t_k, I_{(i,j,k)})$ for range, azimuth, elevation, timestamp, and intensity, respectively. The sets of azimuth and elevation angles are $\{\theta_i \ | \ i=0,1,\dots,n\}$ and $\{\phi_j \ | \ j=0,1,\dots,m\}$, where the cross-product forms a matrix of (range, intensity) pairs. $|\cdot|$ gives the number of elements.

\subsubsection{LiDAR Data Structures}
\begin{figure}[!t]
    \centering
    \includegraphics[width=0.96\linewidth]{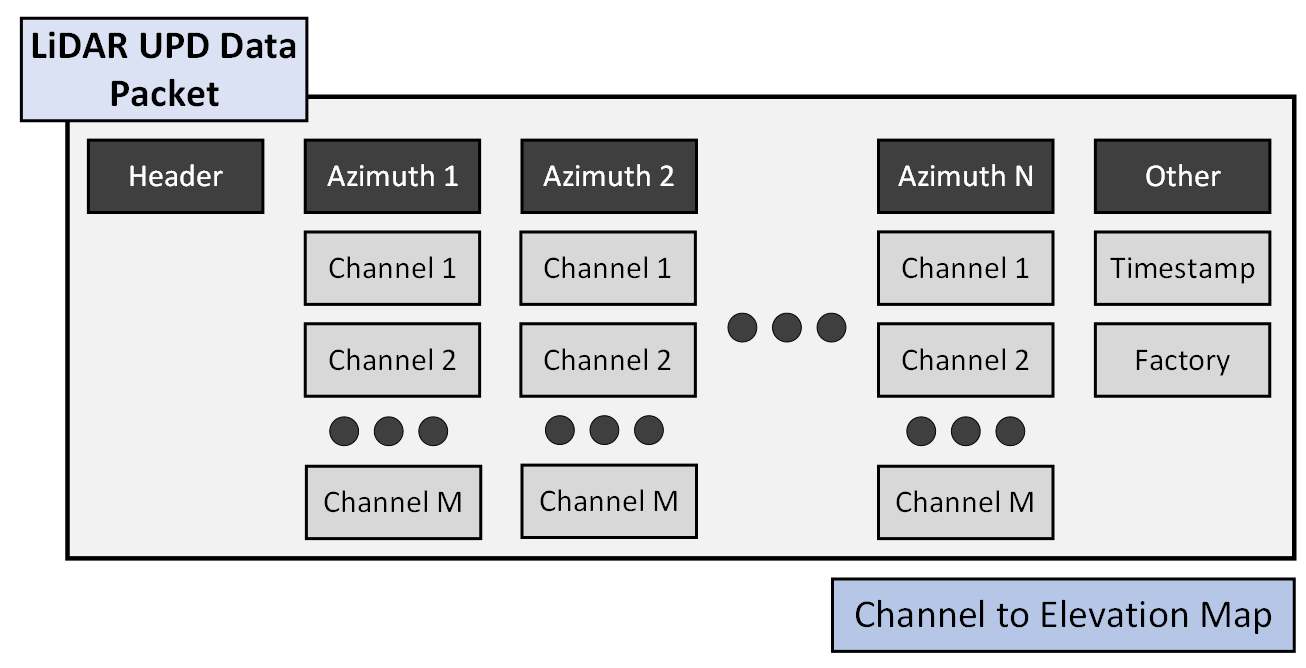}
    \caption{Datagrams follow a consistent structure across sensors within the same family. For example, the Velodyne VLP and HDL series (16, 32, and 64) have nearly identical structures (see~\cite{velodynemanual}). An attacker targeting LiDAR in a cyber level \emph{accesses only the LiDAR datagrams} and can manipulate the data \emph{only within the constraints of the datagram}.}
    \label{fig:lidar-datagram}
\end{figure}

Fig.~\ref{fig:lidar-datagram} shows LiDAR datagram structure. Datagrams are sent over UDP, each containing points from multiple azimuth and all elevation angles. The Velodyne-HDL32E, for example, requires about 175 datagrams for a complete sweep. Unlike datagrams seen by attackers, datasets and simulators save point clouds in matrix format. 

\renewcommand{\subfigwidth}{0.6}
\renewcommand{\graphicswidth}{50mm}
\renewcommand{\graphicsheight}{40mm}


\subsubsection{First Order Data Integrity} \label{sec:integrity}

Using the regular structure of the datagrams, a victim can implement some basic, ``first-order'' integrity checks on the LiDAR. We formalize the integrity mathematically.

\begin{enumerate}[leftmargin=20pt,labelsep=\labelspace pt,align=left,label=\textbf{INT.\arabic*},topsep=0pt,itemsep=-1ex,partopsep=1ex,parsep=1ex]
    \item \label{integ:max-points} \textbf{Max Points:} The receiver knows a maximum of 1 point per angle (ppa) and 2 ppa should appear for single and dual modes, respectively. The receiver thus a-priori knows the maximum possible number of points (in the point cloud set PC) returned by the sensor. Denoting this maximum value as $\alpha$, we can introduce an integrity indicator function
    \begin{align} \zeta_{\alpha} = \left[|PC| \leq \alpha\right]. \end{align}
    
    \item \label{integ:min-points} \textbf{Min Points:} Although it is possible for a laser to not register a return (e.g.,~no strong reflection), the majority of lasers will return. Hence, the receiver can a-priori set~a bound on the minimum number of points (in the point cloud) returned by the sensor. Denoting this minimum value~as $\beta$, the corresponding integrity indicator function~is 
    \begin{align}\zeta_{\beta} = \left[|PC| \geq \beta\right].\end{align}
    
    \item \label{integ:timestamp} \textbf{Packet Timestamps:} The time interval between packets may exhibit Gaussian noise, making the residual (difference between estimated and measured intervals) follow a $\chi^2$ distribution with one degree of freedom. This residual serves as an integrity indicator for the azimuth angle; a normalized residual, $\frac{y_i}{\sigma_{\gamma}}$, is compared against a confidence threshold, $\gamma$, i.e.:
    \begin{align}\zeta_{\gamma} = \left[ \left(\frac{y_i}{\sigma_{\gamma}}\right)^2 \leq \gamma \right].\end{align}
    
    \item \label{integ:dual} \textbf{Dual Verification:} If the sensor is in \texttt{dual} mode, the order of the returns is (strongest, last) unless the strongest is the last return in which case the order is (second strongest, last). Thus, the second return will always have larger range than the first, i.e.:
    \begin{align}
    \zeta_{\rho} \coloneqq \begin{cases}
        \texttt{False} & \texttt{mode}==\text{dual},\ \rho^{(1)}_{i,j,k} > \rho^{(2)}_{i,j,k} \\
        \texttt{True} & \text{otherwise}
    \end{cases}
    \end{align}
\end{enumerate}

\noindent Finally, an aggregate receiver integrity indicator function $\zeta$ is defined as the conjunction of indicators; specifically,
\begin{align}
    \zeta \coloneqq \zeta_{\alpha} \land \zeta_{\beta} \land \zeta_{\gamma} \land \zeta_{\rho}
\end{align}
with integrity passing if $\zeta ==\texttt{True}$. This helps limit spoofing and mitigate nonsensical data manipulation.

\subsection{Attacker Capability} \label{sec:cyber-attack-capabilities}
We now outline attacker capabilities that maintain stealthiness against datagram and timing-related integrity.

\subsubsection{General attack capability.}
Given only access to datagrams, an attacker may perform the following manipulations at a cyber level.
\begin{enumerate} [label=\textbf{C.\arabic*},topsep=0pt,itemsep=-1ex,partopsep=1ex,parsep=1ex]
    \item \label{capab:time} \textbf{Timestamp}: Modify the datagram timestamp.
    \item \label{capab:drop-data} \textbf{Drop datagrams}: Drop data, e.g.,~denial of service.
    \item \label{capab:new-data} \textbf{Add datagrams}: Send additional (fake) datagrams.
\end{enumerate}

\vspace{2pt}
If the sensor driver is monitoring the sensing rate, then significant modifications to timestamps and dropping/adding new datagrams will trigger integrity warnings.

\subsubsection{Stealthy capability for camera attacks.} \label{sec:camera-capability} Since the image is a fixed-size amount of data, a camera attacker can only modify the pixel intensity.

\begin{enumerate} [label=\textbf{C.\arabic*},leftmargin=1.5\parindent,resume,topsep=0pt,itemsep=-1ex,partopsep=1ex,parsep=1ex]
    \item \label{capab:intensity-mod} \textbf{Intensity}: Modify the 8-bit intensity for pixels.
\end{enumerate}

\subsubsection{Stealthy capability for \lidar\ attacks.} \label{sec:lidar-capability} The \lidar\ datagram is consistent across sensors between a vendor. The attacker can perform the following stealthy operations to the datagrams.
\begin{enumerate} [label=\textbf{C.\arabic*},resume,leftmargin=1.5\parindent,topsep=0pt,itemsep=-1ex,partopsep=1ex,parsep=1ex]
    \item \label{capab:mod} \textbf{Range modification}: Modify range in the datagrams.
    \item \label{capab:null} \textbf{Range nullification}: Set some range entries to \texttt{NULL}.
    \item \label{capab:spoof} \textbf{Range spoofing}: Set \texttt{NULL} range entries in the datagram to some floating point values.
    \item \label{capab:todual} \textbf{Mode manipulation + spoofing}: Switch mode from \texttt{single} to \texttt{dual} and add fake points.
    \item \label{capab:tosingle} \textbf{Mode manipulation + dropping}: Switch mode from \texttt{dual} to \texttt{single} and drop points.
\end{enumerate}

Pushing these capabilities to their limits can still trigger integrity checks. For instance, capability \ref{capab:null} should avoid nullifying excessive LiDAR points, and the use of \ref{capab:todual} must align with dual integrity. The attacker \emph{cannot} alter azimuth or elevation angles (aside from minor adjustments in \ref{capab:time} due to the rigid datagram structure and fixed geometry.

This realistic threat model underscores practical, limited-access attacks that exploit known vulnerabilities in cyber-physical sensors, highlighting the need for robust, multi-layered defenses as addressed in our proposed security enhancements for AV perception and fusion modules.
\section{Conditions for Attack Effectiveness} \label{sec:6-evaluations}

Attacks are only relevant if they affect \emph{safety-critical} AV modules. We define three conditions necessary for attacks to produce safety-critical outcome.

\vspace{4pt}
\noindent \textbf{Condition 1: Perception.} We analyze single-frame metrics, focusing on false positives (FP), false negatives (FN), and translation outcomes as in \cite{2019cao-spoofing, 2022hally-frustum}. We consider only \emph{per-frame-increment} FPs and FNs resulting from attacks, not those seen in a baseline.

\vspace{4pt}
\noindent \textbf{Condition 2: Longitudinal.} We assess longitudinal outcomes in target tracking, identifying \textit{per-frame-increment} changes in false track (FT) and missed track (MT) incidences, similar to perception.

\vspace{4pt}
\noindent \textbf{Condition 3: Safety-Critical.}
Using the responsibility-sensitive safety (RSS) metric, we classify an outcome as safety-critical if it impacts scene safety at the prediction, planning, or control levels. For prediction/planning, safety-critical outcomes occur if the victim’s perception mismatches reality (e.g., perceiving safe when unsafe). At the control level, outcomes are critical if actions transition from safe to unsafe or decrease quantitative safety.
\section{Autonomy Algorithms} \label{sec:3-av-model}

\subsection{AV Analysis Framework}
Testing a real AV in a physical environment poses safety and resource challenges. Instead, we use (1) real sensing datasets (KITTI~\cite{2013kittidataset}, nuScenes~\cite{2020nuscenesdataset}) for module-level analysis, and (2) the Carla simulator~\cite{2017carla} with the Baidu Apollo stack~\cite{BaiduApollo} for high-fidelity case studies.

\vspace{4pt}
\noindent \textbf{Classical AV designs.} 
\labeltext{\textbf{AV~1}}{av:lidar}
\labeltext{\textbf{AV~2}}{av:cam-lidar-v1}
Fig.~\ref{fig:av-algs} presents the AV architectures from a high level. Our first two AV implementations follow classical LiDAR-based and centralized camera-LiDAR fusion. \ref{av:lidar} (Fig.~\ref{fig:av-lidar}) is a LiDAR-based architecture where PointPillars~\cite{2019pointpillars} perception runs on LiDAR data and detections are passed to a 10-state (3-position, 3-velocity, 4-bounding-box) Kalman filter tracker. \ref{av:cam-lidar-v1} (Fig.~\ref{fig:av-cam-lidar-v1}) is camera-LiDAR fusion where PointPillars~\cite{2019pointpillars} runs on LiDAR and FasterRCNN~\cite{2015fasterrcnn} runs on camera data. Detections are passed to a centralized fusion tracker (e.g.,~\cite{2021eagermot}).

The two security-aware AV implementations are presented in Section~\ref{sec:6-defenses}. In summary, they are:
\begin{itemize}
    \item \ref{av:cam-lidar-v2} (Fig.~\ref{fig:av-cam-lidar-v2}) extends centralized camera fusion by adding a data asymmetry monitor.
    \item \ref{av:ttt} (Fig.~\ref{fig:av-ttt}) is a distributed fusion algorithm using DDF concepts~\cite{2013ddfreview}. Detections from both 3D LiDAR and monocular 3D data are tracked with 10-state Kalman filters and fused via classical assignment (e.g., JVC~\cite{1986JVC}) and covariance intersection~\cite{2017ddfwithCI}.
\end{itemize}

\renewcommand{\subfigwidth}{0.45}
\renewcommand{\graphicswidth}{84mm}
\renewcommand{\graphicsheight}{16mm}
\renewcommand{\graphicsheightt}{24mm}

\begin{figure*}[!t]
     \centering
     \begin{subfigure}[b]{\subfigwidth\linewidth}
         \centering
         \includegraphics[width=0.9\linewidth]{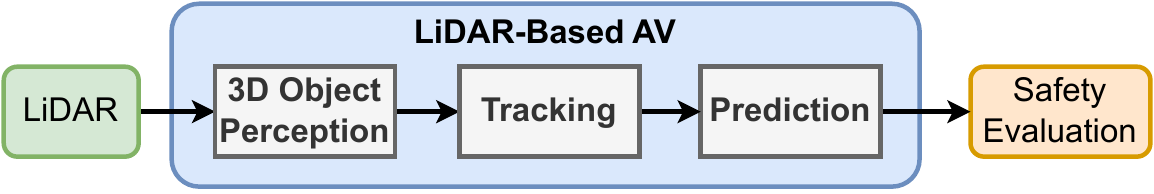}
         \caption{LiDAR-based detection and tracking is a common single-sensor approach to AV decision.}
         \label{fig:av-lidar}
     \end{subfigure}
     \hfill
     \begin{subfigure}[b]{\subfigwidth\linewidth}
         \centering
         \includegraphics[width=0.9\linewidth]{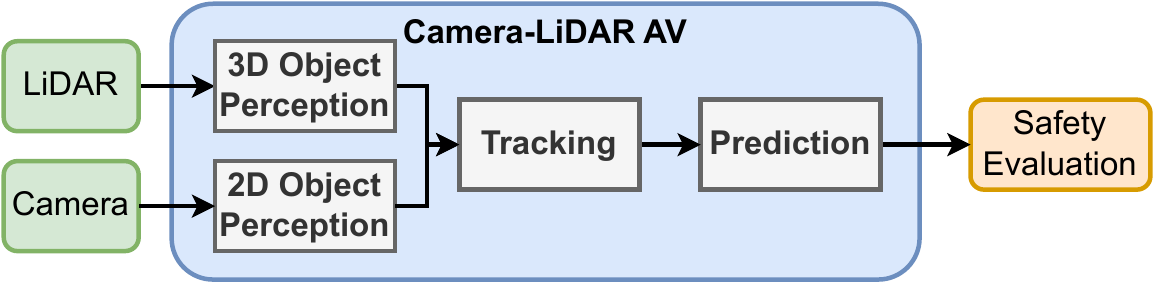}
         \caption{Fusing data jointly at tracking can lead to improved performance over a single-modality.}
         \label{fig:av-cam-lidar-v1}
     \end{subfigure}
     \begin{subfigure}[b]{\subfigwidth\linewidth}
         \centering
         \includegraphics[width=0.9\linewidth]{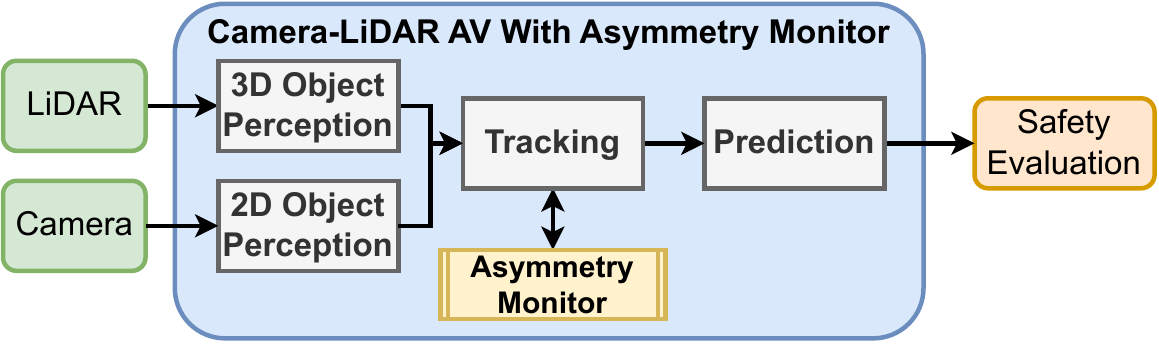}
         \caption{Centralized fusion with data-asymmetry monitor to catch sensor mismatches.}
         \label{fig:av-cam-lidar-v2}
     \end{subfigure}
     \hfill
     \begin{subfigure}[b]{\subfigwidth\linewidth}
         \centering
         \includegraphics[width=0.9\linewidth]{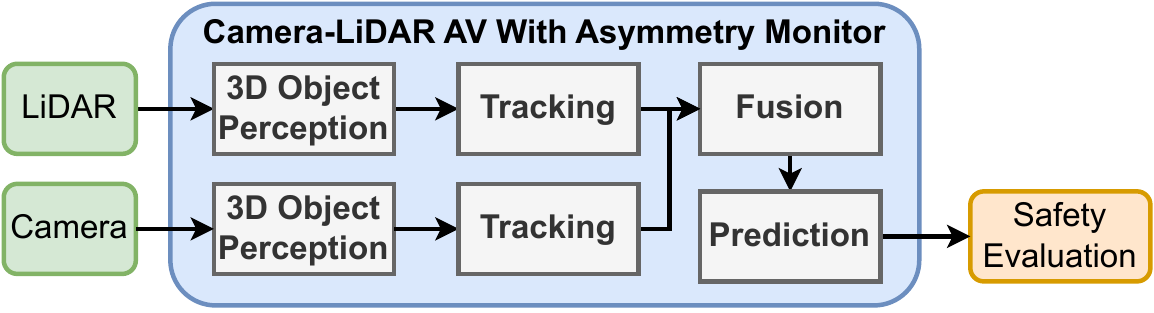}
         \caption{T2T-3DlM: post-tracking fusion can help mitigate differences in FP/FN rates between sensor modalities.}
         \label{fig:av-ttt}
     \end{subfigure}
    \caption{Fusing information from multiple sensors can~be achieved using \textit{many} different architectures/designs;~thus, it is critical to evaluate several common implementations.}
    \label{fig:av-algs}
\end{figure*}

\subsection{Vulnerability of Sensor Fusion}
\begin{figure}
    \centering
    \begin{subfigure}[b]{.48\linewidth}
        \centering
        \includegraphics[width=\linewidth]{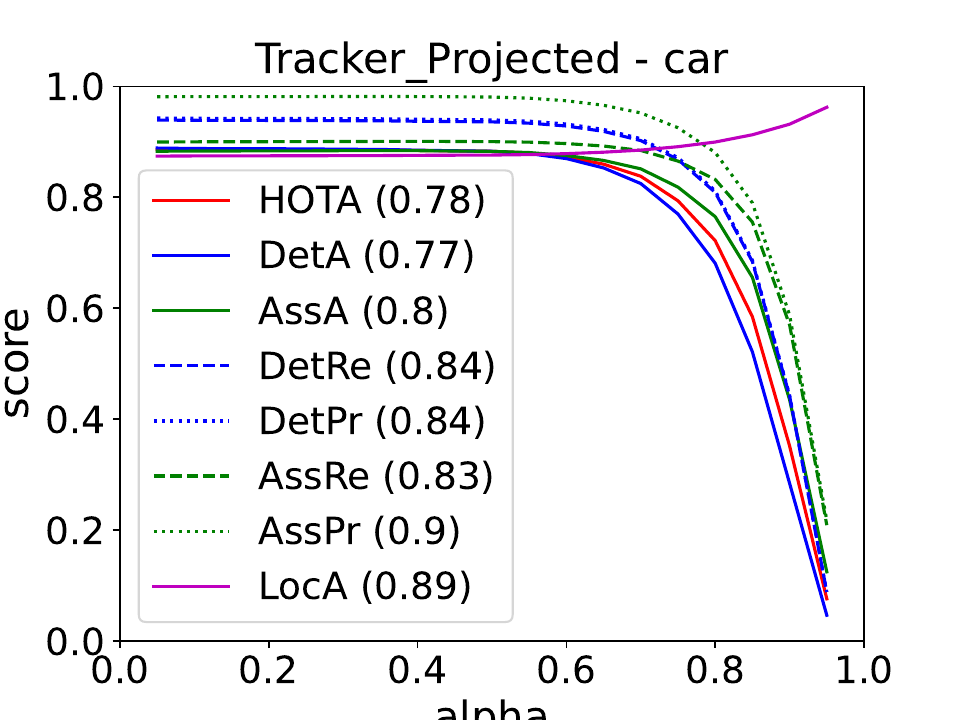}
        \caption{Adding noise to 2D dets.}
        \label{fig:track-fusion-camera}
    \end{subfigure}        
    \centering
    \begin{subfigure}[b]{.48\linewidth}
        \centering
        \includegraphics[width=\linewidth]{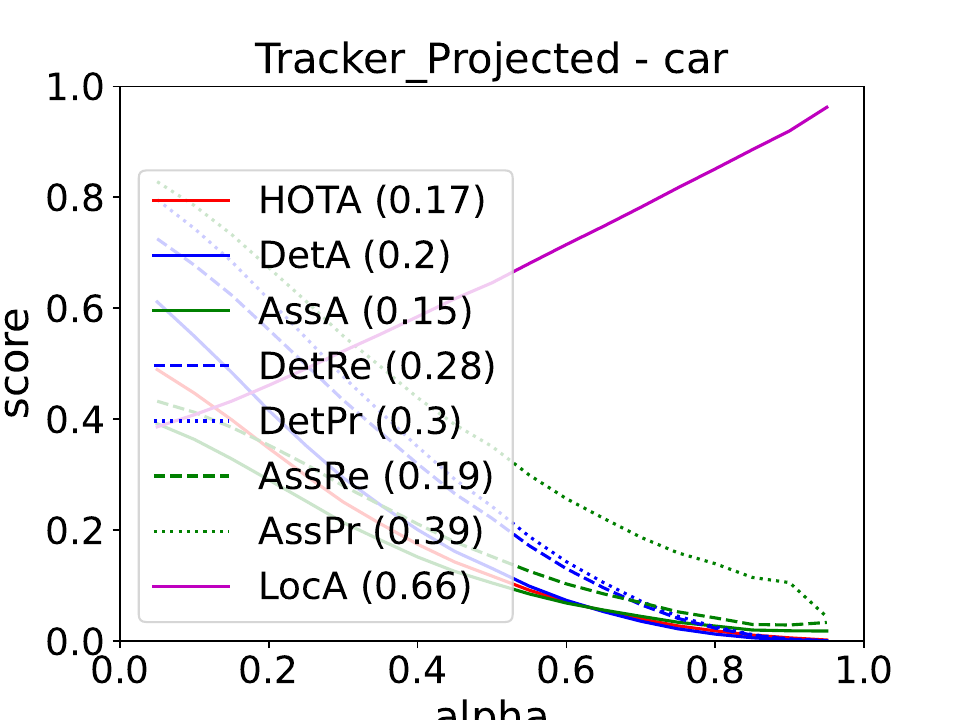}
        \caption{Adding noise to 3D dets.}
        \label{fig:track-fusion-lidar}
    \end{subfigure}
    \caption{Track fusion performance metrics via~\cite{2021hotametric}; better performance is towards the upper right. (a) Adding significant noise ($\sigma=100~pix$) to 2D detections results in minimal tracking performance change compared to the baseline in~\cite{2021eagermot}, indicating reliance on 3D data. (b) Adding minor noise ($\sigma=0.5~m$) to 3D \lidar\ detections significantly degrades tracking, suggesting \lidar\ is a TCB component.}
    \label{fig:track-fusion}
\end{figure}

Accurate 3D object state data from \emph{sensor fusion} is crucial for motion prediction and safe path planning. We analyze vulnerabilities in multi-sensor fusion, observing that state-of-the-art algorithms heavily rely on LiDAR-based 3D detections, making them susceptible to attacks on LiDAR data, while 2D camera data attacks pose less risk. Relevant sensor fusion algorithms include,

\vspace{4pt}
\noindent \textbf{EagerMOT~\cite{2021eagermot}}: A 2D + 3D fusion algorithm with detection-to-track matching. LiDAR-only tracking performs within 8.5\% of camera+LiDAR fusion and even exceeds it in precision. Significant noise in 2D detections has minimal impact (Fig.~\ref{fig:track-fusion-lidar}), while degradation in 3D detections greatly reduces performance, showing vulnerability to LiDAR-based attacks like the \emph{frustum attack}~\cite{2022hally-frustum}.

\vspace{4pt}
\noindent \textbf{DeepFusionMOT~\cite{wang2022deepfusionmot}}: Tracks 3D and 2D objects, with initial associations relying on 3D detections. Altering 2D detections without affecting 3D has little effect, but manipulating 3D data significantly degrades performance due to its critical role in initial associations.

\vspace{4pt}
\noindent \textbf{Joint Multi-Object Det. and Tracking (JMODT)~\cite{huang2021joint}}: Runs joint detection and tracking, using camera data to augment LiDAR features. Since tracking is primarily based on 3D LiDAR data, disrupting 2D data impacts outcomes less than compromising 3D LiDAR data.

\section{Attack Designs With Limited Information} \label{sec:5-attack-designs}

The prior analysis indicates that LiDAR in multi-sensor AVs is critical to the vehicle's Trusted Computing Base (TCB) for safety. We therefore examine LiDAR-based attacks within a restricted threat model where attackers can only access raw LiDAR data. Using the knowledge and capabilities defined in Sec.\ref{sec:4-cyber-model}, we design both "context-aware" attacks (requiring situational awareness) and "context-unaware" attacks. All attacks maintain first-order receiver integrity as outlined in Sec.~\ref{sec:cyber-attack-capabilities}.

\subsection{Attack Algorithms} \label{sec:attack-definitions}


\noindent \textbf{Context-unaware attacks:}
\begin{enumerate}[label=\textbf{ATT.\arabic*},leftmargin=1\parindent,topsep=0pt,itemsep=-1ex,partopsep=1ex,parsep=1ex,itemindent=15pt]
    \item \label{att:fp} \textbf{False Positive:} Manipulate the range within an angular subset (i.e., $(\theta_i,\ \phi_j)$ pairs) into some desired shape that is likely to be detected as an object by the AV.
    \item \label{att:fp-dual} \textbf{False Positive (Dual):} Change the mode bit from \texttt{single} to \texttt{dual}. Duplicate datagrams and manipulate the second copy, staying consistent with integrity~\ref{integ:dual}.
    \item \label{att:forreplay} \textbf{Forward Replay:} Duplicate existing datagrams; store the copy in a buffer. After some delay, attack by replaying from the buffer with consistent timestamps.
    \item \label{att:revreplay} \textbf{Reverse Replay:} \ref{att:forreplay}, but in reverse order.
\end{enumerate}

\vspace{6pt}
\noindent \textbf{Context-aware attacks} (i.e., with situational awareness):
\begin{enumerate}[label=\textbf{ATT.\arabic*},resume,leftmargin=1\parindent,topsep=0pt,itemsep=-1ex,partopsep=1ex,parsep=1ex,itemindent=15pt]
    \item \label{att:clean} \textbf{Clean-Scene:} Monitor the range of each $(\theta_i,\ \phi_j)$ pair. Find ranges that change between frames (e.g.,~may be an object). Replace those with ranges such that no object will be detected (i.e., act as `\emph{background}' ranges).
    \item \label{att:remove} \textbf{Object Removal:} Determine which $(\theta_i,\ \phi_j)$ pairs encapsulate an object. Over these angles, replace the range entries with estimated \emph{background} ranges.
    \item \label{att:frust-trans} \textbf{Frustum Translation:} Determine which $(\theta_i,\ \phi_j)$ pairs encapsulate an object. Over these angles, apply a shift function to points to translate the object forward or backward within the frustum (see Fig.~\ref{fig:attack-examples}, Appendix~\ref{app:attack_visualisations});
    \item \label{att:frust-fp-dual} \textbf{Frustum False Positive (Dual):} Determine which $(\theta_i,\ \phi_j)$ pairs encapsulate an object. Change from \texttt{single} to \texttt{dual}. Duplicate datagrams and use the second copy for a false positive behind the true object.
\end{enumerate}

\noindent We do not evaluate all proposed attacks in this work. Attacks relying on \texttt{dual} mode (e.g., \ref{att:fp-dual} and \ref{att:frust-fp-dual}) are excluded since \texttt{dual} mode is uncommon in AVs, as well as the \emph{clean scene} attack~\ref{att:clean}, which requires further investigation for monitoring solutions.

Hence, from context-unaware attacks, we consider the \emph{false positive}~\ref{att:fp}, \emph{forward replay}~\ref{att:forreplay}, and \emph{reverse replay}~\ref{att:revreplay} attacks. From context-aware attacks, we consider the \emph{object removal}~\ref{att:remove} and \emph{frustum translation}~\ref{att:frust-trans} attacks. These attacks are well-motivated from prior security studies; \ref{att:fp} is an instance of~\cite{2019cao-spoofing, 2020sun-spoofing}; replay attacks have been studied in estimation applications (e.g.,~\cite{2009mo_replay, 2013stochasticreplay}); and \ref{att:frust-trans} borrows ideas from~\cite{2022hally-frustum}.

\subsection{Attack Implementations} \label{sec:attack-imp}

We partition the attacker's decision process into three sequential modules, executed at every time step:
\begin{enumerate}[topsep=0pt,itemsep=-1ex,partopsep=1ex,parsep=1ex]
    \item \textbf{Monitor:} Gain insight on scene details (e.g., ground plane, objects) from raw sensor (i.e., LiDAR) data.
    \item \textbf{Schedule:} Use monitoring information to plan attack over both spatial and temporal dimensions.
    \item \textbf{Execute:} Compose attack operations against LiDAR data to best instantiate the scheduled plan.
\end{enumerate}

\begin{figure}[!t]
    \centering
    \includegraphics[trim={0 0 2cm 0},width=0.6\linewidth]{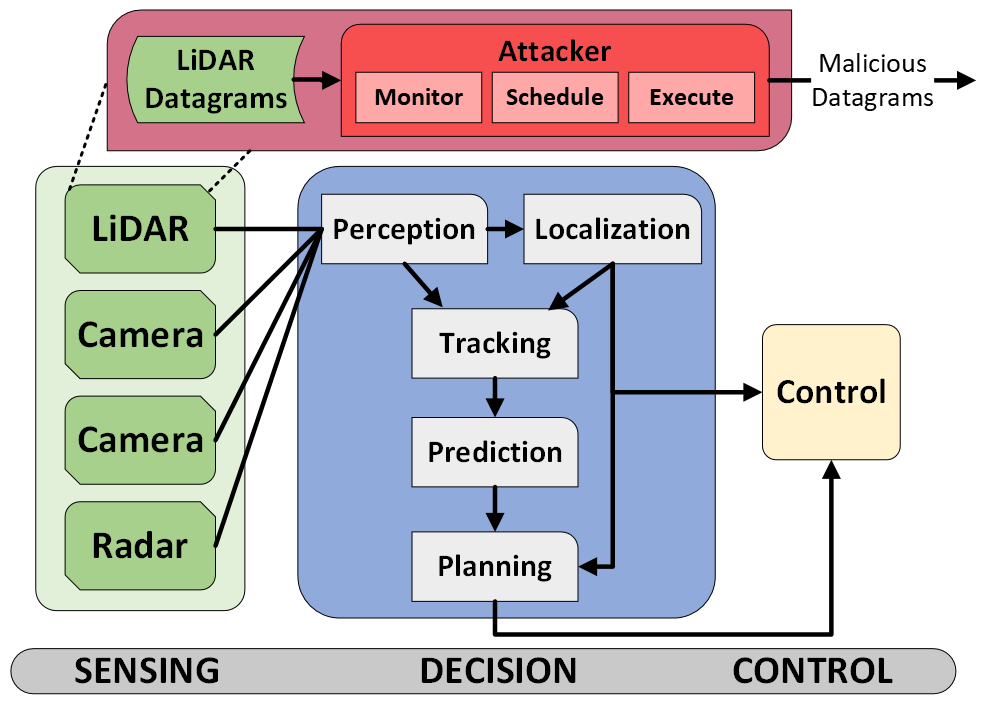}
    \caption{Even  compromising only LiDAR data percolates through AV pipeline despite the multiple sensors' data use. Incoming data packets are the attacker's only knowledge.}
    \label{fig:av-attack-model}
\end{figure}

Algorithm~\ref{alg:attacker} shows the interfaces between the partitioned modules; partitioning is a notational convenience that allows for clearer description of attack implementations. Such attacks take the form summarized in Fig.~\ref{fig:av-attack-model} with the incoming data packets the attacker's only knowledge. We describe attack module implementation (Monitor, Schedule, Execute) for the five selected attacks.

\begin{algorithm}[!t]
    \caption{LiDAR-level attacker specified in the module-based framework. Attacker may not need to monitor ``obj\_tracks'' for some attacks (only for some ``context-aware'' attacks). T\_L2G is the transformation from LiDAR sensor to ground plane.}
    \label{alg:attacker}
    \begin{algorithmic}[1]
    \renewcommand{\algorithmicrequire}{\textbf{Input:}}
    \renewcommand{\algorithmicensure}{\textbf{Output:}}
    \REQUIRE LiDAR datagram $D_k$, frame index $k\geq0$, obj\_tracks
    \ENSURE Compromised LiDAR datagram $D_k^a$, obj\_tracks
    \STATE T\_L2G,\, obj\_tracks $\leftarrow$ \texttt{MONITOR}($D_k$, obj\_tracks, k)
    \IF {T\_L2G is not \texttt{NULL}}
        \STATE attack\_sched $\leftarrow$ \texttt{SCHEDULE}(T\_L2G, obj\_tracks,\, k)
        \IF {attack\_sched is not \texttt{NULL}}
            \STATE $D_k^a$ $\leftarrow$ \texttt{EXECUTE}($D_k$,\, attack\_sched)
        \ENDIF
    \ENDIF
    \RETURN $D_k^a$, obj\_tracks
    \end{algorithmic} 
\end{algorithm}

\subsubsection{Monitor}

LiDAR monitoring enables attackers to gain situational awareness, optimizing attack placement and timing.

\begin{figure}[!t]
    \centering
    \includegraphics[width=\linewidth]{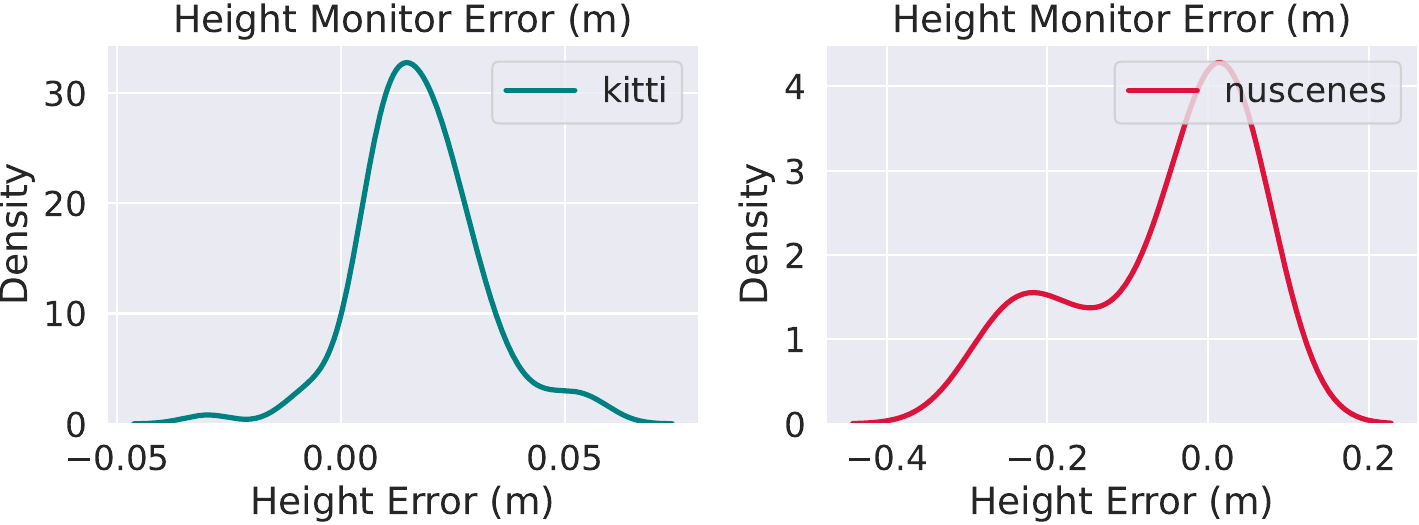}
    \caption{Attacker can use simple geometry on the point cloud or datagrams to estimate the sensor height: the height error distribution on KITII (left) and nuScenes (right) datasets. Estimate has low error on average and can be computed quickly without prior information.}
    \label{fig:sensor_height_monitor}
\end{figure}

\vspace{4pt}
\noindent \textbf{Sensor Height/Ground Plane} \ The attacker doesn’t know the sensor height initially. For attacks adding false objects, estimating sensor height via LiDAR monitoring ensures points follow physical constraints. The attacker assumes that many lowest-elevation-angle returns come from the ground plane. Sensor height $(h)$ is calculated as $h = \rho \sin{\phi}$ from elevation ($\phi$) and range ($\rho$). Errors are minimal, as in Fig.\ref{fig:sensor_height_monitor}, and applying this method to KITTI and nuScenes datasets does not affect attack performance.

\vspace{4pt}
\noindent \textbf{Object Detection} \ Context-aware attacks require monitoring objects. Once a high-confidence scenario is detected, the attacker acts. We use the lightweight SECOND algorithm~\cite{2018second} for LiDAR detection, though real-time DNN execution may be limited without parallel hardware. Model compression (e.g.,~\cite{2022deeplearningonmobile}) can reduce computation 16-fold with minimal performance loss on KITTI and nuScenes. Alternatively, attackers can monitor simplified 2D or top-view projections.

\subsubsection{Schedule}

We outline key scheduling tools for attack planning and the specific scheduling methods for each attack. Attack schedules include a \emph{stable} phase, where the attacker establishes an effect (e.g., starting a false track), and an \emph{attacking} phase, where this effect is exploited for impact (e.g., moving a false track toward a target).

\vspace{4pt}
\noindent \textbf{Kinematics Models} \ To simulate imminent collisions, we design false object paths using constant-jerk kinematics, which has proven effective for longitudinal attacks. Under this model, objects will be accelerating at an increasing rate, causing the appearance of imminent danger for the victim; the kinematic parameters evolve as:
\begin{equation*}
    \begin{aligned}
    j & \coloneqq 6\frac{\rho_n - \rho_0}{\Delta T_{\text{attack}}^3},\ \quad     a_k = a_{k-1} + j \Delta t\\
    v_k &= v_{k-1} + \frac{a_k + a_{k-1}}{2} \Delta t + \frac{1}{2} j \Delta t^2\\
    r_k &= r_{k-1} + \frac{v_k + v_{k-1}}{2} \Delta t + \frac{1}{2} \frac{a_k + a_{k-1}}{2} \Delta t^2 + \frac{1}{6} j \Delta t^3,
\end{aligned}
\end{equation*}
where $\rho_0$, $\rho_n$ are the attacker's choice of initial/final ranges of the false object, $\Delta t$ is time interval, $\Delta T_\text{attack}$ is the duration of the entire attack, $k$ is the index of the current frame, and $j,\,a_k\,v_k\,r_k$ are the jerk, acceleration, velocity, and range for the false object.

\vspace{4pt}
\noindent \textbf{Target Object Selection} \ Attacks like \emph{object removal} (\ref{att:remove}) and \emph{frustum translation} (\ref{att:frust-trans}) focus on single-object targeting. Despite reduced model accuracy, monitoring-based features (e.g., object lifetime, range, velocity) achieve a 90\% true object selection rate. The attacker bypasses perception noise, selecting targets based on favorable conditions. Context-aware attacks score objects on tracked features, scaling and applying a sigmoid to obtain a final score; the object with the lowest score is targeted. If no target is found, scores are recalculated in the next frame. Once selected, if the target is lost, its location is estimated until the attack.

\begin{itemize}[leftmargin=12pt,topsep=0pt,itemsep=-1ex,partopsep=1ex,parsep=1ex]
    \item \textbf{Object lifetime}: Last min of 4 frames to avoid false pos.
    \item \textbf{Angular offset}: Within $[-15^{\circ}, 15^{\circ}]$ from forward direction for safety impact.
    \item \textbf{Range}: Within $[5, 40]$ meters for likely safety relevance.
    \item \textbf{Lateral velocity}: Between $[-1, 1]$ m/s relative to victim to avoid perpendicular traffic.
    \item \textbf{Forward velocity}: Between $[-2, 5]$ m/s to avoid targets soon exiting.
\end{itemize}

\begin{figure}[!t]
    \centering
    \begin{subfigure}[b]{0.58\linewidth}
        \centering
        \includegraphics[width=0.80\linewidth]{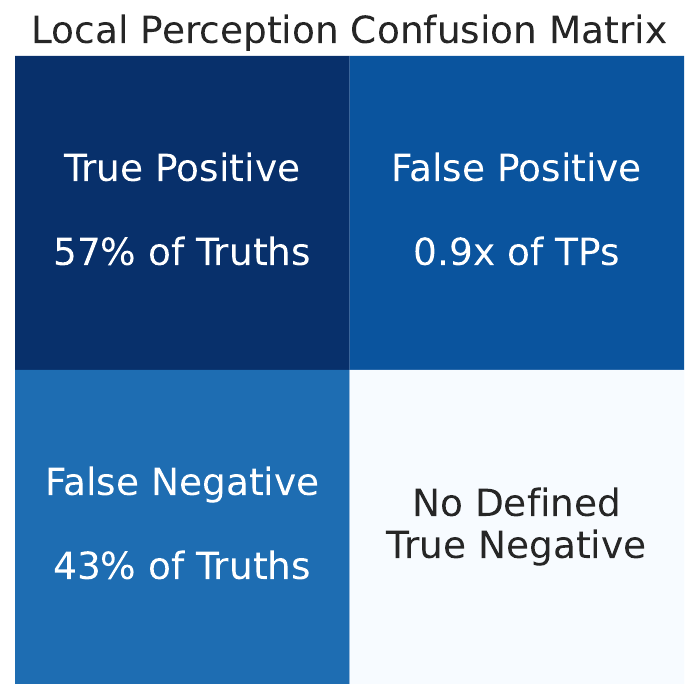}
        \caption{Attacker only has access to one sensor (LiDAR) for building scene knowledge, leading to noisier situational awareness.}
        \label{fig:attacker-local-percep}
    \end{subfigure}
    \hfill
    \begin{subfigure}[b]{0.368\linewidth}
        \centering
        \includegraphics[width=0.8\linewidth]{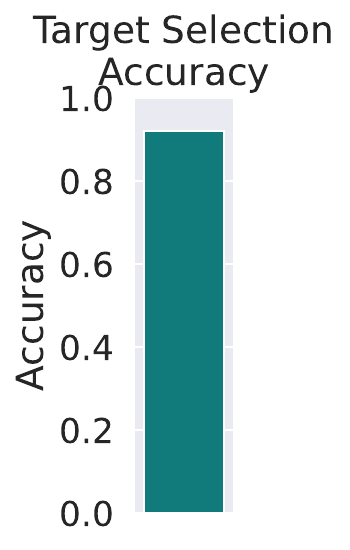}
        \caption{Despite noisy object detection, tracking and scoring lead to accurate target selection.}
        \label{fig:attacker-local-selection}
    \end{subfigure}
    \caption{The attacker uses only a single sensor (LiDAR) with lightweight perception due to computational limits. (a) This results in noisy perception, but perfect object detection is not required. (b) The attacker focuses on \emph{waiting for the ``right'' moment to attack}, using tracking and scoring to ensure targeted objects (e.g., for object removal, frustum translation) are \emph{real} and not false positives.}
    \label{fig:attacker-local}
\end{figure}

In longitudinal cases with noisy perception algorithms, the attacker’s perception achieves a 57\% true positive rate, with many false positives. However, by using tracking and scoring, the attacker selects a true target with 95\% accuracy, as shown in Fig.~\ref{fig:attacker-local-selection}.

\noindent \textbf{Scheduling For False Positive~\ref{att:fp}} \ Requires ``stable'' and ``attack'' durations, plus initial and final positions, and a kinematics model. Parameters: $\Delta T_{\text{stable}}=2.5$ s, $\Delta T_{\text{attack}}=4.5$ s, with ranges $\rho_0=15$ m to $\rho_n=1$ m.

\noindent \textbf{Scheduling For Forward Replay Attack~\ref{att:forreplay}} \ Uses a 40-frame buffer in KITTI and a 15-frame buffer in nuScenes, playing back from the start of the buffer until it exhausts.

\noindent \textbf{Scheduling For Reverse Replay Attack~\ref{att:revreplay}} \ Uses the same buffer but replays frames in reverse, avoiding discontinuities from starting at the beginning of the buffer.

\noindent \textbf{Scheduling For Object Removal Attack~\ref{att:remove}} \ Selects a target based on monitoring scores, initiating removal based on the highest feature score.

\noindent \textbf{Scheduling For Frustum Translation Attack~\ref{att:frust-trans}} \ Combines \emph{object removal} and \emph{FP} schedules, using a jerk model to move the false object in alignment with the true object’s angular consistency.

\subsection{Execute}

Attack execution is decomposed into reusable subroutines that act as building blocks. Table~\ref{table:execute-algorithms} illustrates these compositions. \emph{Notation:} ``Missing angles'' refer to angular pairs that register as \texttt{NULL} in the point cloud matrix. The ``get point mask from X' subroutine creates a Boolean mask for points meeting criterion X. The ``inpaint mask as X from Y'' subroutine uses information from Y to modify masked points to mimic X. A ``trace'' is a subset of points representing an object or shape.

\begin{table}[!t]
    \centering
    \caption{Attack executions are constructed from subroutines. Frustum-type attacks use other attacks as subroutines.}
    \label{table:execute-algorithms}
    \resizebox{\columnwidth}{!}{
    \begin{tabular}{lll}
    \toprule
    Num. & Att. Case Name & Subroutines \\ 
    \midrule \midrule
    \ref{att:fp} & False Positive & \tworowsubtableleft{\tworowsubtableleft{\texttt{FindMissingAngles}}{\texttt{GetPointMaskFromTrace}}}{\texttt{InpaintMaskAsObjectFromTrace}} \\ 
    \midrule
    \ref{att:fp-dual} & Dual False Positive & \tworowsubtableleft{\tworowsubtableleft{\texttt{FindMissingAngles}}{\texttt{GetPointMaskFromTrace}}}{\texttt{InpaintMaskAsObjectFromTrace}} \\
    \midrule
    \ref{att:forreplay} & Forward Replay & N/A \\
    \midrule
    \ref{att:revreplay} & Reverse Replay & N/A \\
    \midrule
    \ref{att:clean} & Clean Scene & \texttt{InpaintMaskAsBackgroundFromContext}\\
    \midrule
    \ref{att:remove} & Object Removal & \tworowsubtableleft{Object Detection, Tracking}{\tworowsubtableleft{\texttt{GetPointMaskFromObject}}{\texttt{InpaintMaskAsBackgroundFromContext}}} \\
    \midrule
    \ref{att:frust-trans} & \tworowsubtableleft{Frustum}{Translation} & \tworowsubtableleft{Object Removal}{False Positive} \\
    \midrule
    \ref{att:frust-fp-dual} & \tworowsubtableleft{Dual Frustum}{False Positive} & \tworowsubtableleft{Object Removal}{False Positive} \\
    \bottomrule
    \end{tabular}
    }
\end{table}

\noindent \labeltext{\texttt{SUB.1}}{sub:missing-angles} \texttt{SUB.1: FindMissingAngles} \ Identify missing angular pairs in a noisy point cloud by creating a bivariate grid of angles. Points are assigned to cells via L2 norm, and the grid is dilated to address noise-related gaps, leaving a matrix of missing (non-returned) measurements.

\noindent \labeltext{\texttt{SUB.2}}{sub:point-mask-trace} \texttt{SUB.2: GetPointMaskFromTrace} \ Identify indices in the point cloud that align with a trace to replace them, using azimuth and elevation bounds and Delaunay triangulation to define the point mask.

\noindent \labeltext{\texttt{SUB.3}}{sub:point-mask-object} \texttt{SUB.3: GetPointMaskFromObject} \ Identify points within a 3D object bounding box using Delaunay triangulation over the bounding box, excluding points outside the convex hull in angular space.

\noindent \labeltext{\texttt{SUB.4}}{sub:inpaint-mask-trace} \texttt{SUB.4: InpaintMaskAsObjectFromTrace} \ Set ranges in the masked points to approximate a trace by querying a bivariate B-spline function over the angles, replacing original ranges with those that form the trace shape.

\noindent \labeltext{\texttt{SUB.5}}{sub:inpaint-mask-background} \texttt{SUB.5: InpaintMaskAsBackgroundFromContext} \ Mimic the background for masked points using a kNN regression model built from surrounding non-masked points to fill ranges at masked $(\theta, \phi)$ pairs.

\begin{figure}[!t]
    \centering
    
    \suppsubfig{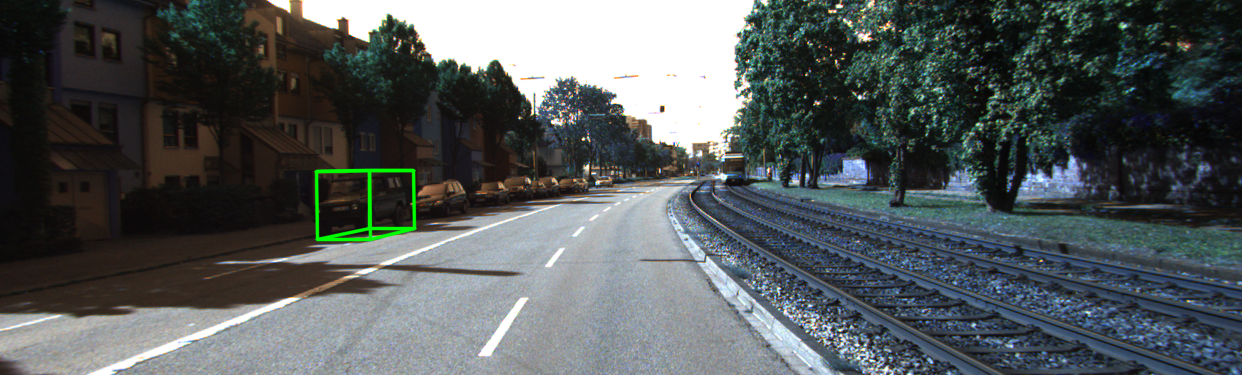}{Original image\\with object box.}
    \suppsubfig{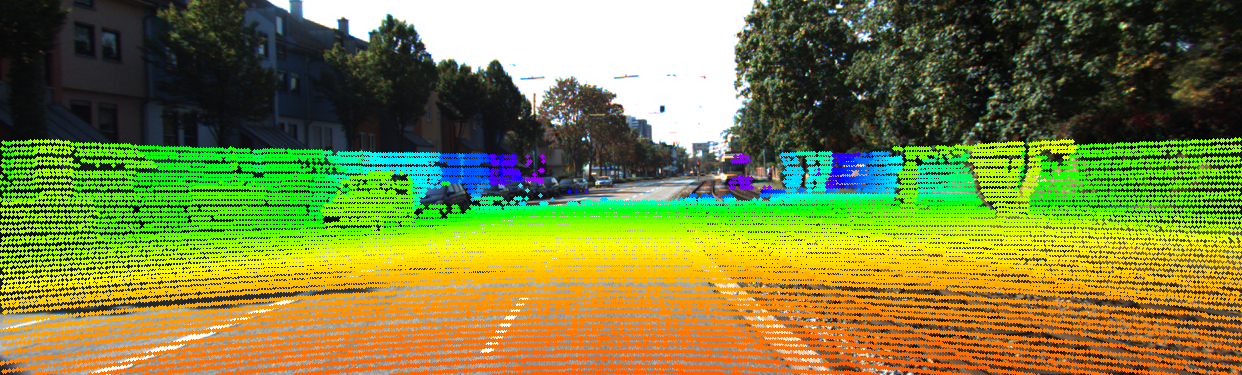}{Original image\\with original point\\cloud.}

    \suppsubfigwithtikz{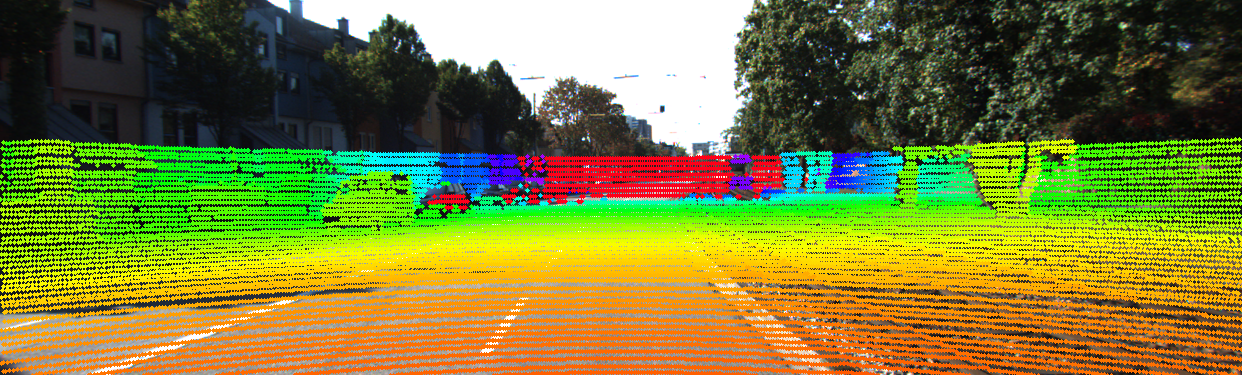}{\ref{sub:missing-angles} Missing\\points (angular pairs)\\identified in red.}{70}{18}{0}{0}
    \suppsubfigwithtikz{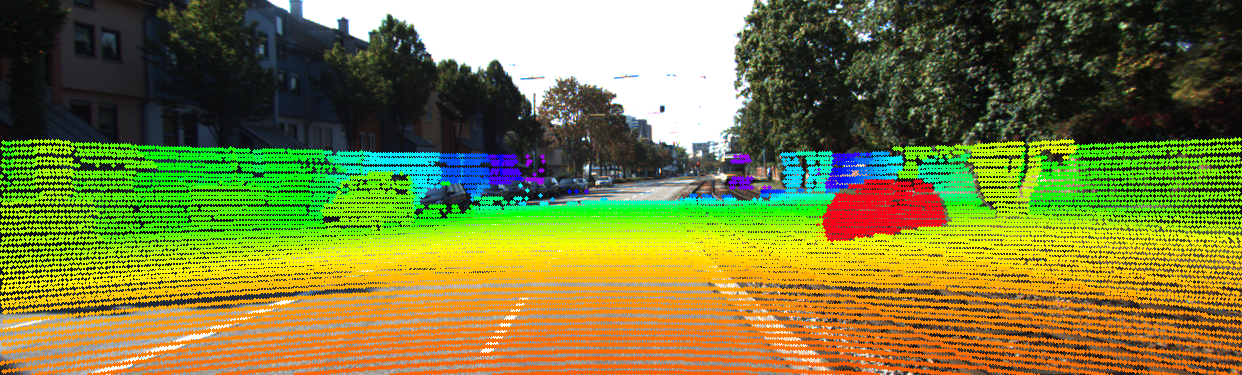}{\ref{sub:point-mask-trace} External\\trace masked into\\point cloud in red.}{30}{20}{28}{-3}
    \suppsubfigwithtikz{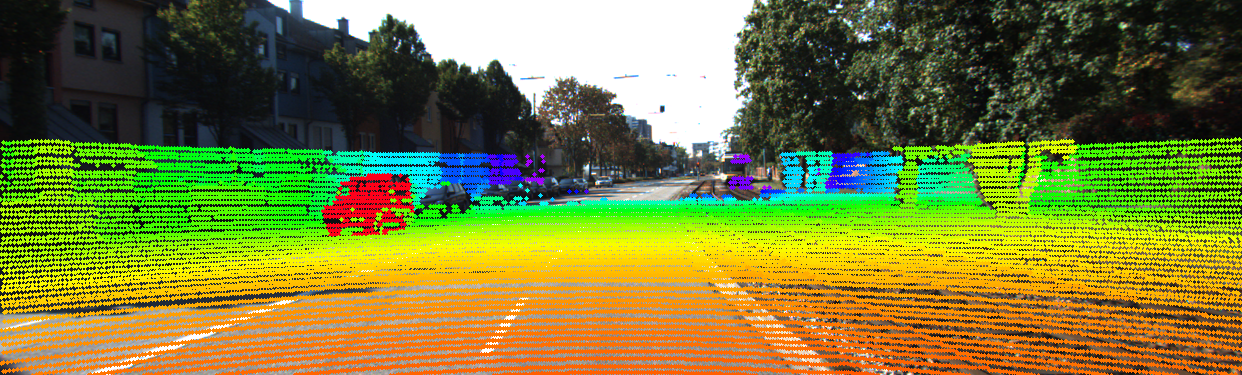}{\ref{sub:point-mask-object} Object\\points masked in\\point cloud in red.}{30}{20}{-28}{-3}
    \suppsubfigwithtikz{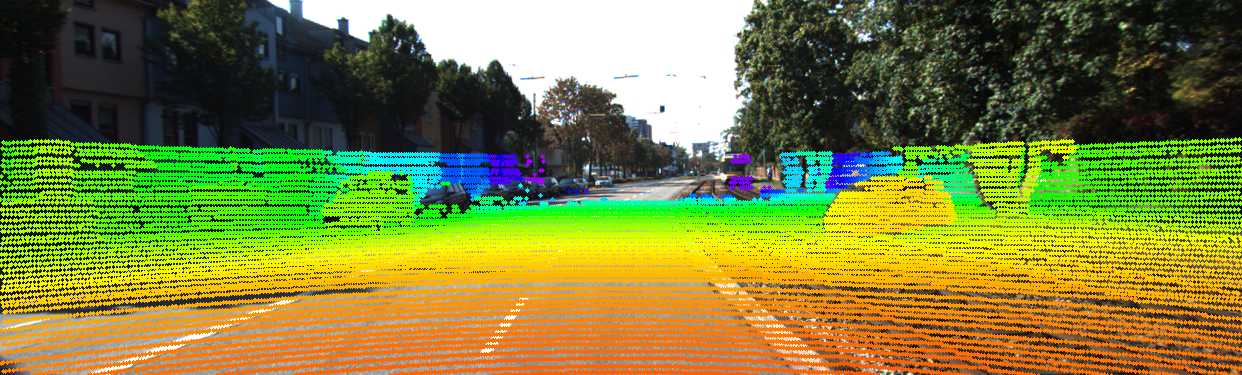}{\ref{sub:inpaint-mask-trace} External\\trace inpainted into\\point cloud.}{30}{20}{28}{-3}
    \suppsubfigwithtikz{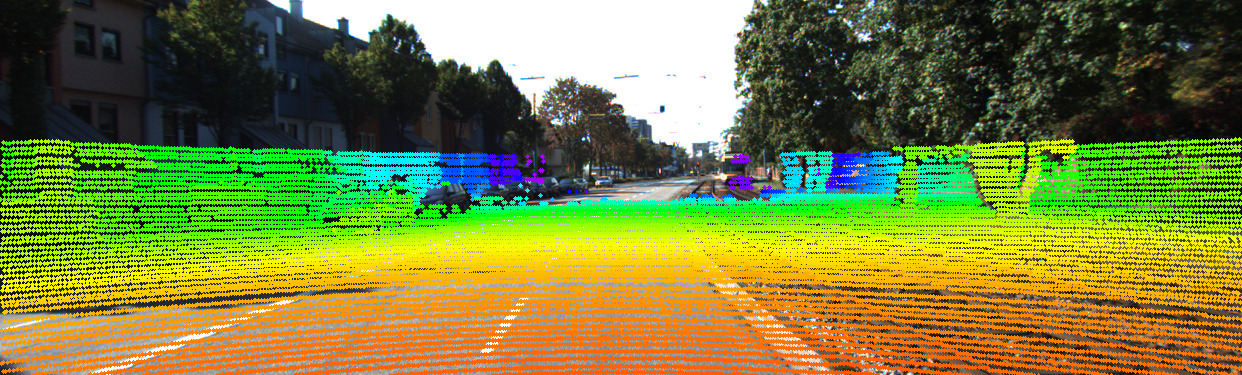}{\ref{sub:inpaint-mask-background} Object\\points inpainted as\\background.}{30}{20}{-28}{-3}
    \caption{Example of attack subroutines in (c)-(g).}
    \label{fig:execute-subroutines}
\end{figure}

Examples of these subroutines are shown in Fig.\ref{fig:execute-subroutines}. For instance, if injecting a fake object, the attacker first locates points to replace using\ref{sub:point-mask-trace} and then modifies their ranges to match the fake object shape via~\ref{sub:inpaint-mask-trace}.

\captionsetup{position=top}


\subsection{Practical Considerations} \label{sec:practical-considerations}

We test attacks on datasets and simulators instead of physical platforms due to the high cost of LiDAR. Commonly used 32 and 64-line scanners, such as Velodyne HDL-32E and HDL-64E, cost over \$30,000 and \$75,000, making them inaccessible for research. The affordable 16-line Velodyne Puck is not representative of typical AVs.

\section{Case Studies} \label{sec:7-case-studies}

On several case studies, we evaluate the effectiveness of attacks on LiDAR given different AV architectures.

\subsection{Baidu Apollo \& Carla Simulator}

We test our attacks on Baidu's industry-grade Apollo AV software~\cite{BaiduApollo}, configured with the latest LiDAR-based and camera-LiDAR fusion v7.0 vehicles. We evaluate two context-unaware attacks on Apollo: the \emph{false positive} and \emph{reverse replay} attacks (\ref{att:revreplay} and \ref{att:fp}). The attacks are executed running in a \texttt{docker} container with the \texttt{socket} library. The attacker intercepts point clouds over the network, modifies them according to their capabilities, and sends them back to the Apollo agent for processing.

Results are presented in Fig.\ref{fig:baidu-apollo}, demonstrating the attacks on the camera-LiDAR fusion agent. Both attacks succeed despite limited prior knowledge. The FP attack triggers unnecessary emergency braking (Fig.~\ref{fig:baidu-apollo-2}), while the reverse replay attack results in a collision due to object detection errors (Fig.~\ref{fig:baidu-apollo-4}).

\renewcommand{\subfigwidth}{0.24}
\renewcommand{\graphicswidth}{0.8}

 \begin{figure*}[!t]
    \centering
    \begin{subfigure}[b]{\subfigwidth \linewidth}
        \centering
        \includegraphics[trim={10cm 5cm 10cm 5cm},clip,width=\graphicswidth \linewidth]{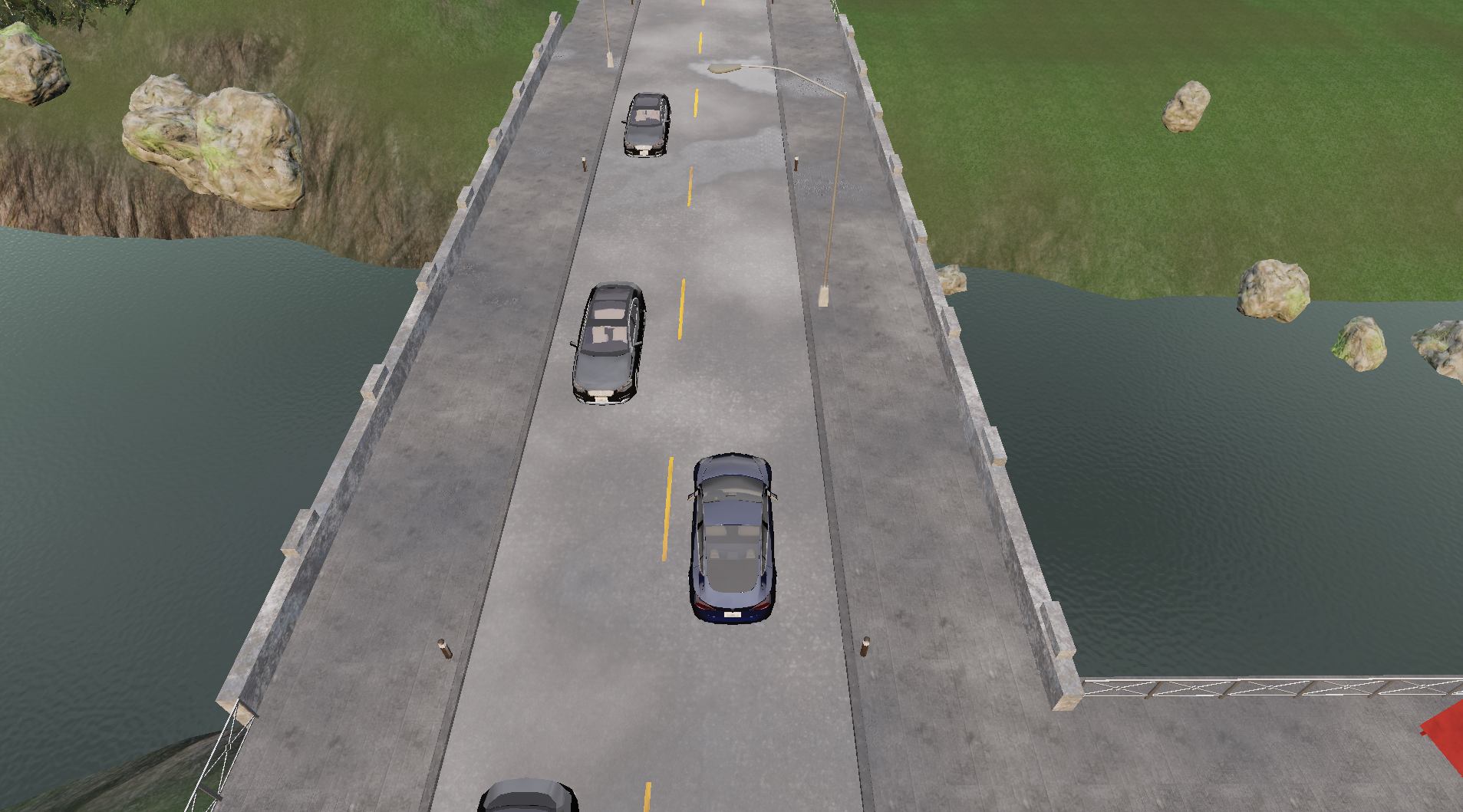}
    \end{subfigure}
    \begin{subfigure}[b]{\subfigwidth \linewidth}
        \centering
        \includegraphics[trim={10cm 5cm 10cm 5cm},clip,width=\graphicswidth \linewidth]{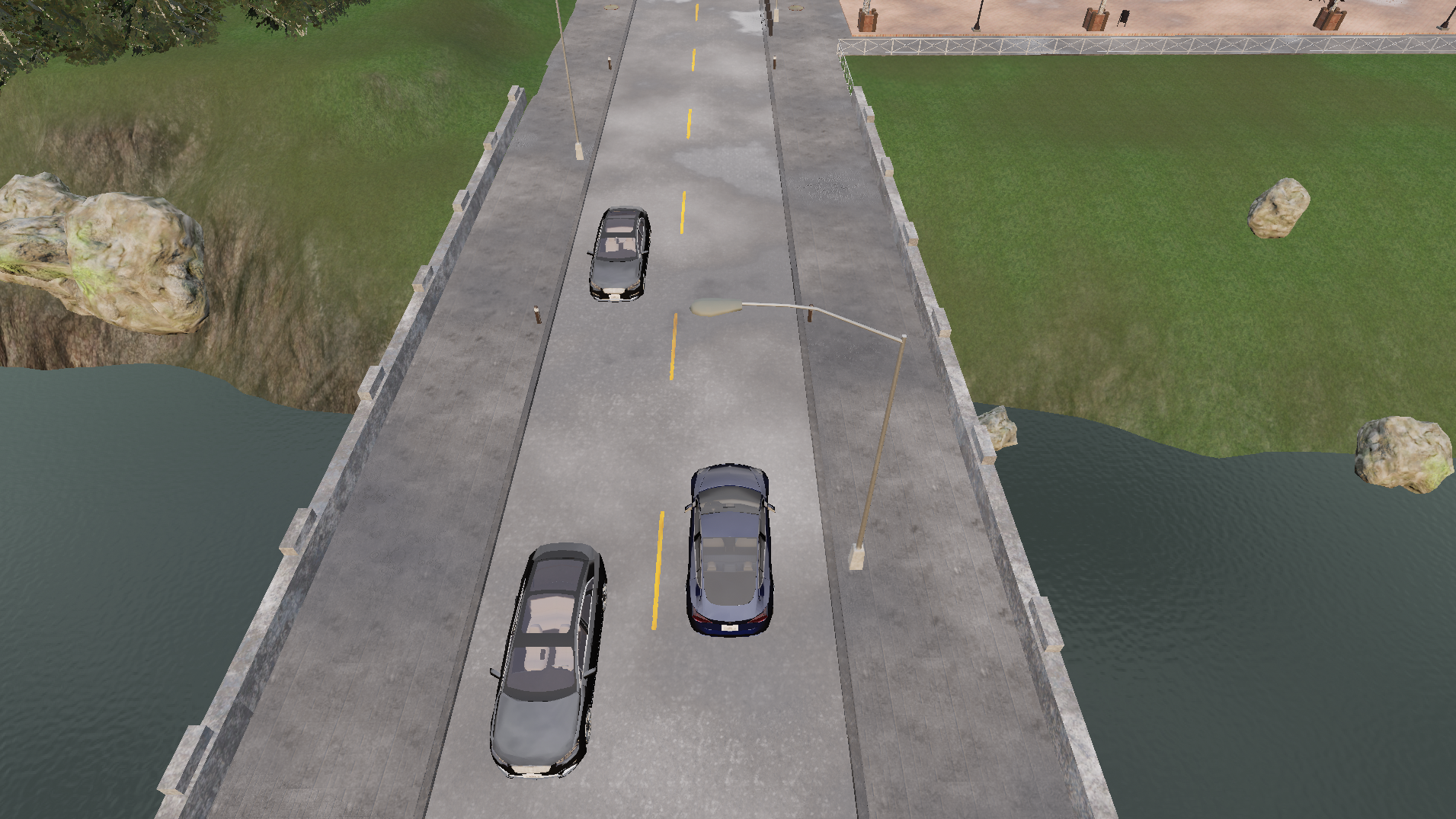}
    \end{subfigure}
    \begin{subfigure}[b]{\subfigwidth \linewidth}
        \centering
        \includegraphics[trim={10cm 5cm 10cm 5cm},clip,width=\graphicswidth \linewidth]{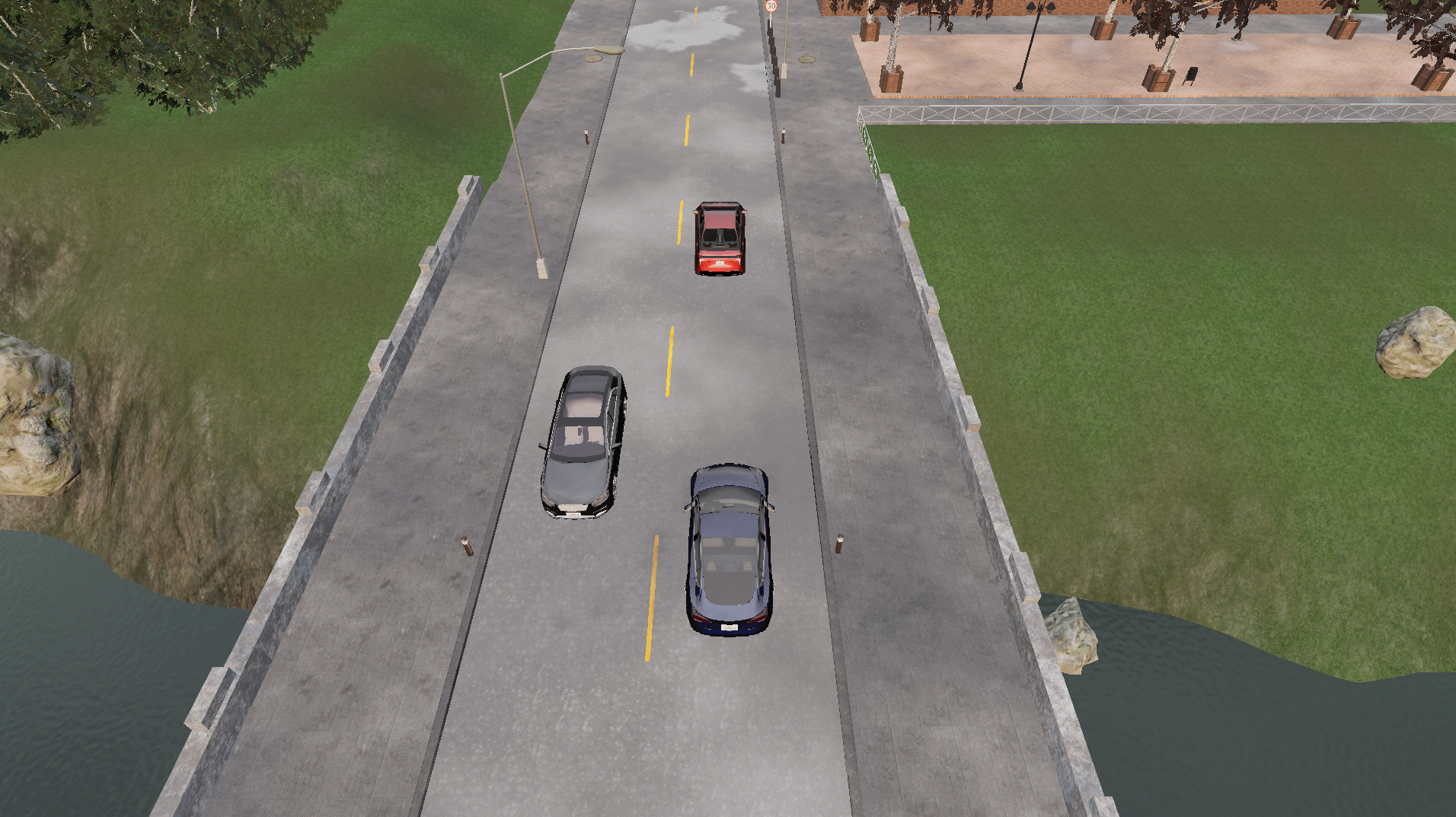}
    \end{subfigure}
    \begin{subfigure}[b]{\subfigwidth \linewidth}
        \centering
        \includegraphics[trim={10cm 5cm 10cm 5cm},clip,width=\graphicswidth \linewidth]{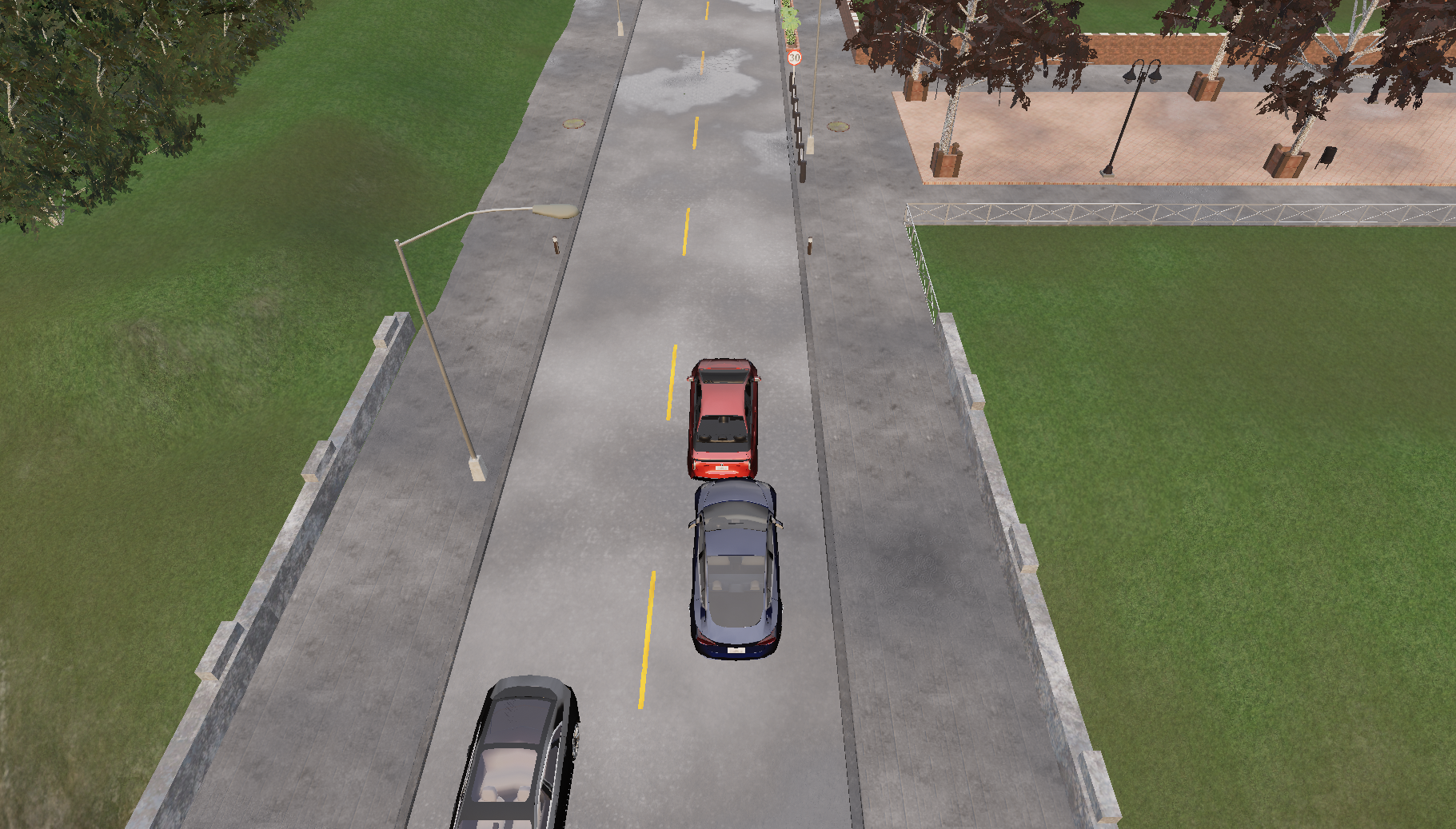}
    \end{subfigure}
    \begin{subfigure}[b]{\subfigwidth \linewidth}
        \centering
        \includegraphics[trim={20cm 5cm 20cm 15cm},clip,width=\graphicswidth \linewidth]{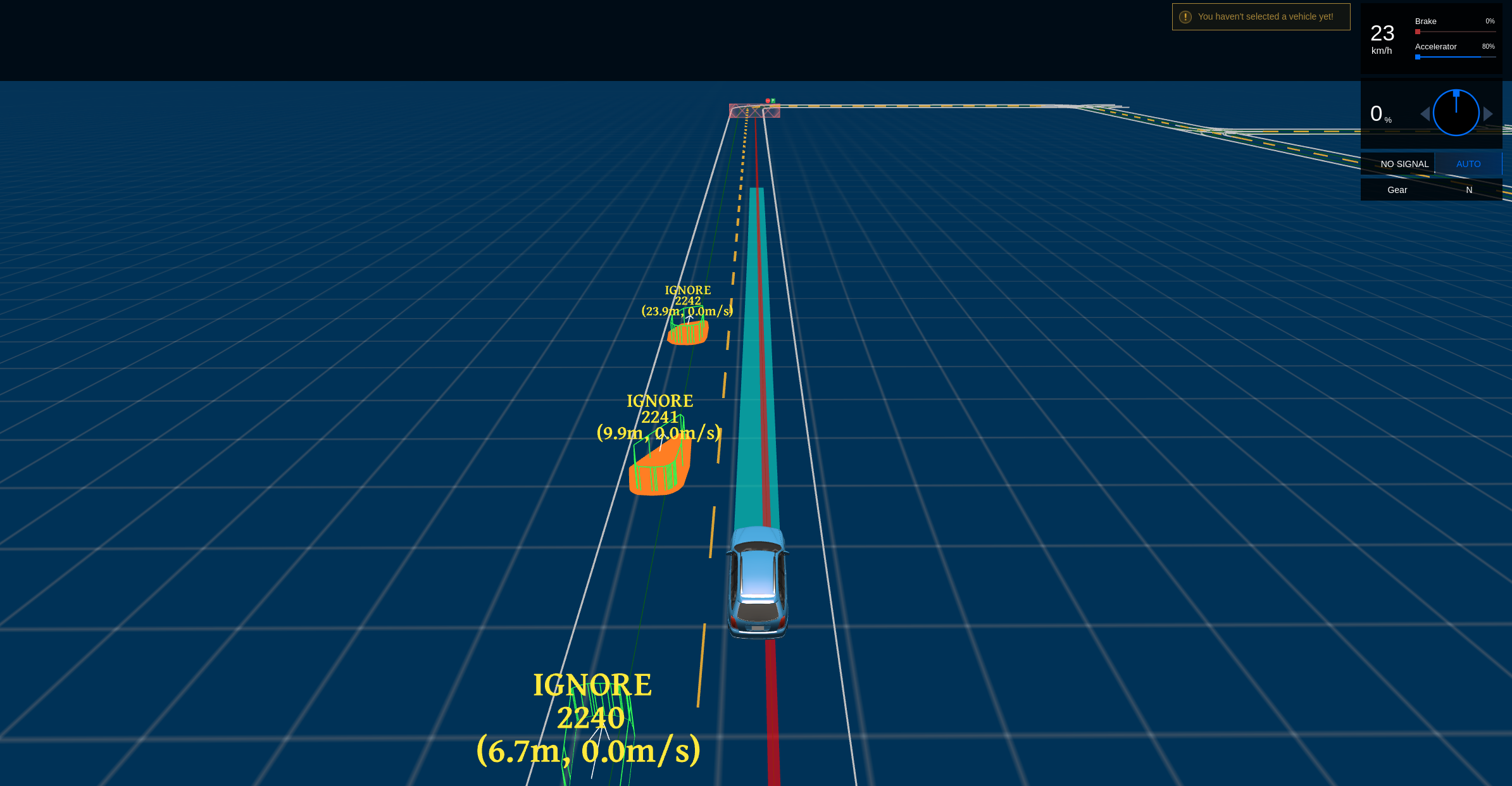}
        \caption{Unattacked. Ego sees lane is clear and plans straight path.}
        \label{fig:baidu-apollo-1}
    \end{subfigure}
    \begin{subfigure}[b]{\subfigwidth \linewidth}
        \centering
        \includegraphics[trim={20cm 5cm 20cm 15cm},clip,width=\graphicswidth \linewidth]{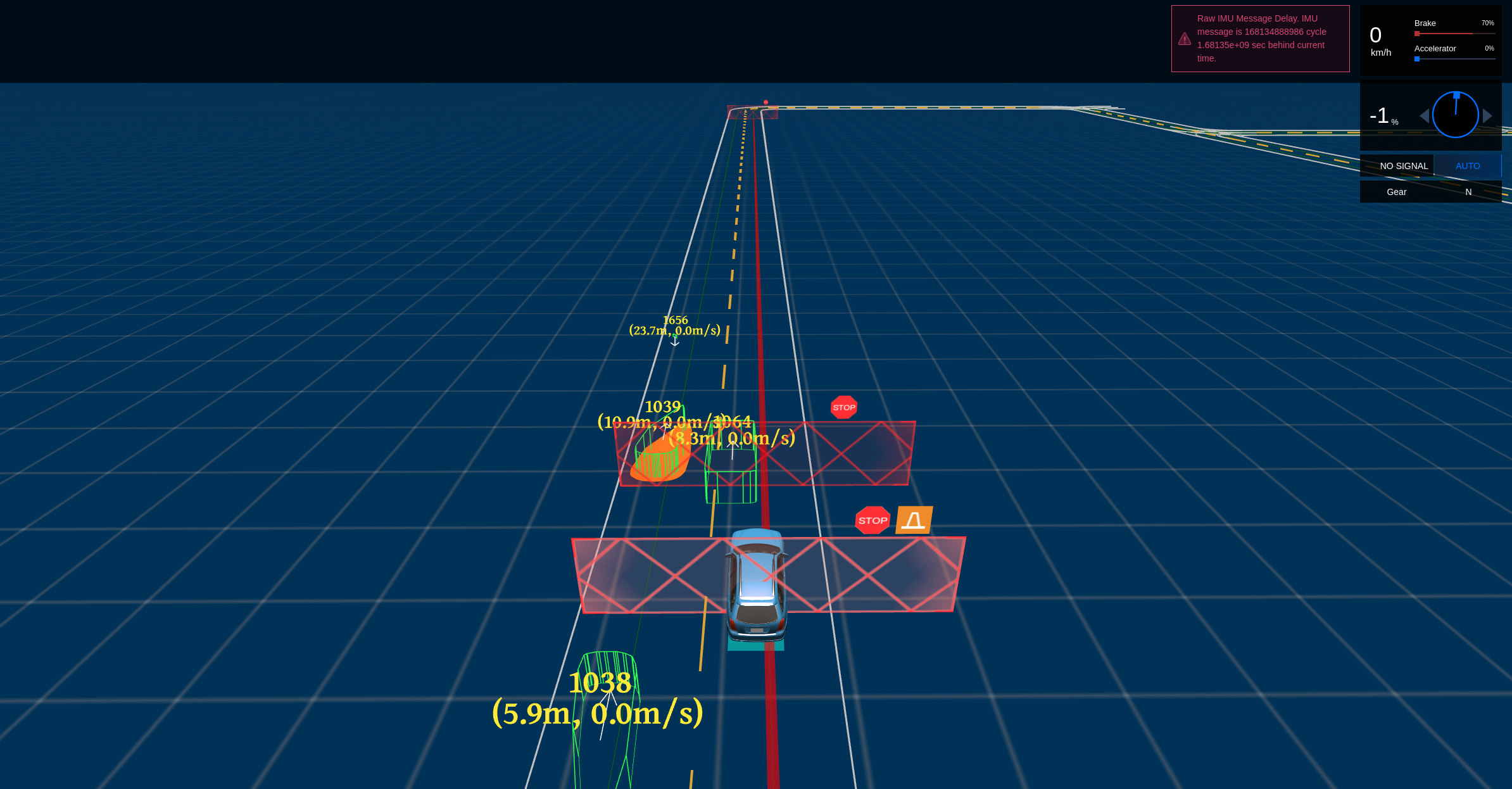}
        \caption{FP attack~\ref{att:fp}. Ego emergency brakes to avoid fake obj.}
        \label{fig:baidu-apollo-2}
    \end{subfigure}
    \begin{subfigure}[b]{\subfigwidth \linewidth}
        \centering
        \includegraphics[trim={20cm 5cm 20cm 15cm},clip,width=\graphicswidth \linewidth]{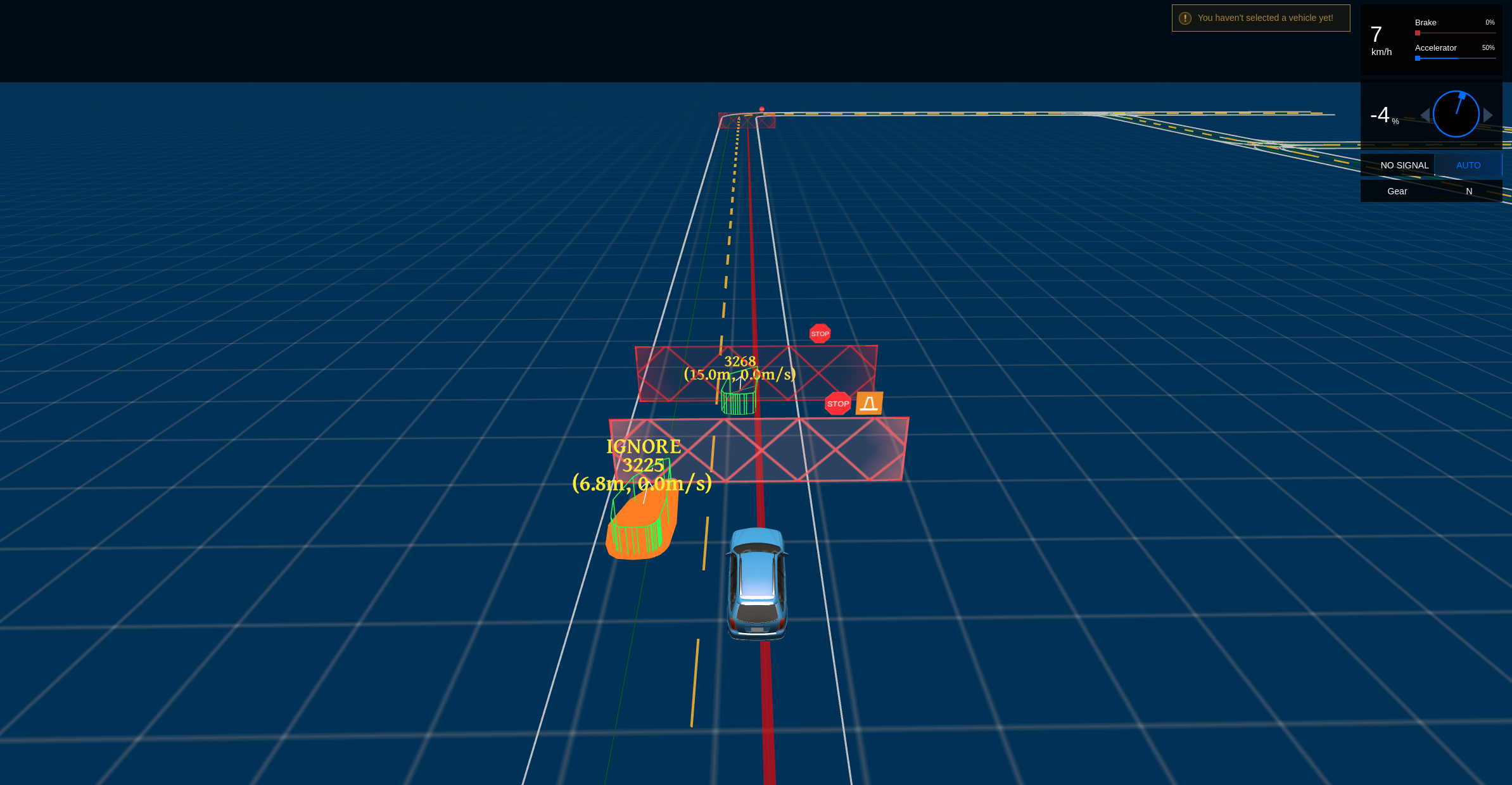}
        \caption{Unattacked. Ego stops ahead of existing stopped car.}
        \label{fig:baidu-apollo-3}
    \end{subfigure}
    \begin{subfigure}[b]{\subfigwidth \linewidth}
        \centering
        \includegraphics[trim={20cm 5cm 20cm 15cm},clip,width=\graphicswidth \linewidth]{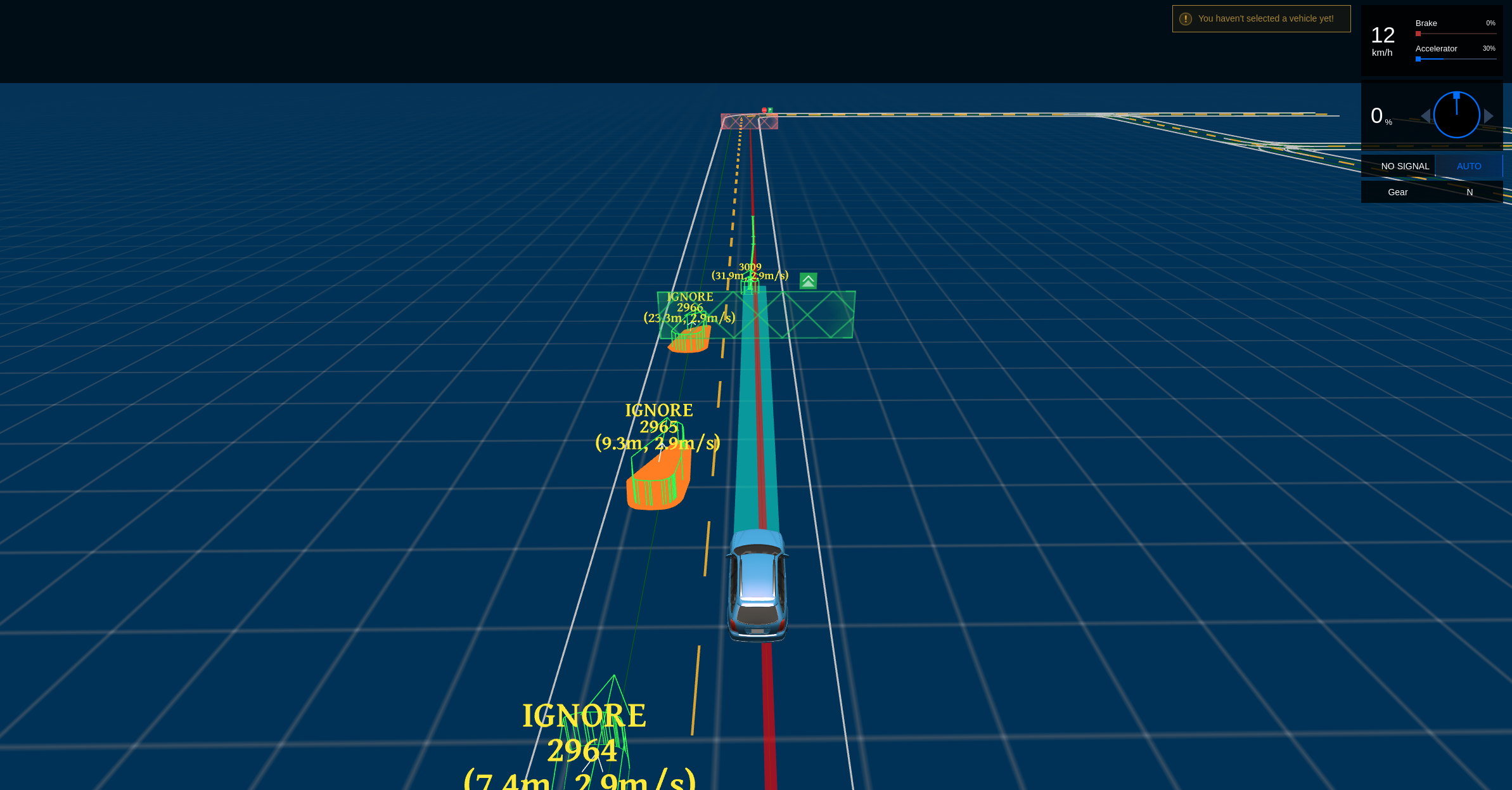}
        \caption{Rev-replay attack~\ref{att:revreplay}. Ego crashes into stopped car.}
        \label{fig:baidu-apollo-4}
    \end{subfigure}
    \caption{Cyber attacks~\ref{att:fp} and~\ref{att:revreplay} are successful against the industry-grade Apollo driving software. Our attacks run in a \texttt{docker} container and intercept datagrams with the \texttt{socket} library as they travel from sensing to computing. (top) simulation view, (bottom) camera-LiDAR fusion perception, tracking. Results against fusion suggest Apollo places large amount of confidence on LiDAR rather than camera for 3D perception.}
    \label{fig:baidu-apollo}
\end{figure*}

\subsection{KITTI \& nuScenes Datasets}
We use longitudinal scenes from KITTI and nuScenes to test attack effectiveness. Prior works have explored the vulnerability of LiDAR-only perception and primitive camera-LiDAR fusion on naive and frustum attacks~\cite{2020sun-spoofing, 2022hally-frustum}. Our case studies focus on novel contributions in understanding defenses. The results are summarized below.
\begin{enumerate}[leftmargin=10pt,labelsep=\labelspace pt,align=left,label=\textbf{Case \Roman*},topsep=0pt,itemsep=-1ex,partopsep=1ex,parsep=1ex]
    \item Reverse replay \ref{att:revreplay} against data asymmetry \ref{av:cam-lidar-v2} architecture. Data asymmetry may fail because camera still partially consistent with reversed LiDAR.
    \item Reverse replay \ref{att:revreplay} against T2T-3DLM \ref{av:ttt} design. T2T-3DLM will succeed because object 3D positions will disagree.
    \item Frustum translation \ref{att:frust-trans} against data asymmetry \ref{av:cam-lidar-v2} architecture. Data asymmetry may fail because camera will be consistent with frustum LiDAR attack.
    \item Frustum translation \ref{att:frust-trans} against T2T-3DLM \ref{av:ttt}. T2T-3DLM will succeed because object 3D positions will disagree.
\end{enumerate}

\subsection{Cases I and II: Reverse Replay~\ref{att:revreplay}}
The attacker exploits the 2D projection from camera data, achieving this solely by manipulating LiDAR data without situational awareness. Since AVs typically follow low-curvature paths, replaying data in reverse creates significant overlap between the 2D projections of an object’s forward and reverse paths. Results are shown in Figs.~\ref{fig:case_AV3_ATT3_track},~\ref{fig:case_AV4_ATT3_track}, with visualizations in Figs.~\ref{fig:case_AV3_ATT3_track_percep} and~\ref{fig:case_AV4_ATT3_track_percep}.

\subsubsection{Case I: Camera-LiDAR Fusion with~\ref{av:cam-lidar-v2}.}
In Fig.~\ref{fig:case_AV3_ATT3_track}, the attacker induces FTs and MTs using only raw LiDAR data without altering camera data. The attack exploits the likely alignment between camera 2D detections and reversed LiDAR 3D detections. While safety-critical incidents are not guaranteed, a lightweight monitoring algorithm could trigger the attack in high-risk situations, such as at intersections, to increase impact.

\renewcommand{\subfigwidth}{0.85}
\renewcommand{\graphicswidth}{0.825}
\renewcommand{\graphicswidthh}{0.75}

\begin{figure}[!t]
    \centering
    \centering
    \begin{subfigure}[b]{.39\linewidth}
        \centering
        \includegraphics[trim={0cm 0cm 12cm 1.5cm},clip,width=\linewidth]{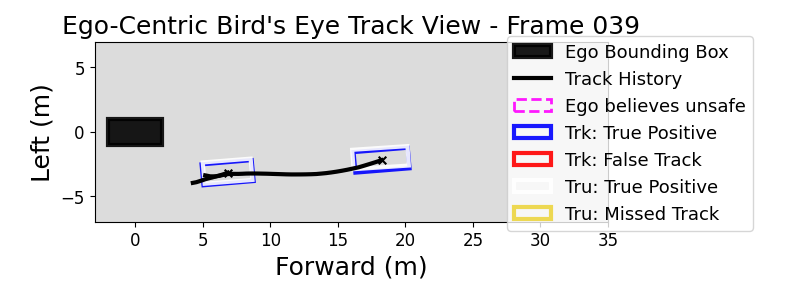}
        \caption{Frame 39 (BEV)}
    \end{subfigure}
    \begin{subfigure}[b]{.58\linewidth}
        \centering
        \includegraphics[trim={3cm 0cm 1.25cm 1.5cm},clip,width=\linewidth]{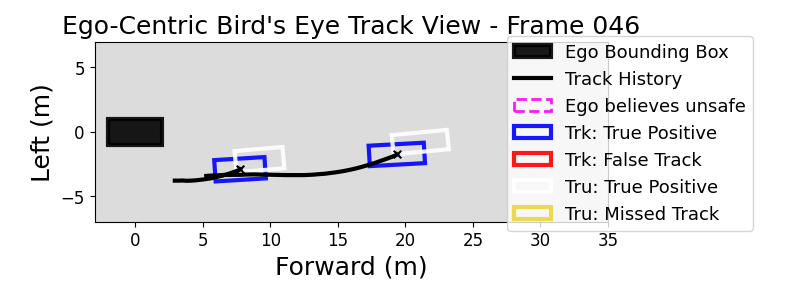}
        \caption{Frame 46 (BEV)}
    \end{subfigure}
    \begin{subfigure}[b]{\subfigwidth\linewidth}
        \centering
        \includegraphics[trim={0cm 0cm 0cm 1.5cm},clip,width=\graphicswidth\linewidth]{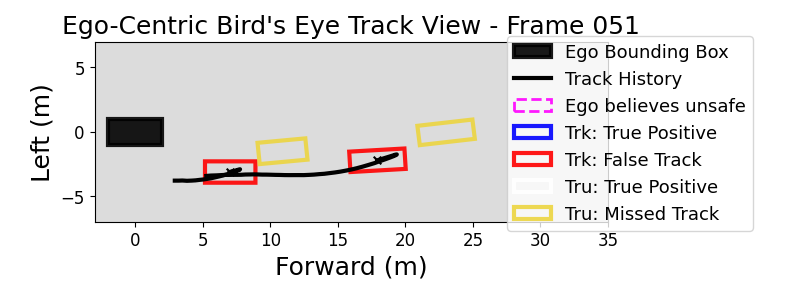}
        \caption{Frame 51 (BEV)}
    \end{subfigure}
    \suppsubfig{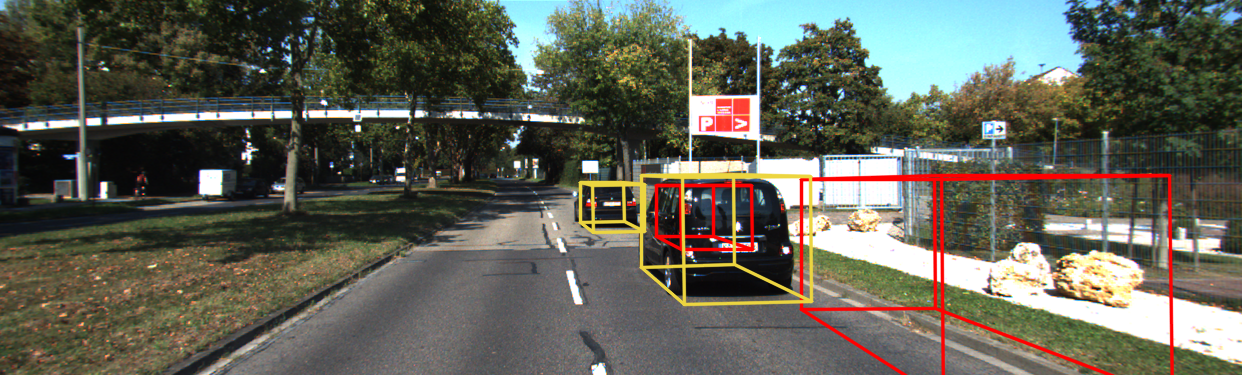}{\emph{Left:} Frame 51 of the rev. replay attack with object tracks overlaid in camera image.}
    \caption{\ref{av:cam-lidar-v2} design is susceptible to reverse replay~\ref{att:revreplay} with both FTs and MTs. Limited-capability attacker needs no estimate of scene-specific information. No safety-critical incident occurs because object changes lanes.}
    \label{fig:case_AV3_ATT3_track}
    \vspace{-6pt}
\end{figure}



\vspace{-2pt}
\subsubsection{Case II: T2T-3DLM design~\ref{av:ttt}.}
\begin{figure}[!t]
    \centering
    \begin{subfigure}[b]{.39\linewidth}
        \centering
        \includegraphics[trim={0cm 0cm 12cm 1.5cm},clip,width=\linewidth]{charts/case4-att3-attack-1-kitti-track-movie_frame_039.png}
        \caption{Frame 39 (BEV)}
    \end{subfigure}
    \begin{subfigure}[b]{.58\linewidth}
        \centering
        \includegraphics[trim={3cm 0cm 1.25cm 1.5cm},clip,width=\linewidth]{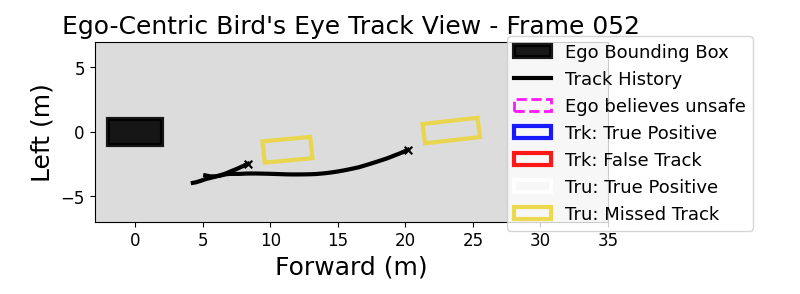}
        \caption{Frame 52 (BEV)}
    \end{subfigure}
    \caption{T2T-LM3D~\ref{av:ttt} mitigates FTs (from Fig.~\ref{fig:case_AV3_ATT3_track}) under a reverse replay attack~\ref{att:revreplay}. MTs still occur.}
    \label{fig:case_AV4_ATT3_track}
\end{figure}

Under a reverse replay attack, T2T-3DLM reduces FTs by detecting inconsistencies between camera and LiDAR data. Monocular 3D detection improves multi-sensor data association, reducing frustum ambiguity~\cite{2022hally-frustum}. An attacker can cause disagreement on an object by either \textbf{(dis. a)} adding a fake object to one sensor or \textbf{(dis. b)} removing a true object’s evidence from one sensor. Currently, neither \ref{av:cam-lidar-v2}'s monitor nor T2T-3DLM (\ref{av:ttt}) distinguishes these cases, treating both as ``fake objects'' and removing them. This leaves both AVs vulnerable to MTs if the attack is \textbf{(dis. b)}, as shown in Fig.~\ref{fig:case_AV4_ATT3_track}; the replay attack causes both an FT (recognized by T2T-3DLM) and an MT (unrecognized by T2T-3DLM). Future defenses could use statistical methods to detect replay attacks, while a ``saturation attack'' would be easily identifiable as it would invalidate all objects.

\renewcommand{\subfigwidth}{0.85}
\renewcommand{\graphicswidth}{0.825}
\renewcommand{\graphicswidthh}{0.75}

\subsection{Cases III, IV: Frust. Translation~\ref{att:frust-trans}}
A frustum translation attack can have more safety consequences than a reverse replay attack. The reverse replay attack manipulates LiDAR data \emph{in the reverse direction of travel}. By contrast, a frustum-type attack manipulates vehicles/LiDAR data \emph{along the line of sight to the victim}. As a result, the attacker can more easily create the appearance of a head-on collision or the illusion of a safe scene in the presence of an imminent collision.

\subsubsection{Case III: Camera-LiDAR Fusion with~\ref{av:cam-lidar-v2}.}
The frustum translation attack is shown in Fig.\ref{fig:case_AV3_ATT5_track} (additional visuals in Fig.\ref{fig:case_AV3_ATT5_track_percep}). Leveraging frustum consistency in camera 2D detections~\cite{2022hally-frustum}, the attack successfully overtakes an existing object's track, even with~\ref{av:cam-lidar-v2}'s data-asymmetry monitor. This creates a simultaneous FP for the inserted object and FN for the original, resulting in a \emph{translation} outcome~\cite{2022hally-frustum}. The new track appears to move directly toward the victim, triggering an evasive maneuver under the RSS safety metric~\cite{2017rsssafety}, despite the object being fictitious.

\renewcommand{\subfigwidth}{0.85}
\renewcommand{\graphicswidth}{0.825}
\renewcommand{\graphicswidthh}{0.75}

\begin{figure}[!t]
    \centering
    \begin{subfigure}[b]{.41\linewidth}
        \centering
        \includegraphics[trim={0cm 0cm 10.5cm 1.5cm},clip,width=\linewidth]{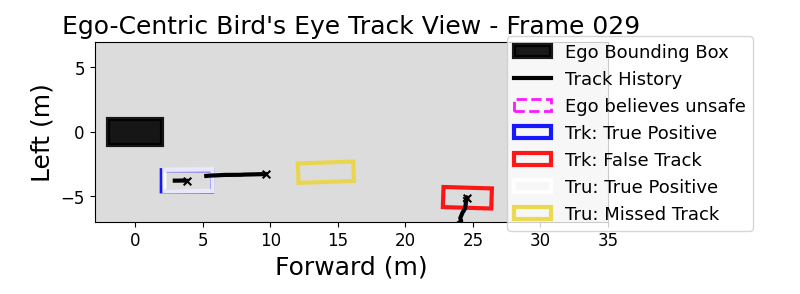}
        \caption{Frame 29 (BEV)}
    \end{subfigure}
    \begin{subfigure}[b]{.56\linewidth}
        \centering
        \includegraphics[trim={3cm 0cm 1.25cm 1.5cm},clip,width=\linewidth]{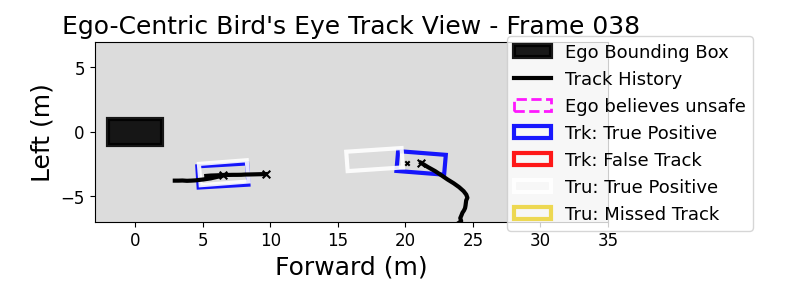}
        \caption{Frame 38 (BEV)}
    \end{subfigure}
    \begin{subfigure}[b]{\subfigwidth\linewidth}
        \centering
        \includegraphics[trim={0cm 0cm 0cm 1.5cm},clip,width=\graphicswidth\linewidth]{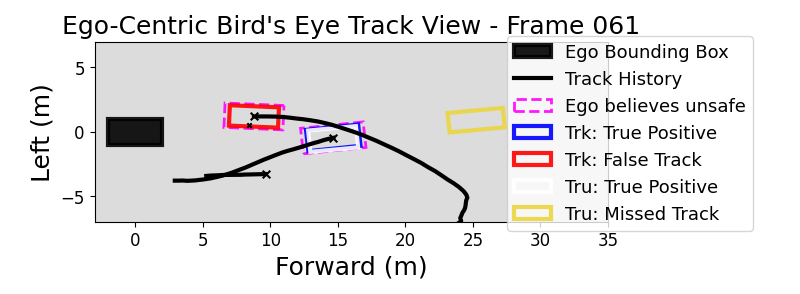}
        \caption{Frame 61 (BEV)}
   \end{subfigure}
   \suppsubfig{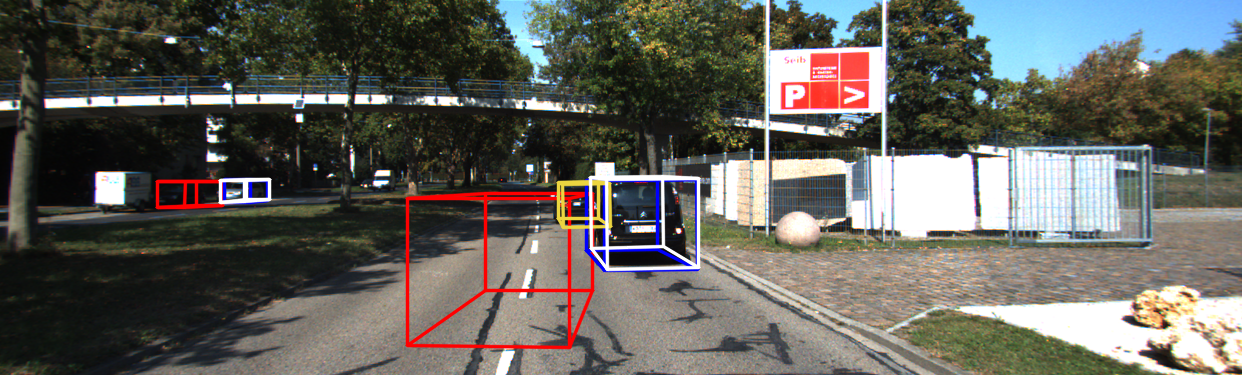}{\emph{Left:} Frame 61 of the frust. translation attack with object tracks overlaid in camera image.}
    \caption{\ref{av:cam-lidar-v2} is vulnerable to frustum translation attacks~\ref{att:frust-trans} with FT and MT outcomes. Attacker monitors objects with local perception, picks a target, and translates LiDAR points to create an unsafe scene.}
    \label{fig:case_AV3_ATT5_track}
\end{figure}


\subsubsection{Case IV: T2T-3DLM Design~~\ref{av:ttt}.}
\begin{figure}[!t]
    \centering
    \begin{subfigure}[b]{.39\linewidth}
        \centering
        \includegraphics[trim={0cm 0cm 12cm 1.5cm},clip,width=\linewidth]{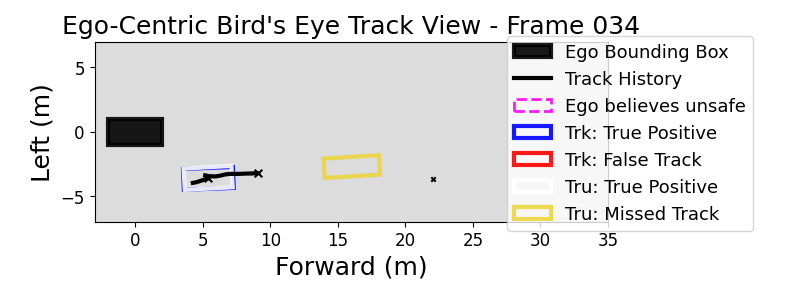}
        \caption{Frame 34 (BEV)}
    \end{subfigure}
    \begin{subfigure}[b]{.58\linewidth}
        \centering
        \includegraphics[trim={3cm 0cm 1.25cm 1.5cm},clip,width=\linewidth]{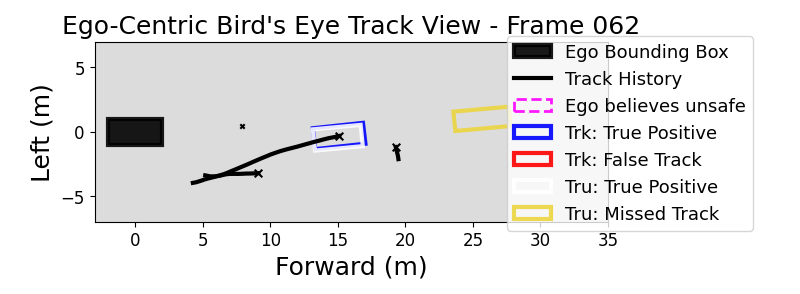}
        \caption{Frame 62 (BEV)}
    \end{subfigure}
    \caption{T2T-LM3D~\ref{av:ttt} mitigates incidence of FTs (from Fig.~\ref{fig:case_AV3_ATT5_track}) under a frustum trans. attack~\ref{att:frust-trans}. MTs still occur, but with NO falsely-identified unsafe~objects.}
    \label{fig:case_AV4_ATT5_track}
\end{figure}

T2T-3DLM mitigates frustum translation attacks by fusing 3D LiDAR with monocular 3D camera data, offering better data association than 2D camera detections. As shown in Fig.\ref{fig:case_AV4_ATT5_track} (and\ref{fig:case_AV4_ATT5_track_percep}), T2T-3DLM avoids tracking false objects, though MTs still occur since it cannot distinguish between \textbf{(dis. a)} and \textbf{(dis. b)} and must assume \textbf{(dis. a)}. A potential defense against MTs would involve using nearby data for local consistency to detect translation attacks.

\renewcommand{\subfigwidth}{0.85}
\renewcommand{\graphicswidth}{0.825}
\renewcommand{\graphicswidthh}{0.75}

\section{Two Security-Aware Architectures} \label{sec:6-defenses}

Traditional LiDAR-based and centralized camera-LiDAR fusion are vulnerable to adversarial manipulation of the inputs~\cite{2020sun-spoofing, 2022hally-frustum}. We introduce two new security-aware realizations of camera-LiDAR fusion in an effort to build resiliency to threats.

\subsection{\textbf{AV 3}: Monitoring Data-Asymmetries} \label{sec:av-3-defense}

\labeltext{\textbf{AV~3}}{av:cam-lidar-v2} 

While~\ref{av:cam-lidar-v1} uses camera and LiDAR data for tracking, it lacks longitudinal consistency between sensors. Our data-asymmetry monitor flags persistent sensor disagreements, such as when updates rely solely on LiDAR or camera data.

In tracking, estimation combines target existence and state estimation. Trackers estimate the number of tracks and their states maintaining tracks on new measurements but only passing confirmed tracks to downstream modules. A per-track "score" classically determines confirmation or false alarm status, adjusting each frame based on measurement consistency~\cite{1964sittlerAssociation}. The likelihood ratio is:
\begin{align}
    \text{LR} = \frac{\Pr(D|H_1) \Pr_0(H_1)}{\Pr(D|H_0) \Pr_0(H_0)} \coloneqq \frac{P_T}{P_F}
\end{align}
where the two hypotheses, $H_1$ and $H_0$, are ``track is a real target'' and ``track is a false alarm'' hypotheses with probabilities $P_T$ and $P_F$, respectively. $D$ is the measurement data, meaning that $\Pr(D_j|H_i)$ is the probability density function of the received data given that $H_i$ is assumed correct. Finally, $\Pr_0(H_i)$ is the prior probability of the hypothesis.

When the log is taken, the likelihood ratio becomes the log-likelihood, commonly known as the ``track score''
\begin{align}
    LLR \coloneqq L = \log \frac{P_T}{P_F}.
\end{align}
Maintaining an estimate of the track score, we can quickly revert back to the true track probability as:
\begin{align}
    P_T = \frac{e^{L}}{1 + e^{L}}.
\end{align}

Conveniently, the track score can be placed in a recursive format given a simple decomposition that follows
\begin{align}
    L_{k} &= L_{k-1} + \Delta L_{k} \\ 
    \Delta L_{k} &= \begin{cases}
        \log (1-\hat{P}_D) \quad \text{if no update} \\
        \Delta L_{k,u} \quad \text{if update.}
    \end{cases}
\end{align}
The track score has two components:

\begin{itemize}
    \item \textbf{Loss:} When a track is not updated, the penalty is based on the expected probability of detection, $\hat{P}_D$, with a higher $\hat{P}_D$ resulting in a larger penalty for missed detections (since $\log$ is negative for $x \in (0, 1)$).
    \item \textbf{Gain:} When a track is updated, the score is based on how closely the measurement matches the track state. Specifically, $\Delta L_u$ is the likelihood of the measurement.
\end{itemize}

However, adversaries can exploit this, creating a fictitious track with a high score using only one sensor~\cite{hallyburton2024bayesian}. In a two-sensor system, if the attacker's inputs align well with the dynamics model, the measurement likelihood gain ($\Delta L_{u,\text{sensor}_1}$) could outweigh the loss from missed detections on the second sensor, as $\Delta L{u,\text{sensor}_1} > |\log \left[1 - \Pr{D,\text{sensor}_2}\right]|$. In AVs, where dynamics are simple, prediction errors remain low, allowing the track score to increase over time without affecting the second sensor.

We improve track scoring by maintaining per-track scores for each sensor. If the score discrepancy exceeds a threshold, the track is flagged as \texttt{invalid}, indicating possible clutter or adversary manipulation, warranting caution.

Track scoring is enhanced for AV-3 by maintaining separate per-sensor track scores alongside the central track score, forming a robust defense protocol such as:
\begin{align*}
    L_{k,\text{sensor-i}} &= L_{k-1,\text{sensor-i}} + \Delta L_{k,\text{sensor-i}} \\ 
    \Delta L_{k,\text{sensor-i}} &= \begin{cases}
        \log (1-\hat{P}_{D,\text{sensor-i}}) & \text{if no update} \\
        \Delta L_{k,u,\text{sensor-i}} & \text{if update}
    \end{cases} \\
    L_{k,\text{central}} &= L_{k-1,\text{central}} + \Delta L_{k,\text{sensor}_1} + \Delta L_{k,\text{sensor}_2}.
\end{align*}
Then, track confirmation is done in the usual manner where a track is confirmed when its score exceeds a designed threshold, $\tau_c$. Track validation is performed by monitoring the difference in track scores between pairs of sensors and triggering on a designed threshold, $\tau_v$ -- i.e., 
\begin{align*}
    \text{Confirmation} &= \begin{cases}
        \texttt{Confirmed} & L_{k,\text{central}} \geq \tau_{c}\\
        \texttt{Unconfirmed} & \text{otherwise}
    \end{cases}\\
    \text{Validation} &= \begin{cases}
        \texttt{Valid} & \parbox{15em}{$\texttt{Confirmed} \ \text{and} \\ | L_{k,\text{sensor}_1} - L_{k,\text{sensor}_2} | < \tau_{v}$} \\
        \texttt{Invalid} & \text{otherwise}
    \end{cases}
\end{align*}
At each timestep, $k$, tracks are confirmed if the score exceeds a threshold, $\tau_c$. The algorithm monitors the difference in track scores between pairs of sensors (e.g., $L_{k,\text{sensor}_1}$ and $L_{k,\text{sensor}_2}$ in a two-sensor AV) and triggers an alarm when this exceeds a designed threshold, $\tau_v$.

\subsection{\textbf{AV~4}: Fusion of Camera, LiDAR Tracks} \label{sec:av-4-defense}

\labeltext{\textbf{AV~4}}{av:ttt}

\cite{2022hally-frustum} identified a vulnerability in centralized camera-LiDAR fusion: an attacker can execute a ``frustum attack'' by shifting LiDAR point clusters forward or backward while keeping them consistent with 2D camera detections. This occurs because 2D images alone cannot fully resolve 3D object positioning, affecting systems using centralized 3D LiDAR \& 2D camera tracking (e.g.,\ref{av:cam-lidar-v1},\ref{av:cam-lidar-v2}).

To mitigate this, we derive 3D detections from 2D images (monocular 3D). Although 3D detection from images is underdetermined, recent algorithms (e.g.,\cite{2022pgd3d, 2020smoke}) use context for improved stability. We enhance per-sensor filtering with distributed tracking at each sensor, then fuse tracks with decentralized data fusion (DDF) algorithms~\cite{1994ddfframework, 2013ddfreview}. To maintain consistent labels, we track sensor-to-object assignments over time, termed \emph{track-to-track fusion of 3D LiDAR and monocular data} (T2T-3DLM).
\section{Large-Scale Evaluations} \label{sec:8-results}
We conduct a large-scale evaluation of all attacks across all AV designs in Table~\ref{tab:table_case_configs} to assess performance in both \textbf{\emph{unattacked and attacked}} scenarios. This comprehensive analysis provides data-driven insights into the effectiveness of secure sensor fusion architectures. Due to space constraints, we focus on the reverse replay (\ref{att:revreplay}) and frustum translation (\ref{att:frust-trans}) attacks. Complete results for \emph{all five attacks} from Sec.\ref{sec:5-attack-designs} on \textbf{\emph{all validation scenes}} and AV cases are available in Table\ref{tab:results_attack_collection} in Appendix~\ref{appendix:all-results}.

\begin{table}[!t]
    \centering
    \caption{AV Configurations}
    \label{tab:table_case_configs}
    \resizebox{\columnwidth}{!}{%
\begin{tabular}{llllll}
\toprule
                 Num. &                                                       AV Case Name &     LiDAR &   Camera &                              Tracking & Fusion \\ \midrule
\midrule
       \ref{av:lidar} &                                                         Lidar-Only & PtPillars &      N/A &                               AB3DMOT &    N/A \\ \midrule
\ref{av:cam-lidar-v1} &                Cam-LiDAR fusion at detect & PtPillars & FstrRCNN &                              EagerMOT &    N/A \\ \midrule
\ref{av:cam-lidar-v2} & \tworowsubtableleft{Cam-LiDAR Fusion at}{Detection w/ Asym. Check} & PtPillars & FstrRCNN &                              EagerMOT &    N/A \\ \midrule
         \ref{av:ttt} &                 Cam-LiDAR fusion at track & PtPillars &      PGD & \tworowsubtableleft{AB3DMOT}{ (each)} &    T2T \\ \midrule
\bottomrule
\end{tabular}
    }
\end{table}


\vspace{6pt}
\noindent \textbf{Methods:} For each dataset (KITTI, nuScenes), we evaluate all validation scenes (36 for KITTI, 135 for nuScenes) by running each AV design on unattacked scenes to capture metrics at perception, tracking, motion prediction, and safety levels. Then, we run each AV design through the five attacks from Sec.~\ref{sec:5-attack-designs} on every scene, capturing the same metrics. Finally, we compare the baseline and attacked scenarios to isolate the attack's effects on \textit{each module}.

\subsection{Baseline AV Performance} \label{sec:results-av-perfomance}

We characterize the performance of four proposed AV architectures under an unattacked baseline where the AVs are expected to have high performance and mitigate unsafe scenarios. The results of the baseline experiments are in Fig.~\ref{fig:baseline_percep_bars} and~\ref{fig:baseline_track_bars}. T2T-3DLM is on par with FP performance compared to other AVs but it sacrifices a degree of MT performance at the tracking level. The results that follow present the \emph{per-frame increment over baseline} -- the average amount of change from the baseline to the attack case in a given metric on a~per-frame~basis.

\subsection{Attack Effectiveness} \label{sec:results-attack-effectiveness}

\subsubsection{Perception Outcomes}
\begin{figure}[!t]
    \centering
    \begin{subfigure}[b]{.49\linewidth}
        \centering
        \includegraphics[width=1\linewidth]{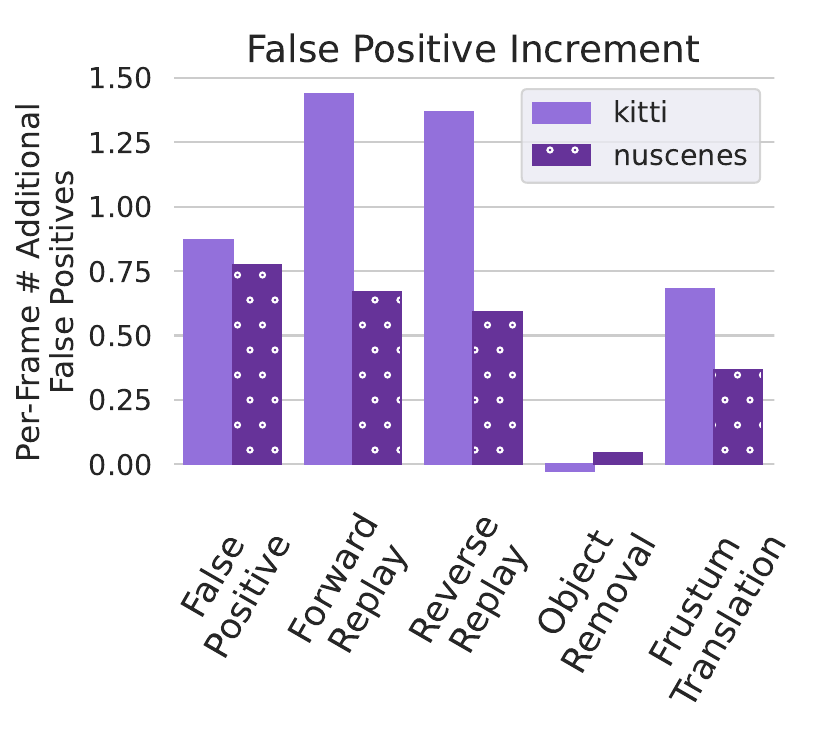}
        \vspace{-20pt}
        \caption{FP inc. for all but removal.}
    \end{subfigure}
    \begin{subfigure}[b]{.49\linewidth}
        \centering
        \includegraphics[width=1\linewidth]{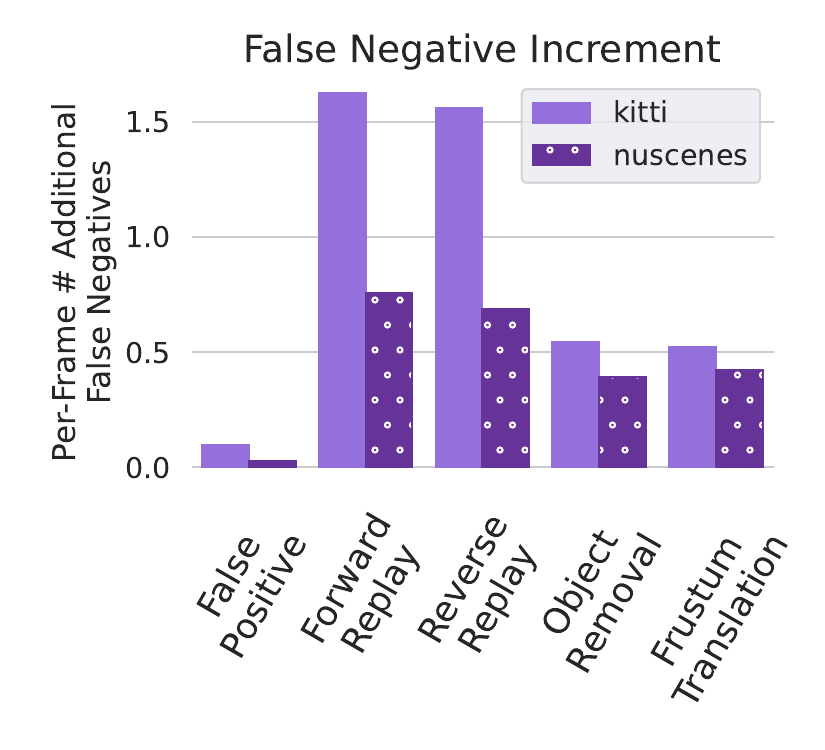}
        \vspace{-20pt}
        \caption{FN inc. for all but false pos.}
    \end{subfigure}
    \caption{Perception increments over baseline (Fig.~\ref{fig:baseline_percep_bars}) per-frame marginalized over AV cases (shared LiDAR-based perception) are intuitive from attack goals.}
    \label{fig:attack_percep_all}
\end{figure}

Since the LiDAR perception algorithm is the same across all AV designs, attack impact at the perception level is consistent across cases. We therefore aggregate results over AV cases and present perception outcomes for each attack in Fig.~\ref{fig:attack_percep_all}. The perception-level outcomes of the attacks are intuitive, as shown moving left to right in the figure.
\begin{enumerate}[label=(\arabic*),topsep=0pt,itemsep=-1ex,partopsep=1ex,parsep=1ex]
    \item False positive att.~\ref{att:fp}: solely FP outcomes.
    \item Forward replay attack~\ref{att:forreplay}: FP and FN identically.
    \item Reverse replay~\ref{att:revreplay}: FP and FN identically.
    \item Object removal attack~\ref{att:remove}: solely FN outcomes.
    \item Frustum translation~\ref{att:frust-trans}: similar FP, FN outcomes.
\end{enumerate}
The security-aware architectures build defense at tracking; attacks will have ``success'' at the perception-level alone.

\subsubsection{Tracking Outcomes}
We only discuss attacks~\ref{att:revreplay} and \ref{att:frust-trans}. Table~\ref{tab:results_attack_collection} in Appendix~\ref{appendix:all-results} summarizes results for all attacks.

\vspace{6pt}

\begin{figure}[!t]
    \centering
    \begin{subfigure}[b]{.49\linewidth}
        \centering
        \includegraphics[width=1\linewidth]{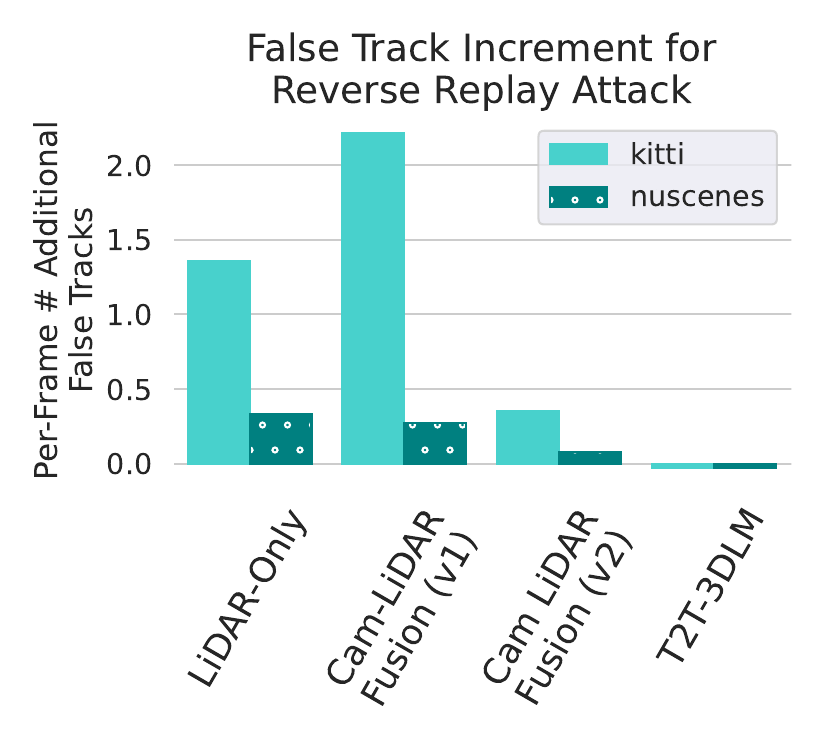}
        \caption{(3,4) reduce per-frame FTs}
        \label{fig:attack_revreplay_bars_a}
    \end{subfigure}
    \begin{subfigure}[b]{.49\linewidth}
        \centering
        \includegraphics[width=1\linewidth]{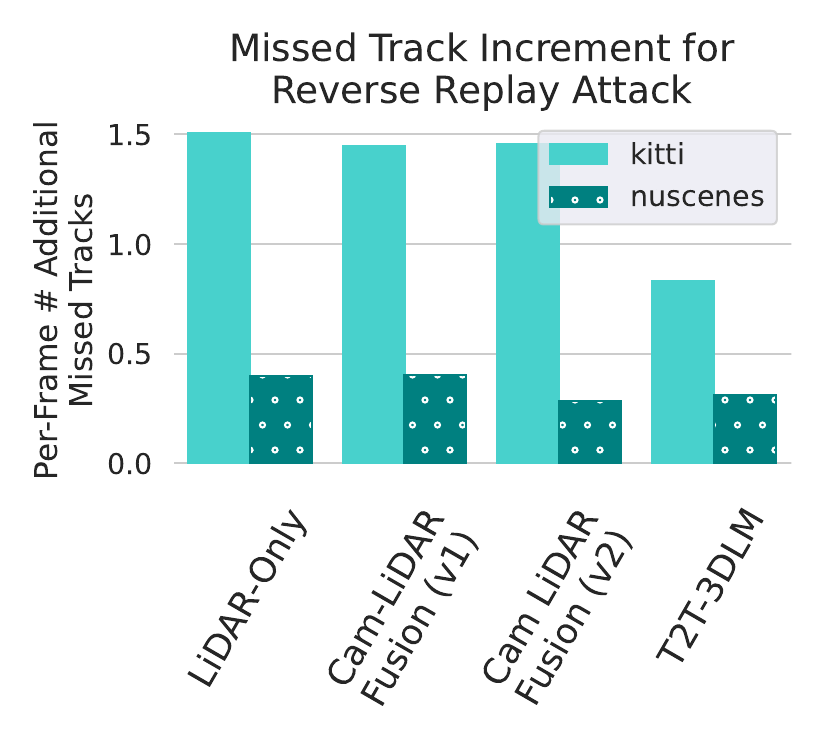}
        \caption{All cases see MT increment.}
        \label{fig:attack_revreplay_bars_b}
    \end{subfigure}
    \caption{Reverse replay attack: Centralized fusion with asymmetry and T2T-3DLM (cases 3, 4) mitigates FT inc. over Fig.~\ref{fig:baseline_track_bars}. T2T-3DLM has best MT performance.}
    \label{fig:attack_revreplay_bars}
\end{figure}

\noindent \textbf{Reverse Replay Attack~\ref{att:revreplay}.}
Replay attacks maintain consistent dynamics on straight/low-curvature paths, affecting both LiDAR-only and camera-LiDAR fusion (Fig.\ref{fig:attack_revreplay_bars}). In camera-LiDAR fusion, the attack maintains alignment with 2D bounding boxes despite manipulating only LiDAR data, reducing replay-induced discontinuities. Fig.\ref{fig:attack_revreplay_bars_a} shows traditional fusion is prone to false and missed tracks under reverse replay, while both the data asymmetry monitor and T2T-3DLM reduce false tracks fivefold.

\begin{figure}[!t]
    \centering
    \begin{subfigure}[b]{.49\linewidth}
        \centering
        \includegraphics[width=1\linewidth]{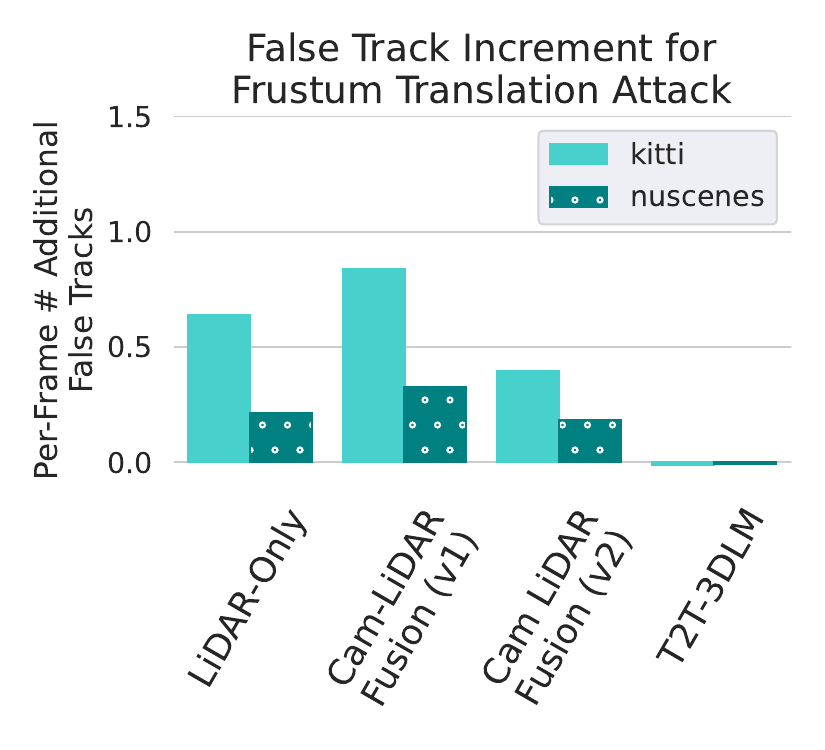}
        \caption{(4) reduces per-frame FT}
        \label{fig:attack_frusttrans_bars_a}
    \end{subfigure}
    \begin{subfigure}[b]{.49\linewidth}
        \centering
        \includegraphics[width=1\linewidth]{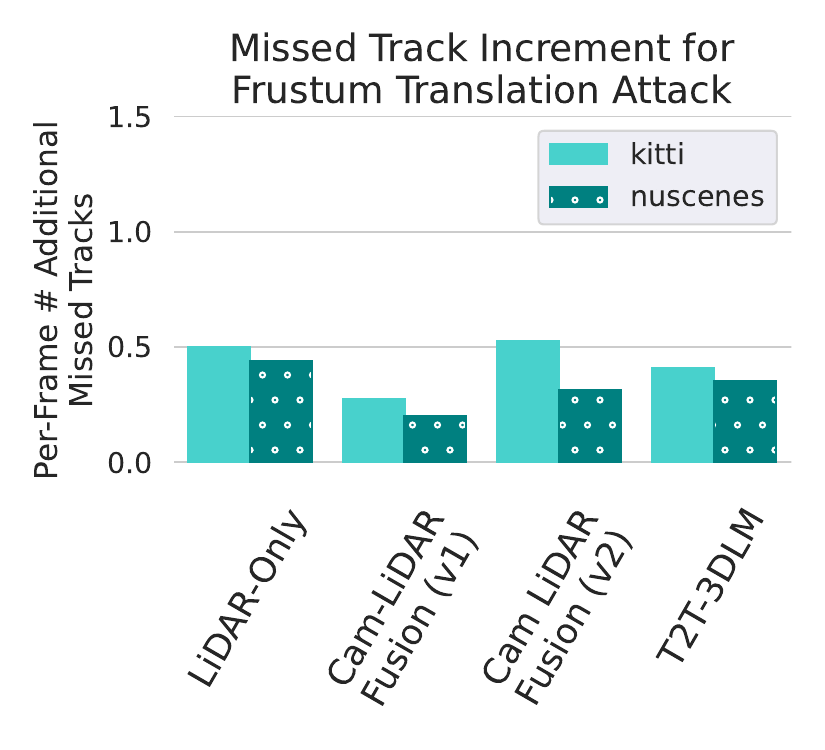}
        \caption{All cases see MT increment.}
        \label{fig:attack_frusttrans_bars_b}
    \end{subfigure}
    \caption{Frustum translation attack: (a) Only T2T-3DLM mitigates per-frame FT increment over baseline from Fig.~\ref{fig:baseline_track_bars}. (b)~All cases have MT increment.}
    \label{fig:attack_frusttrans_bars}
\end{figure}

\noindent \textbf{Frustum Translation Attack~\ref{att:frust-trans}.}
This context-aware attack exploits existing tracks and frustum ambiguity, increasing FT and MT rates for both LiDAR-only and camera-LiDAR fusion (Fig.\ref{fig:attack_frusttrans_bars}). The frustum translation attack has a 60\% success rate against camera-LiDAR fusion, compared to 40\% for context-unaware FP attack (Fig.\ref{fig:results-all-outcomes}). T2T-3DLM was specifically designed to address camera data ambiguity~\cite{2022hally-frustum}, effectively filtering out inconsistent false objects, reducing FT to nearly 1.0\%. Further work is needed to mitigate the increase in missed tracks for T2T-3DLM.

\subsubsection{Safety Outcomes}
\begin{figure}[!t]
    \centering
    \begin{subfigure}[b]{.49\linewidth}
        \centering
        \includegraphics[width=\linewidth]{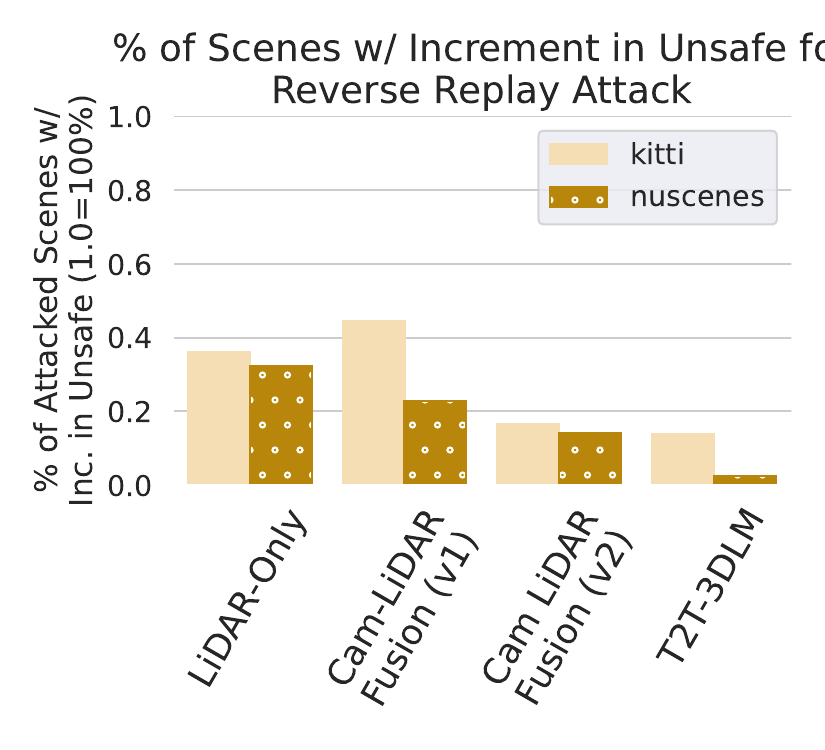}
        \caption{Unsafe scenes in all AVs.}
        \label{fig:attack_unsafe_bars_a}
    \end{subfigure}
    \begin{subfigure}[b]{.49\linewidth}
        \centering
        \includegraphics[width=\linewidth]{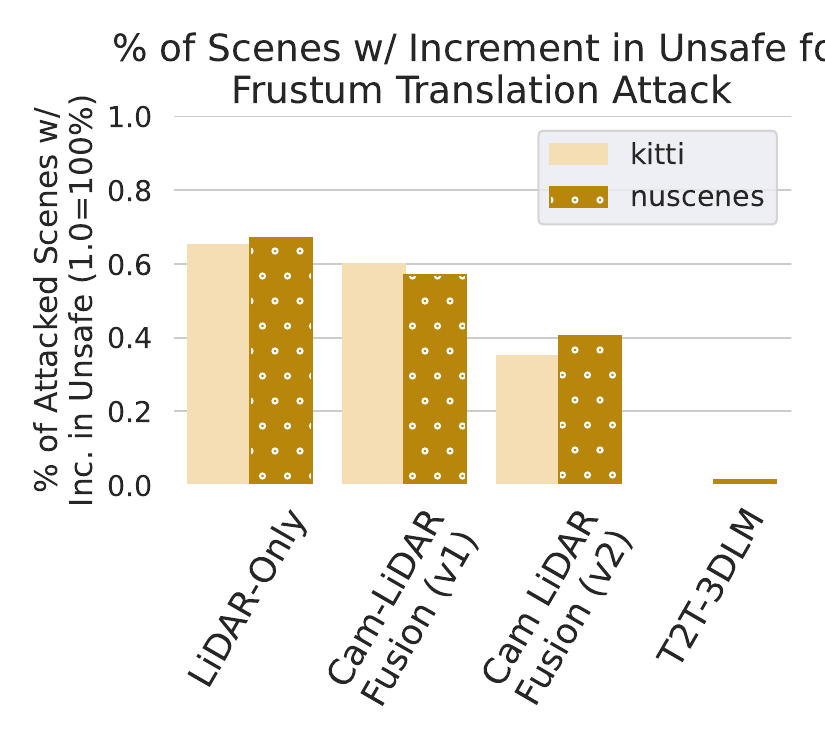}
        \caption{(4) reduces unsafe scenes.}
        \label{fig:attack_unsafe_bars_b}
    \end{subfigure}
    \caption{Safety evaluated via $\%$ of scenes where more frames are unsafe (measured with RSS~\cite{2017rsssafety}) in attack vs. baseline. T2T-3DLM mitigates strongest attack.}
    \label{fig:attack_unsafe_bars}
\end{figure}

Safety outcomes are scenes where the number of \emph{unsafe} objects differs between the baseline and attacked case in at least one frame. Fig.\ref{fig:attack_unsafe_bars_a} shows that the reverse replay attack creates unsafe scenarios in AVs 1-3, as FTs approach the victim AV along reversed paths. Traditional camera-LiDAR fusion (\ref{av:cam-lidar-v1}) suffers safety-critical incidents in nearly 50\% of cases under this attack. T2T-3DLM using 3D object positioning from camera reduces attack success to 10\%. Fig.\ref{fig:attack_unsafe_bars_b} shows that the frustum translation attack~\ref{att:revreplay} induces more unsafe scenarios by targeting line-of-sight to the victim, unlike reverse replay, which follows existing paths. In both attacks, T2T-3DLM significantly reduces unsafe outcomes by lowering FT presence, bringing the unsafe frame increment down to nearly 1.0\%.

\section{Discussion} \label{sec:9-discussion}

\noindent \textbf{Merit of secure architectures.}
An adversary with limited information can successfully launch attacks by modifying raw LiDAR data alone. Secure architectures offer significant improvements to AV sensing, though the increased baseline missed track rate remains a concern. Future enhancements to T2T-3DLM, particularly in 3D camera-based perception and track identification, could further improve performance without compromising security.

\noindent \textbf{Limitations.} \label{sec:discussion-limitations}
For context-aware attacks, the attacker uses the lightweight SECOND~\cite{2018second} LiDAR-based DNN. However, running a DNN as malware may be limited without parallel hardware. Instead, high-confidence detections are sufficient, so state-of-the-art models aren’t essential. While the attacker could adaptively adjust scheduling, we used fixed parameters across runs with minimal tuning. A pre-deployment tuning phase and adaptive runtime adjustments could further enhance attack effectiveness.
\section{Conclusion}

Our vulnerability analysis identified LiDAR as critical to AV safety (in the TCB). We then conducted an in-depth evaluation of LiDAR-based sensing using AV simulators and datasets, considering practical constraints like limited attacker information and datagram integrity. We developed eight context-aware/unaware attacks on LiDAR data, evaluating five on longitudinal datasets and AV simulators. The attacks were largely effective, even compromising Baidu Apollo with context-unaware methods. To counter vulnerabilities, we introduced two security-aware sensor fusion designs—a data asymmetry monitor and track-to-track fusion—which significantly reduced attack effectiveness.

\vspace{4pt}
\noindent \textbf{Ethical Considerations.} We have disclosed all relevant vulnerabilities to a major manufacturer of LiDAR sensors (additional information is omitted for double-blind review) and are actively working with them to ensure security of sensing against the detected vulnerabilities.

\bibliographystyle{IEEEtranS}
\bibliography{references}

\appendix
\subsection{Attack Visualizations} \label{app:attack_visualisations}

To aid the reader, we include visualizations of select attack procedures. Fig.~\ref{fig:attack-examples} illustrates the attacker’s step-by-step decision process. In the frustum translation attack, for instance, the attacker first monitors existing objects, selects a target, determines a translation direction (e.g., towards the victim), and adjusts the range of specific LiDAR points to form the adversarial object.

\renewcommand{\subfigwidth}{0.95}
\renewcommand{\graphicswidth}{0.7}

\begin{figure}[!t]
    \centering
    \begin{subfigure}[b]{\subfigwidth\linewidth}
        \centering
        \includegraphics[width=\graphicswidth\linewidth]{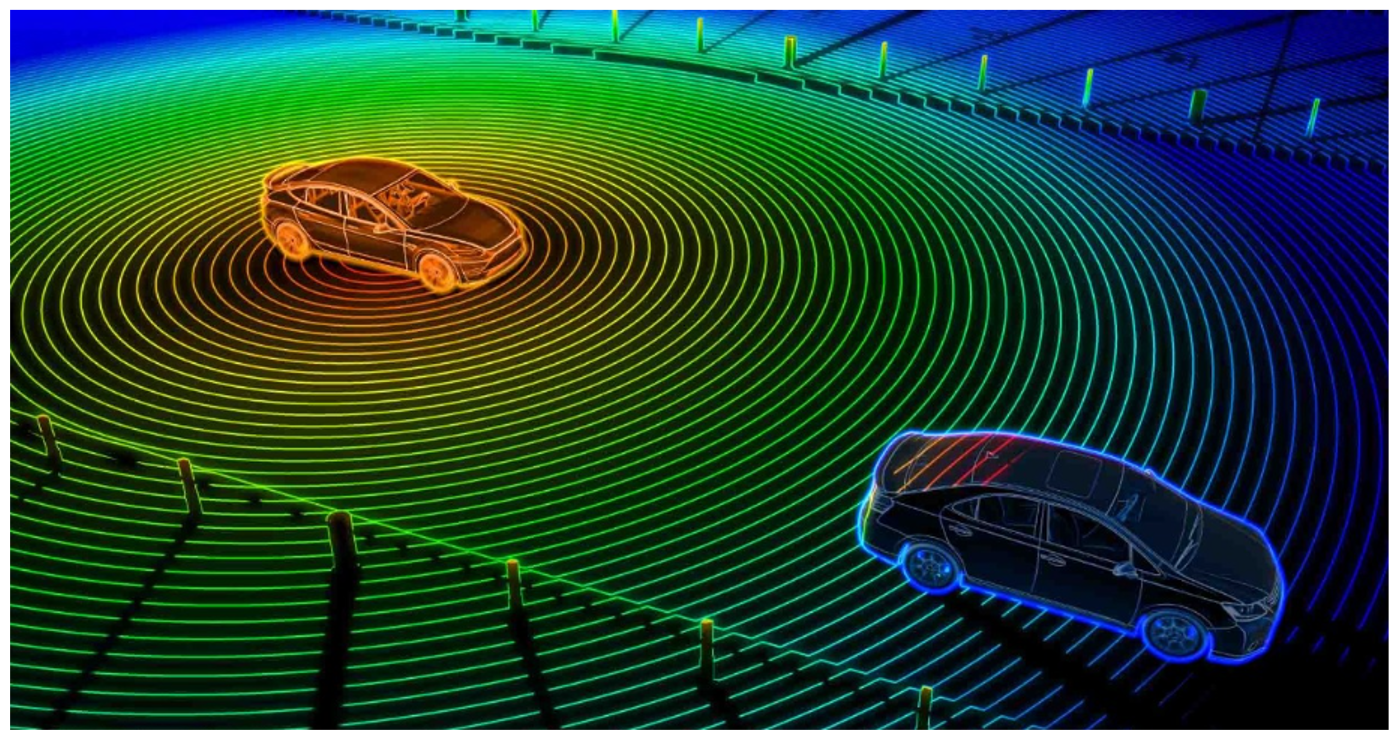}
        \caption{True scene with victim at left (orange) and another object at right (blue). LiDAR sensor generates point cloud capture of scene which includes points reflecting from objects.}
    \end{subfigure}
    \begin{subfigure}[b]{\subfigwidth\linewidth}
        \centering
        \includegraphics[width=\graphicswidth\linewidth]{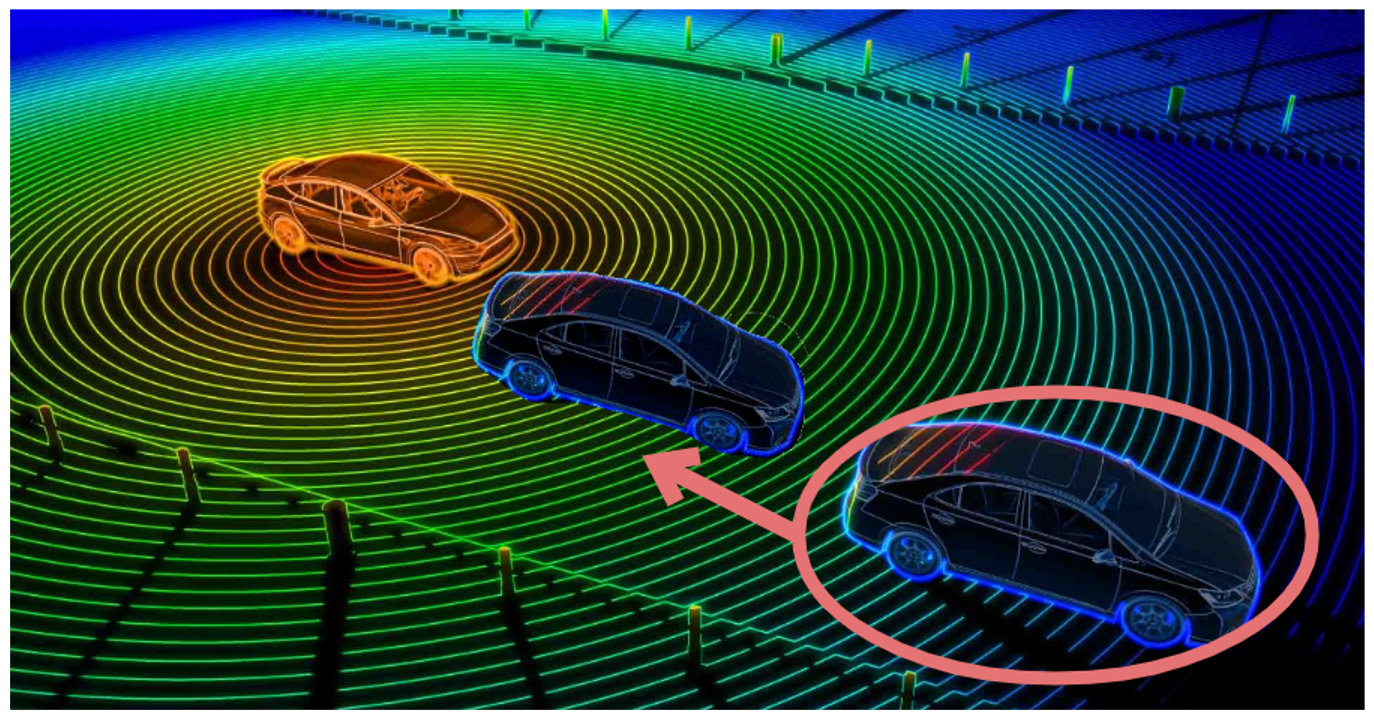}
        \caption{Adversary manipulates other object's points to move them closer to victim following a frustum-translation attack. Only the LiDAR points are manipulated. This can cause a false object to be detected.}
    \end{subfigure}
    \caption{An attacker issuing a frustum translation attack will (a) first find and select an existing object in the scene to attack. (b) Then, the attacker will manipulate the LiDAR points belonging to that object to make it appear closer or farther than the original object. The attack must only scale the range of the points, so that object will then appear to be translated directly toward or away from the victim. Base LiDAR scene image from~\cite{densox}.}
    \label{fig:attack-examples}
\end{figure}
\subsection{Case Study Supplementals}

\begin{figure}[!t]
    \centering
    \begin{subfigure}[b]{0.8\linewidth}
        \includegraphics[width=\linewidth]{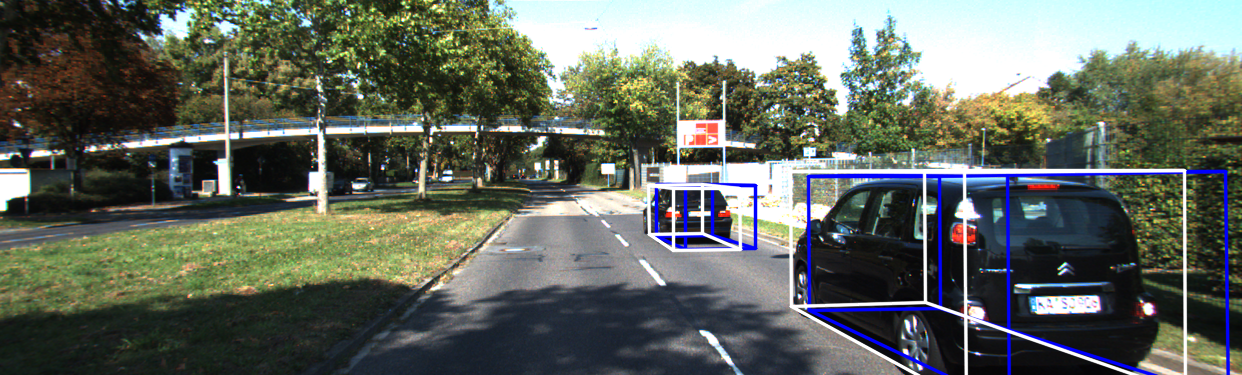}
    \end{subfigure}
    \begin{subfigure}[b]{0.8\linewidth}
        \includegraphics[width=\linewidth]{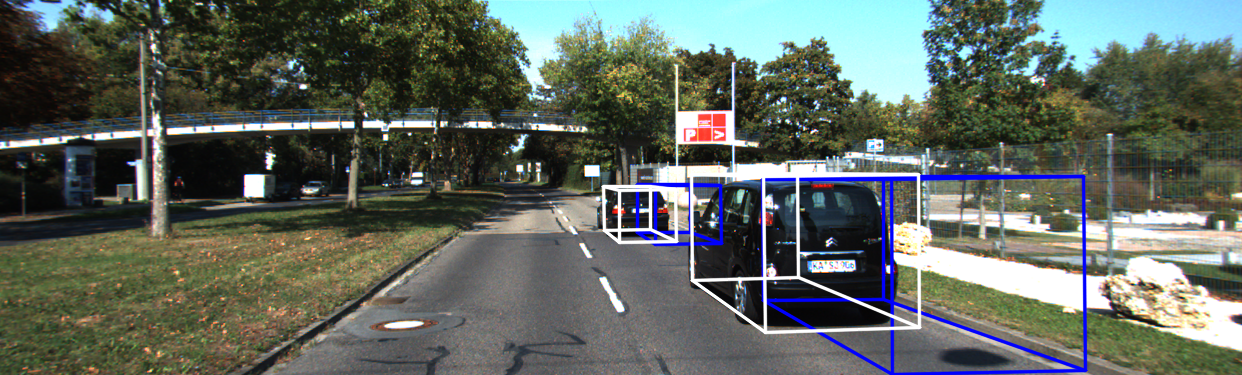}
    \end{subfigure}
    \begin{subfigure}[b]{0.8\linewidth}
        \includegraphics[width=\linewidth]{charts/case3-att3-attack-1-kitti-track_percep-movie_frame_051.png}
    \end{subfigure}
    \caption{Front-view images of the reverse replay attack on~\ref{av:cam-lidar-v2}; companion images with Fig.~\ref{fig:case_AV3_ATT3_track}. Compare to T2T-LM3D under same attack in Fig.~\ref{fig:case_AV4_ATT3_track_percep}.}
    \label{fig:case_AV3_ATT3_track_percep}
\end{figure}

\begin{figure}[!t]
    \centering
    \begin{subfigure}[b]{0.8\linewidth}
        \includegraphics[width=\linewidth]{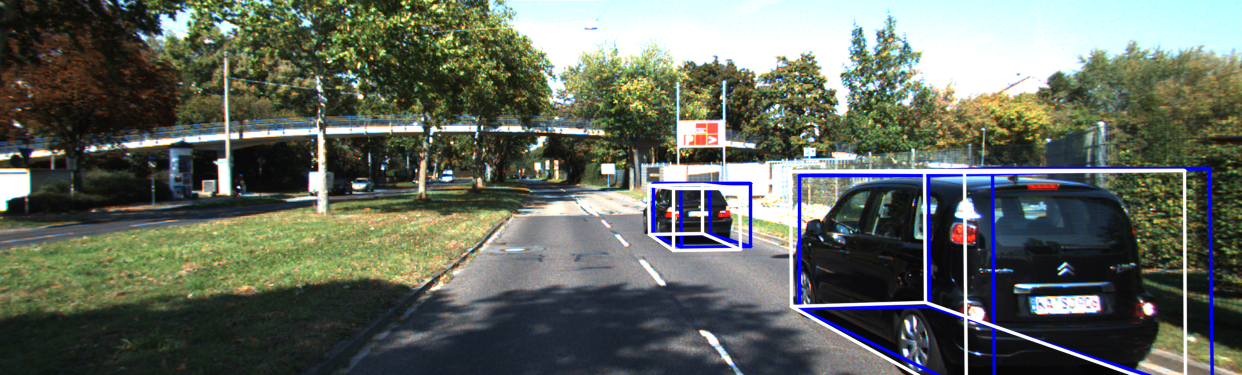}
    \end{subfigure}
    \begin{subfigure}[b]{0.8\linewidth}
        \includegraphics[width=\linewidth]{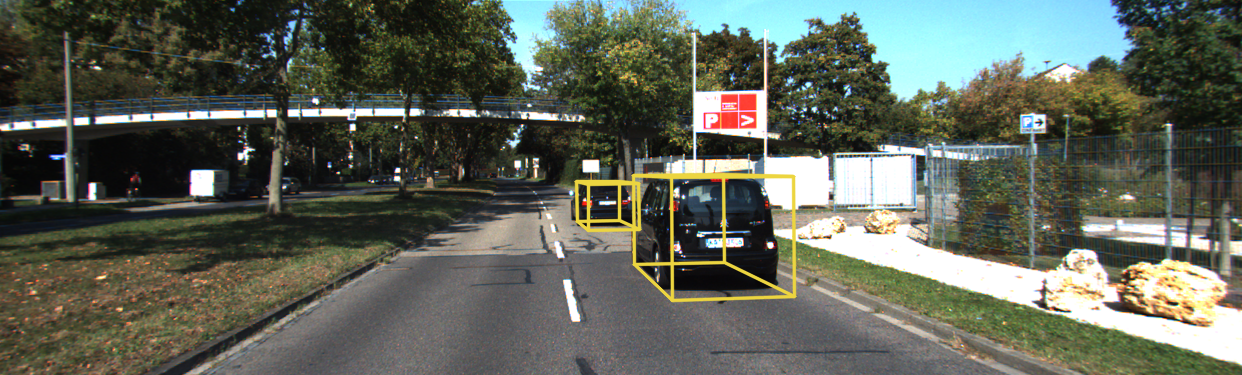}
    \end{subfigure}
    \caption{Front-view images of the frustum trans. attack on \ref{av:cam-lidar-v2}; companion images with Fig.~\ref{fig:case_AV4_ATT3_track}. Compare to camera-lidar detection-level fusion in Fig.~\ref{fig:case_AV3_ATT3_track_percep}.}
    \label{fig:case_AV4_ATT3_track_percep}
\end{figure}

\begin{figure}[!t]
    \centering
    \begin{subfigure}[b]{0.8\linewidth}
        \includegraphics[width=\linewidth]{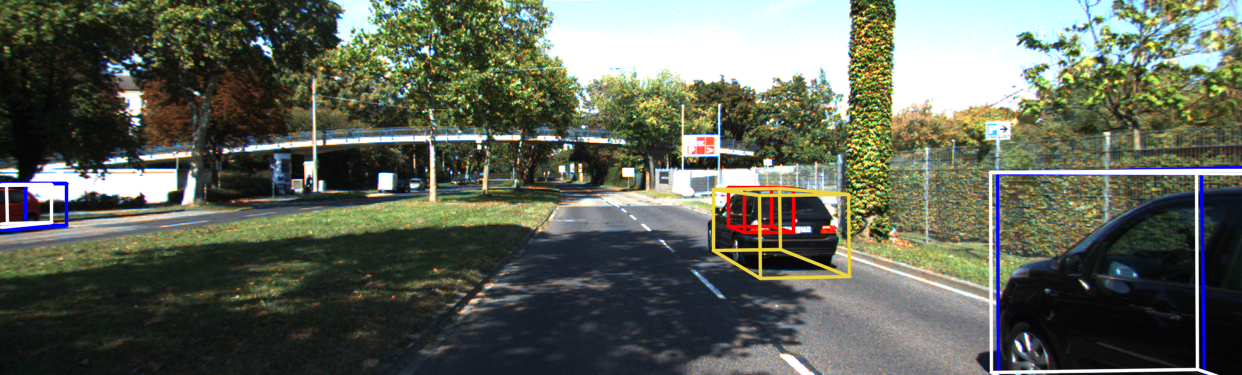}
    \end{subfigure}
    \begin{subfigure}[b]{0.8\linewidth}
        \includegraphics[width=\linewidth]{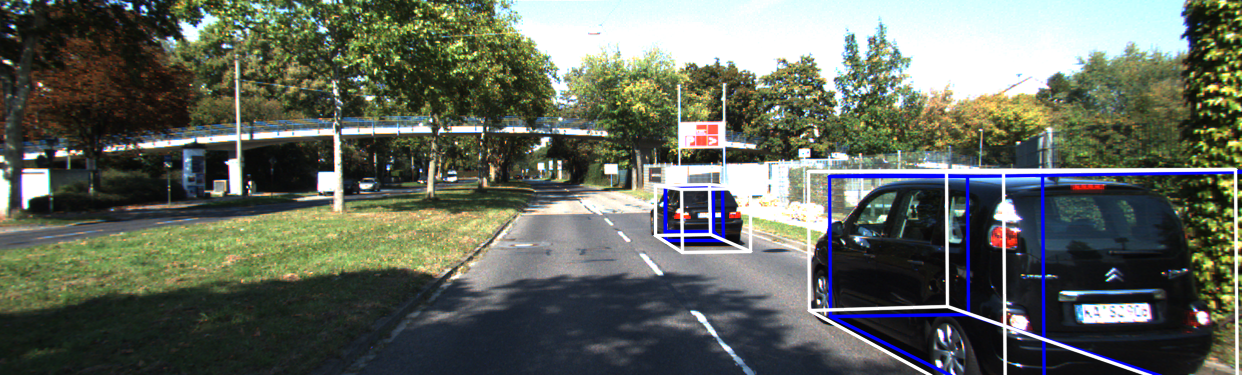}
    \end{subfigure}
    \begin{subfigure}[b]{0.8\linewidth}
        \includegraphics[width=\linewidth]{charts/case3-att5-attack-1-kitti-track_percep-movie_frame_061.png}
    \end{subfigure}
    \caption{Front-view images of the frustum translation attack on~\ref{av:cam-lidar-v2}; companion images with Fig.~\ref{fig:case_AV3_ATT5_track}. Compare to T2T-LM3D under same attack in Fig.~\ref{fig:case_AV4_ATT5_track_percep}.}
    \label{fig:case_AV3_ATT5_track_percep}
\end{figure}

\begin{figure}[!t]
    \centering
    \begin{subfigure}[b]{0.8\linewidth}
        \includegraphics[width=\linewidth]{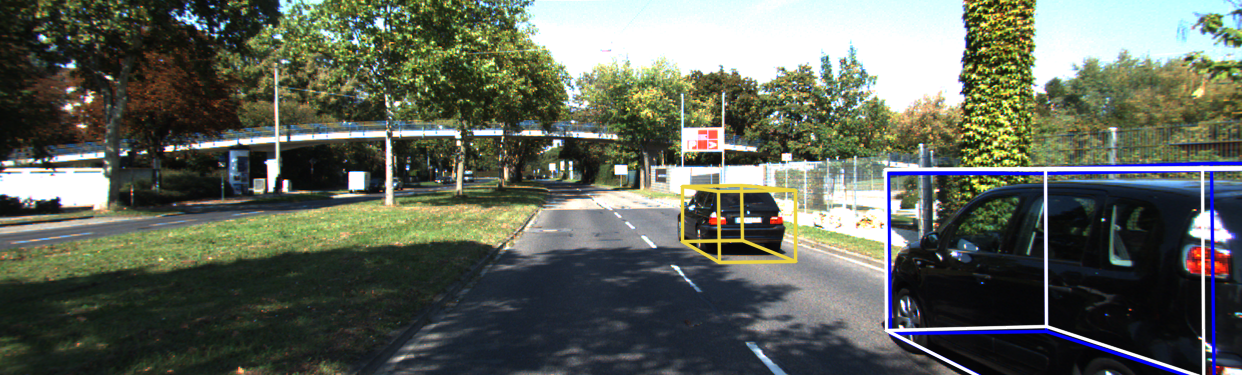}
    \end{subfigure}
    \begin{subfigure}[b]{0.8\linewidth}
        \includegraphics[width=\linewidth]{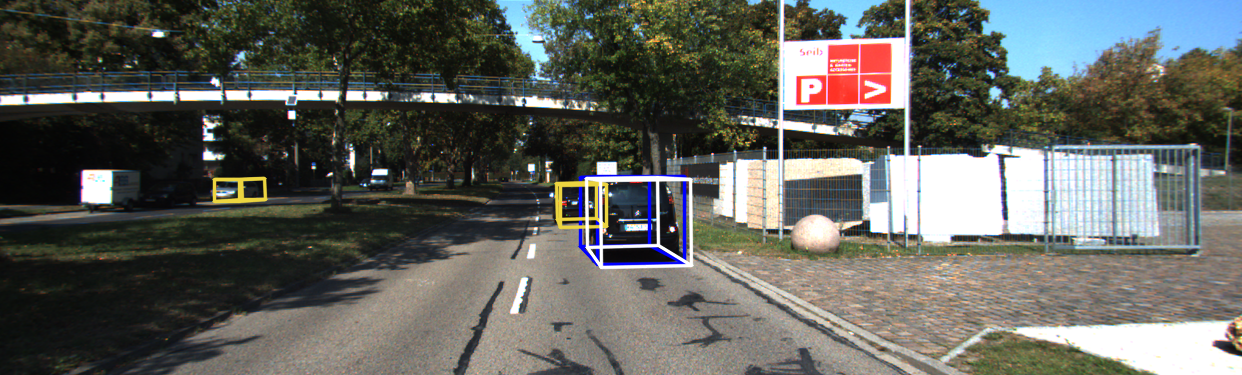}
    \end{subfigure}
    \caption{Front-view images of frustum translation attack on~\ref{av:ttt}; companion images Fig.~\ref{fig:case_AV4_ATT5_track}. Compare to camera-lidar detection-level fusion under same attack Fig.~\ref{fig:case_AV3_ATT5_track_percep}.}
    \label{fig:case_AV4_ATT5_track_percep}
\end{figure}

Supplemental visualizations for case studies are in Figures~\ref{fig:case_AV3_ATT3_track_percep}, \ref{fig:case_AV4_ATT3_track_percep}, \ref{fig:case_AV3_ATT5_track_percep}, \ref{fig:case_AV4_ATT5_track_percep}.

\subsection{All Results} \label{appendix:all-results}

\subsubsection{Baseline Results}
\begin{figure}[!t]
    \centering
    \begin{subfigure}[b]{.45\linewidth}
        \centering
        \includegraphics[width=0.9\linewidth]{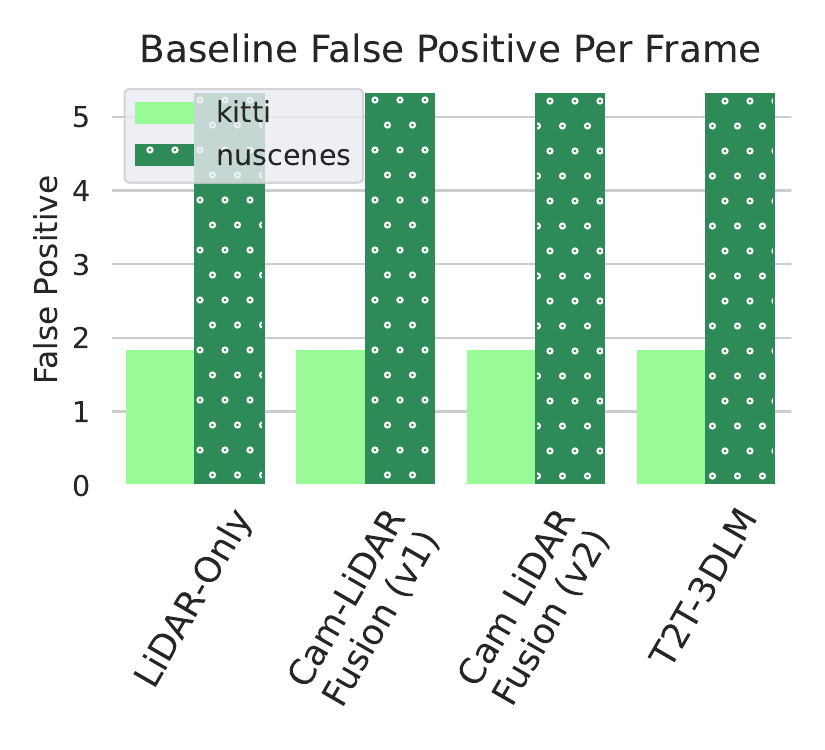}
        \caption{Baseline FPs consistent.}
        \label{fig:baseline_percep_bars_a}
    \end{subfigure}
    \begin{subfigure}[b]{.45\linewidth}
        \centering
        \includegraphics[width=0.9\linewidth]{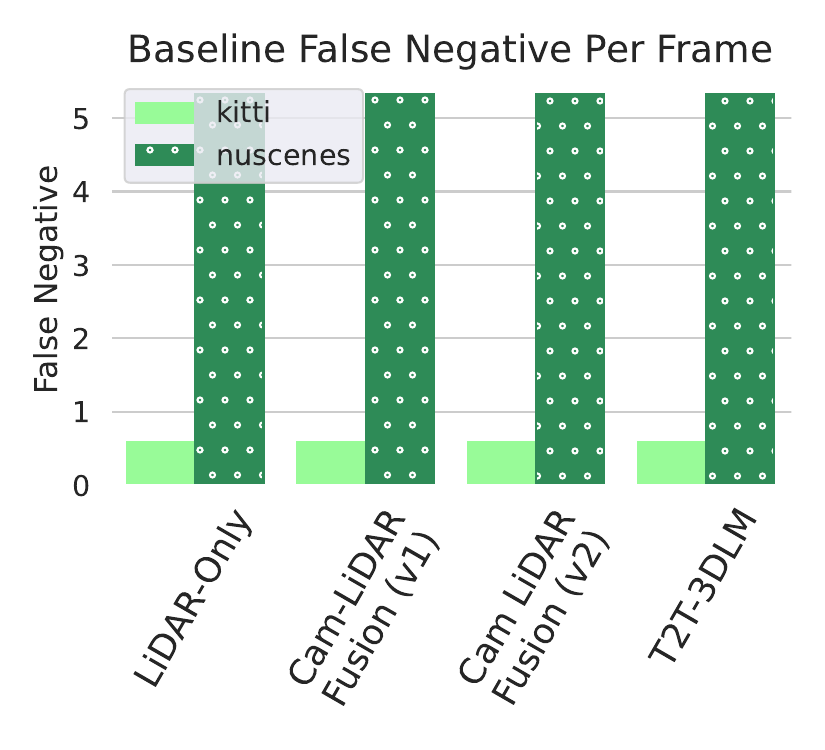}
        \caption{Baseline FNs consistent.}
        \label{fig:baseline_percep_bars_b}
    \end{subfigure}
    \caption{Baseline (unattacked) AV LiDAR perception is constant across all AV cases because all use the same underlying LiDAR perception algorithm. nuScenes has more FP and FN than KITTI because nuScenes scenes include more complex occlusion relationships and temporal dynamics and more objects per scene.}
    \label{fig:baseline_percep_bars}
    \centering
    \begin{subfigure}[b]{.45\linewidth}
        \centering
        \includegraphics[width=0.9\linewidth]{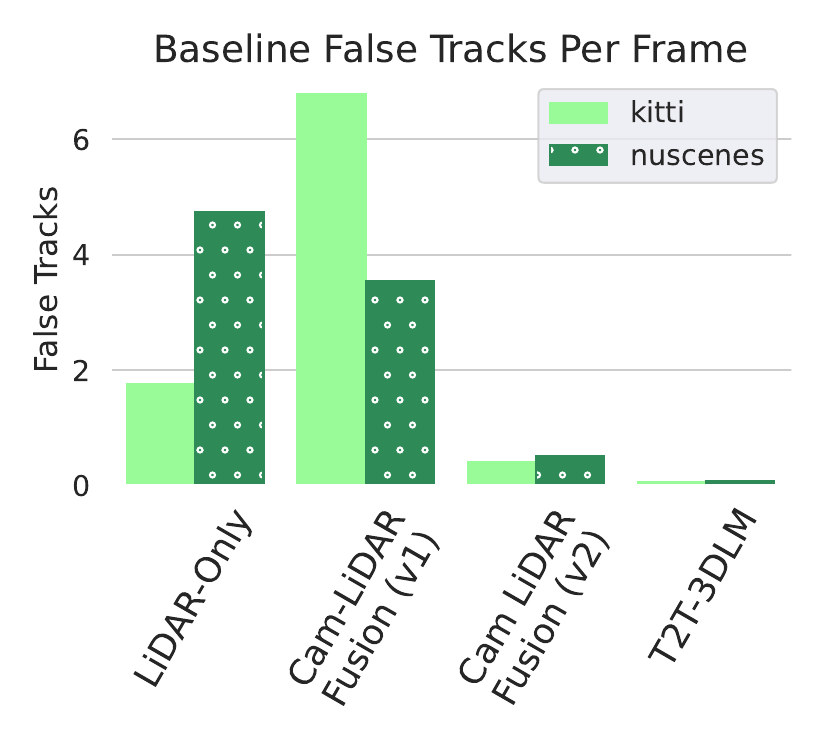}
        \caption{Each AV arch. provides different level of FT protection.}
    \end{subfigure}
    \begin{subfigure}[b]{.45\linewidth}
        \centering
        \includegraphics[width=0.9\linewidth]{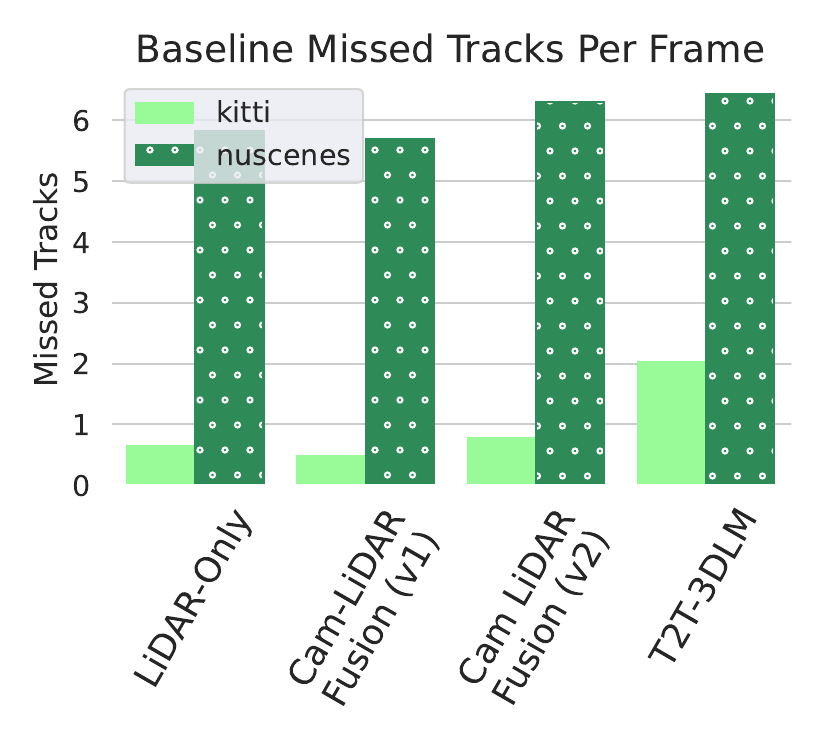}
        \caption{MT outcomes consistent across AV arch.}
    \end{subfigure}
    \caption{Baseline (unattacked) AV tracking across 36 scenes of 150 frames in KITTI (5400 frames) and 135 scenes of 40 frames in nuScenes (5400 frames) illustrates the benefits of sensor fusion. False tracks mitigated with fusion and only small increase in MT observed. Fusion v1 has high FT rate because it is not tuned. nuScenes has higher MT rate due to the larger number of objects in scenes.}
    \label{fig:baseline_track_bars}
\end{figure}

Without attack baseline results are in Figs.~\ref{fig:baseline_percep_bars} and~\ref{fig:baseline_track_bars}.

\subsubsection{Large-scale Evaluations}
Due to space limitations, Section~\ref{sec:results-attack-effectiveness} provided a detailed look at attack outcomes for KITTI and nuScenes datasets against the reverse replay and frustum translation attacks. Here, we present results from \emph{all attacks} in Sec.~\ref{sec:5-attack-designs}. 

Fig.~\ref{fig:results-all-outcomes} and Table~\ref{tab:results_attack_collection} present the same information on the changes due to attack in FT, MT, and safety outcomes.  Table~\ref{tab:results_attack_collection} also includes changes in perception outcomes that were summarized in Fig.~\ref{fig:attack_percep_all}. 

\renewcommand{\subfigwidth}{50mm}
\renewcommand{\subfigheight}{40mm}
\newcommand{\ptskip}{6pt}

\begin{figure*}[!t]
\begin{tabular}{ccc}

    \includegraphics[width=\subfigwidth,height=\subfigheight]{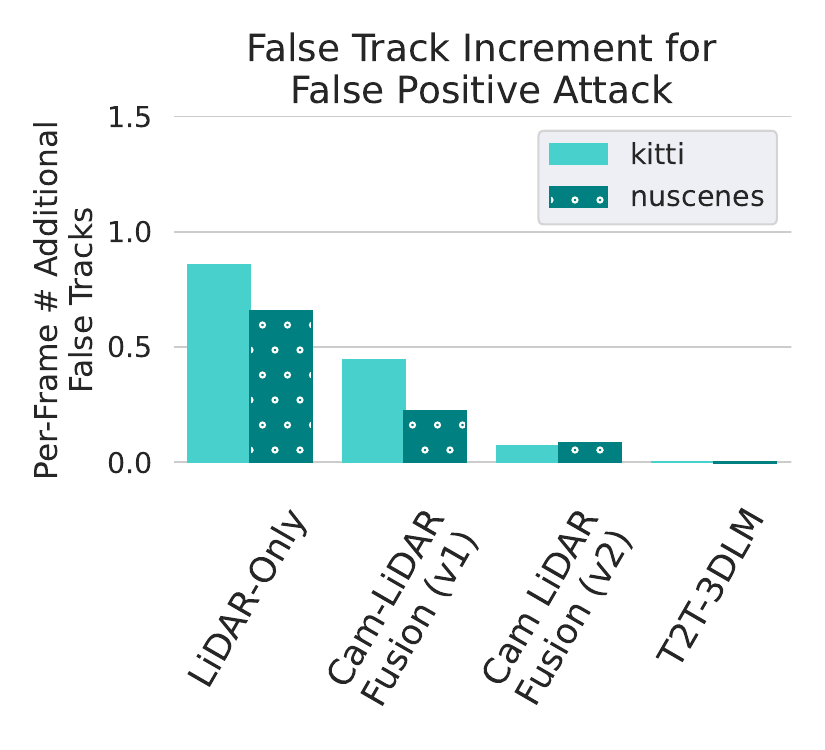} &
    \includegraphics[width=\subfigwidth,height=\subfigheight]{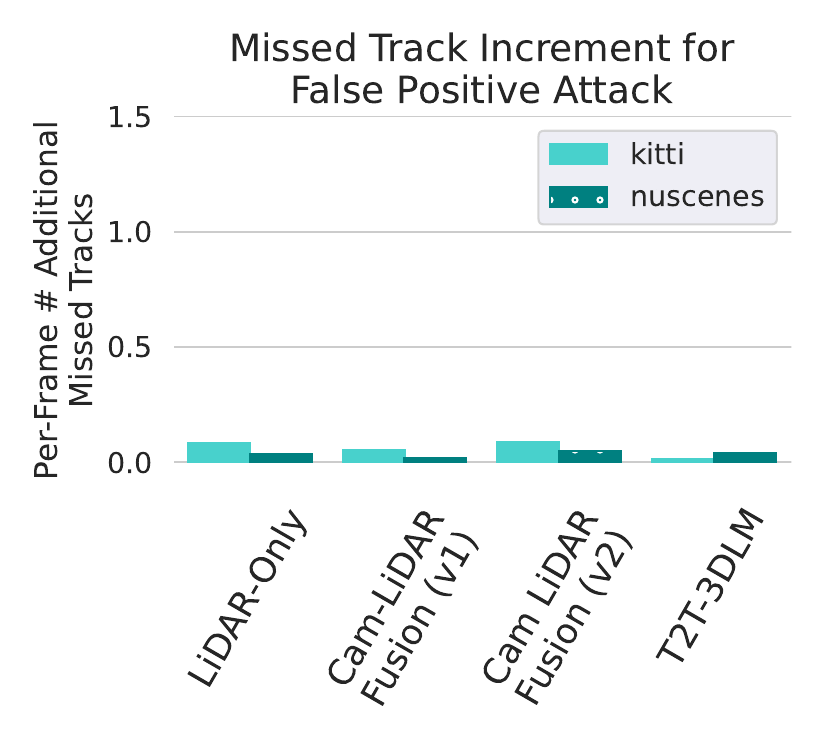} &
    \includegraphics[width=\subfigwidth,height=\subfigheight]{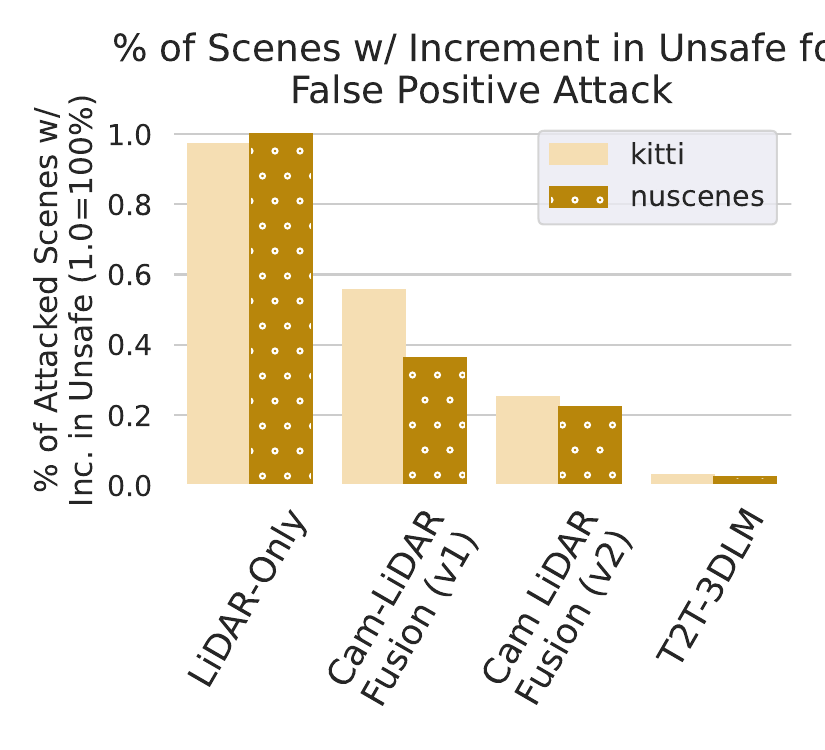}
    \\
    
    \includegraphics[width=\subfigwidth,height=\subfigheight]{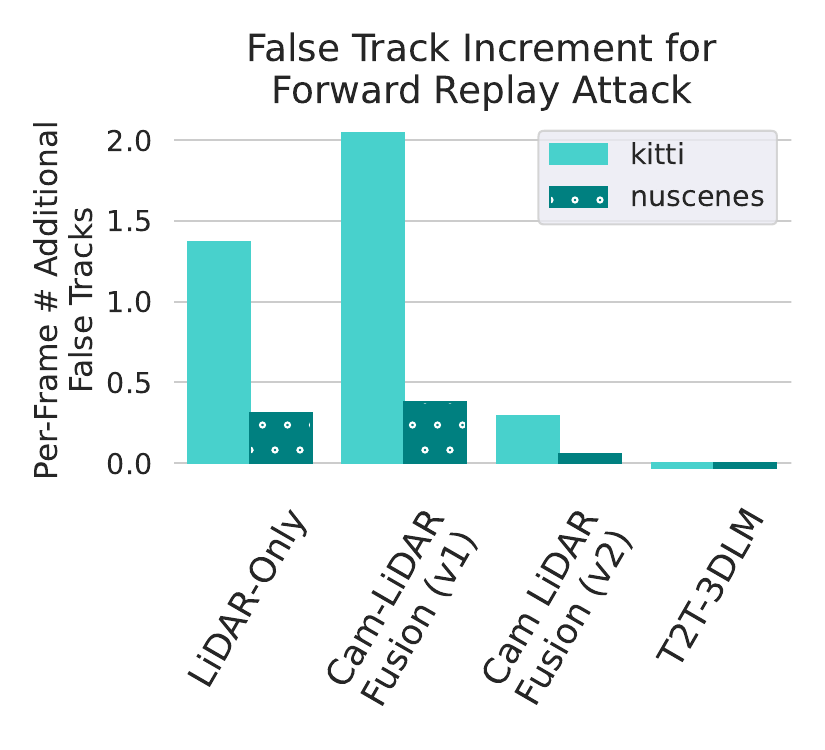} &
    \includegraphics[width=\subfigwidth,height=\subfigheight]{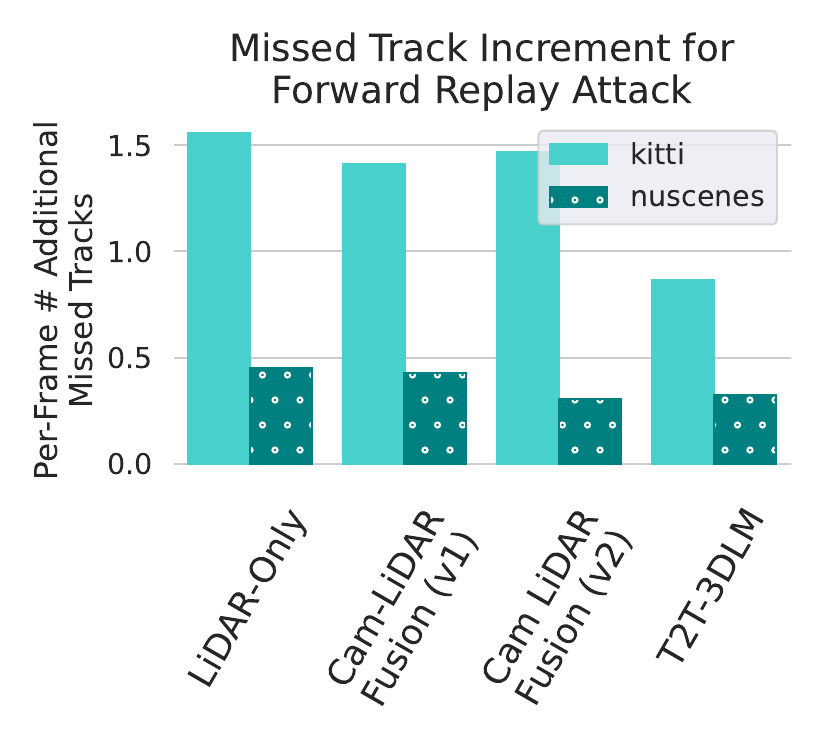} &
    \includegraphics[width=\subfigwidth,height=\subfigheight]{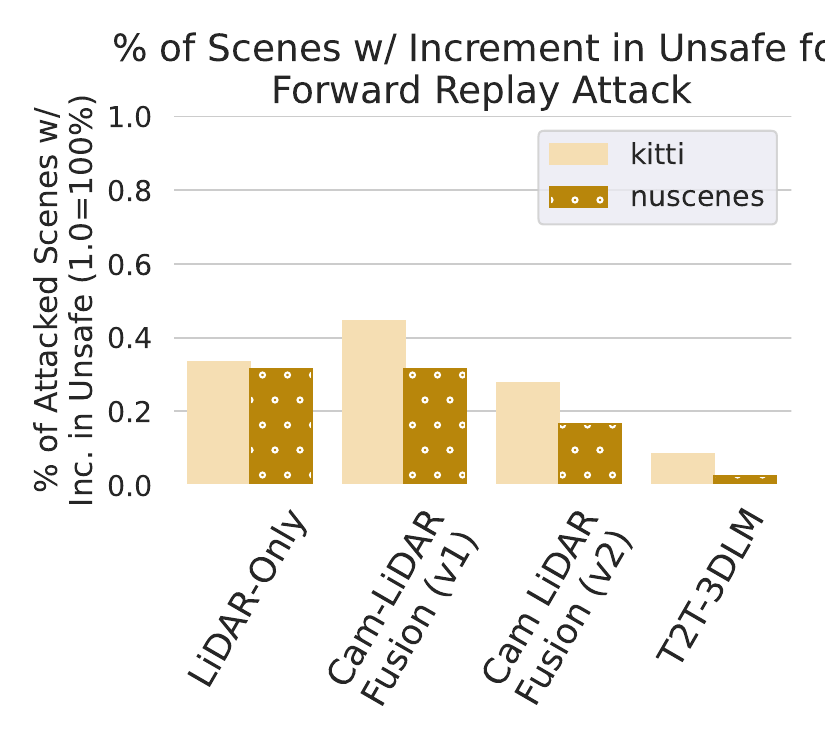}
    \\

    \includegraphics[width=\subfigwidth,height=\subfigheight]{charts/ATT_nFT_inc_per_frame_3.pdf} &
    \includegraphics[width=\subfigwidth,height=\subfigheight]{charts/ATT_nMT_inc_per_frame_3.pdf} &
    \includegraphics[width=\subfigwidth,height=\subfigheight]{charts/ATT_unsafe_existence_bool_3.pdf}
    \\

    \includegraphics[width=\subfigwidth,height=\subfigheight]{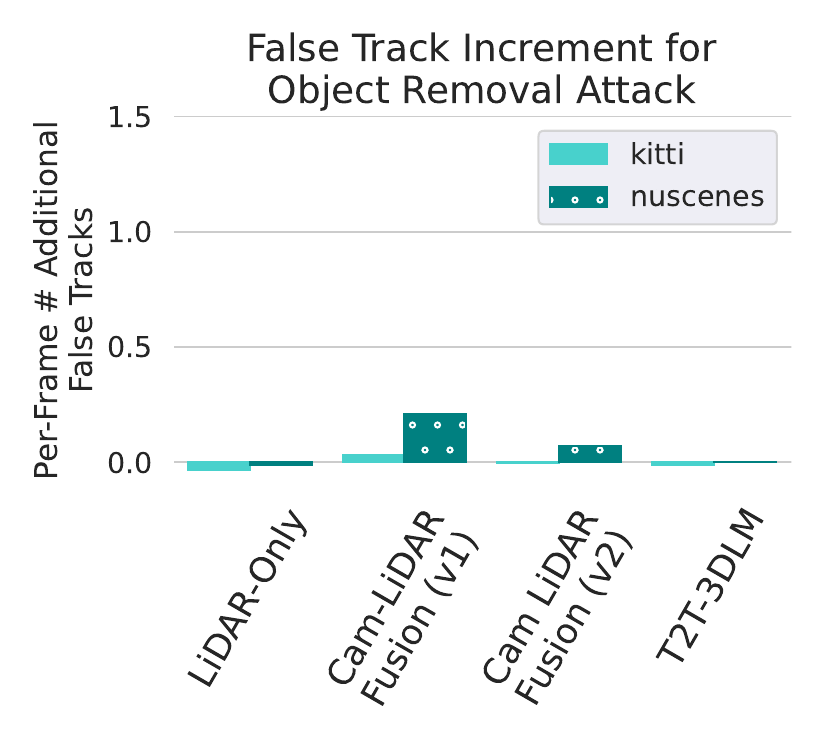} &
    \includegraphics[width=\subfigwidth,height=\subfigheight]{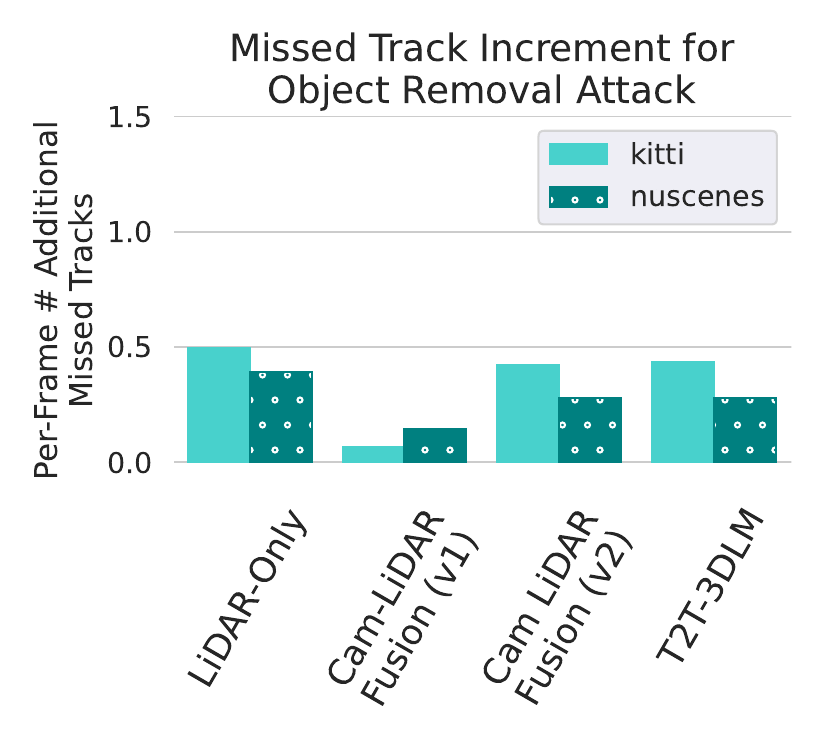} &
    \includegraphics[width=\subfigwidth,height=\subfigheight]{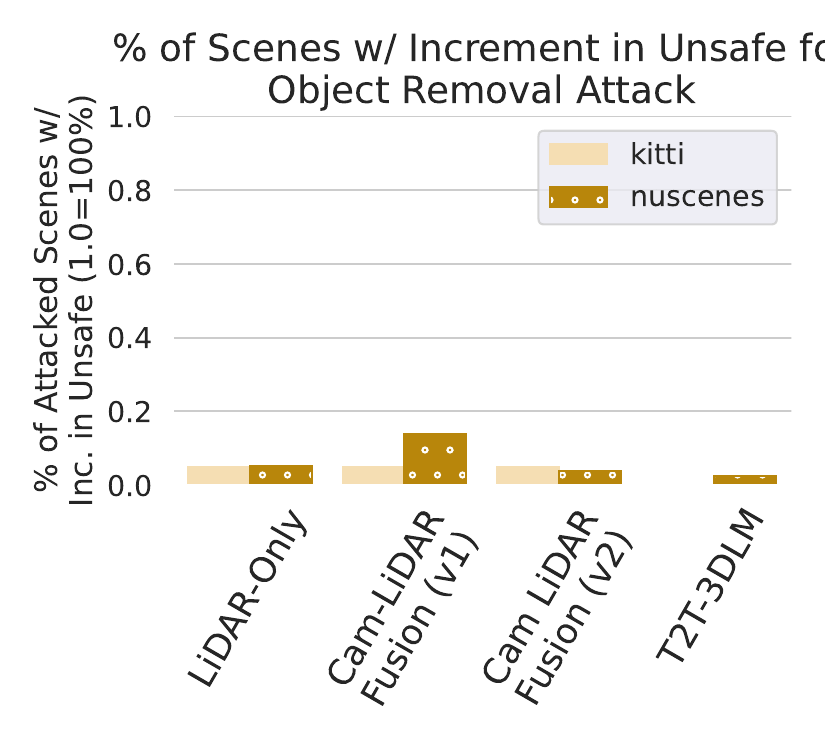}
    \\

    \includegraphics[width=\subfigwidth,height=\subfigheight]{charts/ATT_nFT_inc_per_frame_5.pdf} &
    \includegraphics[width=\subfigwidth,height=\subfigheight]{charts/ATT_nMT_inc_per_frame_5.pdf} &
    \includegraphics[width=\subfigwidth,height=\subfigheight]{charts/ATT_unsafe_existence_bool_5.pdf}
    \\
        
    
\end{tabular}
\caption{\textbf{Column (1):} Avg. per-frame change in the \emph{number} of false tracks (``FT increment'') during the attack compared to baseline run. \textbf{Column (2):} Avg. per-frame change in the \emph{number} of missed tracks (``MT increment'') during the attack compared to baseline run. \textbf{Column (3):} \emph{Fraction} of scenes with safety-critical incidents under attack compared to baseline.}
\label{fig:results-all-outcomes}
\end{figure*}

\begin{table*}[!t]
    \centering
    \caption{``Increment-over-baseline", per-frame outcomes over 36 KITTI scenes of 150 frames and 135 nuScences scences of 40 frames. Red draws attention to poor performance; green to acceptable performance; blank to questionable (in-between) performance. We ignore nuScenes MT number because it is high across the board.
    **Safety is only tested on: ``AV perceives scene unsafe, reality is safe"; important to test safety on ``AV perceives scene safe, reality is unsafe" in AV simulators. e.g.,~all object removal attacks are deemed ``safe" because an object removal attack can only induce the latter condition, not the former (tested).}
    \begin{adjustbox}{max width=\linewidth}
    \begin{tabular}{lllll}
\toprule
                                                                               AV &                                Attacker &                                                           Percep. Outcomes &                                                                                                                          Track Outcomes &                                                                                     Safety Outcomes \\ \midrule
\midrule
                                                        \ref{av:lidar} Lidar-Only &             \ref{att:fp} False Positive &  \tworowsubtableleft{K -- FP:  0.87, FN:  0.10}{N -- FP:  0.77, FN:  0.03} &                           \tworowsubtableleft{K -- \cellcolor{red!25}FT:  0.86, MT:  0.08}{N -- \cellcolor{red!25}FT:  0.66, MT:  0.03} &     \tworowsubtableleft{K -- \cellcolor{red!25}Unsafe:  0.97}{N -- \cellcolor{red!25}Unsafe:  1.00} \\ \midrule
\ref{av:cam-lidar-v1} \tworowsubtablecenter{Cam-LiDAR Fusion}{ at Detection (v1)} &             \ref{att:fp} False Positive &  \tworowsubtableleft{K -- FP:  0.87, FN:  0.09}{N -- FP:  0.77, FN:  0.03} &                                             \tworowsubtableleft{K -- \cellcolor{red!25}FT:  0.44, MT:  0.05}{N -- FT:  0.22, MT:  0.02} &     \tworowsubtableleft{K -- \cellcolor{red!25}Unsafe:  0.56}{N -- \cellcolor{red!25}Unsafe:  0.36} \\ \midrule
\ref{av:cam-lidar-v2} \tworowsubtablecenter{Cam-LiDAR Fusion}{ at Detection (v2)} &             \ref{att:fp} False Positive &  \tworowsubtableleft{K -- FP:  0.87, FN:  0.10}{N -- FP:  0.77, FN:  0.03} &   \tworowsubtableleft{K -- \cellcolor{green!25}FT:  0.07, \cellcolor{green!25}MT:  0.09}{N -- \cellcolor{green!25}FT:  0.08, MT:  0.05} &                                         \tworowsubtableleft{K -- Unsafe:  0.25}{N -- Unsafe:  0.22} \\ \midrule
               \ref{av:ttt} \tworowsubtablecenter{Cam-LiDAR Fusion}{ at Tracking} &             \ref{att:fp} False Positive &  \tworowsubtableleft{K -- FP:  0.87, FN:  0.09}{N -- FP:  0.77, FN:  0.03} & \tworowsubtableleft{K -- \cellcolor{green!25}FT:  -0.00, \cellcolor{green!25}MT:  0.01}{N -- \cellcolor{green!25}FT:  -0.00, MT:  0.04} & \tworowsubtableleft{K -- \cellcolor{green!25}Unsafe:  0.03}{N -- \cellcolor{green!25}Unsafe:  0.02} \\ \midrule
                                                        \ref{av:lidar} Lidar-Only &      \ref{att:forreplay} Forward Replay &  \tworowsubtableleft{K -- FP:  1.44, FN:  1.62}{N -- FP:  0.67, FN:  0.75} &         \tworowsubtableleft{K -- \cellcolor{red!25}FT:  1.37, \cellcolor{red!25}MT:  1.56}{N -- \cellcolor{red!25}FT:  0.31, MT:  0.45} &     \tworowsubtableleft{K -- \cellcolor{red!25}Unsafe:  0.33}{N -- \cellcolor{red!25}Unsafe:  0.31} \\ \midrule
\ref{av:cam-lidar-v1} \tworowsubtablecenter{Cam-LiDAR Fusion}{ at Detection (v1)} &      \ref{att:forreplay} Forward Replay &  \tworowsubtableleft{K -- FP:  1.44, FN:  1.62}{N -- FP:  0.67, FN:  0.75} &         \tworowsubtableleft{K -- \cellcolor{red!25}FT:  2.04, \cellcolor{red!25}MT:  1.41}{N -- \cellcolor{red!25}FT:  0.38, MT:  0.43} &     \tworowsubtableleft{K -- \cellcolor{red!25}Unsafe:  0.44}{N -- \cellcolor{red!25}Unsafe:  0.31} \\ \midrule
\ref{av:cam-lidar-v2} \tworowsubtablecenter{Cam-LiDAR Fusion}{ at Detection (v2)} &      \ref{att:forreplay} Forward Replay &  \tworowsubtableleft{K -- FP:  1.44, FN:  1.62}{N -- FP:  0.67, FN:  0.75} &       \tworowsubtableleft{K -- \cellcolor{red!25}FT:  0.29, \cellcolor{red!25}MT:  1.47}{N -- \cellcolor{green!25}FT:  0.05, MT:  0.30} &                       \tworowsubtableleft{K -- \cellcolor{red!25}Unsafe:  0.28}{N -- Unsafe:  0.17} \\ \midrule
               \ref{av:ttt} \tworowsubtablecenter{Cam-LiDAR Fusion}{ at Tracking} &      \ref{att:forreplay} Forward Replay &  \tworowsubtableleft{K -- FP:  1.44, FN:  1.62}{N -- FP:  0.67, FN:  0.75} &                       \tworowsubtableleft{K -- FT:  -0.03, \cellcolor{red!25}MT:  0.86}{N -- \cellcolor{green!25}FT:  -0.03, MT:  0.32} & \tworowsubtableleft{K -- \cellcolor{green!25}Unsafe:  0.08}{N -- \cellcolor{green!25}Unsafe:  0.02} \\ \midrule
                                                        \ref{av:lidar} Lidar-Only &      \ref{att:revreplay} Reverse Replay &  \tworowsubtableleft{K -- FP:  1.37, FN:  1.56}{N -- FP:  0.59, FN:  0.69} &         \tworowsubtableleft{K -- \cellcolor{red!25}FT:  1.36, \cellcolor{red!25}MT:  1.51}{N -- \cellcolor{red!25}FT:  0.33, MT:  0.40} &     \tworowsubtableleft{K -- \cellcolor{red!25}Unsafe:  0.36}{N -- \cellcolor{red!25}Unsafe:  0.32} \\ \midrule
\ref{av:cam-lidar-v1} \tworowsubtablecenter{Cam-LiDAR Fusion}{ at Detection (v1)} &      \ref{att:revreplay} Reverse Replay &  \tworowsubtableleft{K -- FP:  1.37, FN:  1.56}{N -- FP:  0.59, FN:  0.69} &         \tworowsubtableleft{K -- \cellcolor{red!25}FT:  2.22, \cellcolor{red!25}MT:  1.45}{N -- \cellcolor{red!25}FT:  0.27, MT:  0.40} &                       \tworowsubtableleft{K -- \cellcolor{red!25}Unsafe:  0.44}{N -- Unsafe:  0.23} \\ \midrule
\ref{av:cam-lidar-v2} \tworowsubtablecenter{Cam-LiDAR Fusion}{ at Detection (v2)} &      \ref{att:revreplay} Reverse Replay &  \tworowsubtableleft{K -- FP:  1.37, FN:  1.56}{N -- FP:  0.59, FN:  0.69} &       \tworowsubtableleft{K -- \cellcolor{red!25}FT:  0.35, \cellcolor{red!25}MT:  1.46}{N -- \cellcolor{green!25}FT:  0.08, MT:  0.28} &                     \tworowsubtableleft{K -- Unsafe:  0.17}{N -- \cellcolor{green!25}Unsafe:  0.14} \\ \midrule
               \ref{av:ttt} \tworowsubtablecenter{Cam-LiDAR Fusion}{ at Tracking} &      \ref{att:revreplay} Reverse Replay &  \tworowsubtableleft{K -- FP:  1.37, FN:  1.56}{N -- FP:  0.59, FN:  0.69} &                       \tworowsubtableleft{K -- FT:  -0.03, \cellcolor{red!25}MT:  0.83}{N -- \cellcolor{green!25}FT:  -0.03, MT:  0.31} & \tworowsubtableleft{K -- \cellcolor{green!25}Unsafe:  0.14}{N -- \cellcolor{green!25}Unsafe:  0.02} \\ \midrule
                                                        \ref{av:lidar} Lidar-Only &         \ref{att:remove} Object Removal & \tworowsubtableleft{K -- FP:  -0.02, FN:  0.54}{N -- FP:  0.04, FN:  0.39} &                       \tworowsubtableleft{K -- FT:  -0.03, \cellcolor{red!25}MT:  0.49}{N -- \cellcolor{green!25}FT:  -0.01, MT:  0.39} & \tworowsubtableleft{K -- \cellcolor{green!25}Unsafe:  0.05}{N -- \cellcolor{green!25}Unsafe:  0.05} \\ \midrule
\ref{av:cam-lidar-v1} \tworowsubtablecenter{Cam-LiDAR Fusion}{ at Detection (v1)} &         \ref{att:remove} Object Removal & \tworowsubtableleft{K -- FP:  -0.02, FN:  0.55}{N -- FP:  0.04, FN:  0.39} &                       \tworowsubtableleft{K -- \cellcolor{green!25}FT:  0.03, \cellcolor{green!25}MT:  0.06}{N -- FT:  0.21, MT:  0.15} & \tworowsubtableleft{K -- \cellcolor{green!25}Unsafe:  0.05}{N -- \cellcolor{green!25}Unsafe:  0.14} \\ \midrule
\ref{av:cam-lidar-v2} \tworowsubtablecenter{Cam-LiDAR Fusion}{ at Detection (v2)} &         \ref{att:remove} Object Removal & \tworowsubtableleft{K -- FP:  -0.03, FN:  0.54}{N -- FP:  0.04, FN:  0.39} &                        \tworowsubtableleft{K -- FT:  -0.00, \cellcolor{red!25}MT:  0.42}{N -- \cellcolor{green!25}FT:  0.07, MT:  0.28} & \tworowsubtableleft{K -- \cellcolor{green!25}Unsafe:  0.05}{N -- \cellcolor{green!25}Unsafe:  0.04} \\ \midrule
               \ref{av:ttt} \tworowsubtablecenter{Cam-LiDAR Fusion}{ at Tracking} &         \ref{att:remove} Object Removal & \tworowsubtableleft{K -- FP:  -0.02, FN:  0.55}{N -- FP:  0.04, FN:  0.39} &                       \tworowsubtableleft{K -- FT:  -0.01, \cellcolor{red!25}MT:  0.43}{N -- \cellcolor{green!25}FT:  -0.00, MT:  0.28} & \tworowsubtableleft{K -- \cellcolor{green!25}Unsafe:  0.00}{N -- \cellcolor{green!25}Unsafe:  0.03} \\ \midrule
                                                        \ref{av:lidar} Lidar-Only & \ref{att:frust-trans} Frustum Translate &  \tworowsubtableleft{K -- FP:  0.68, FN:  0.52}{N -- FP:  0.37, FN:  0.42} &                           \tworowsubtableleft{K -- \cellcolor{red!25}FT:  0.64, \cellcolor{red!25}MT:  0.50}{N -- FT:  0.21, MT:  0.44} &     \tworowsubtableleft{K -- \cellcolor{red!25}Unsafe:  0.65}{N -- \cellcolor{red!25}Unsafe:  0.67} \\ \midrule
\ref{av:cam-lidar-v1} \tworowsubtablecenter{Cam-LiDAR Fusion}{ at Detection (v1)} & \ref{att:frust-trans} Frustum Translate &  \tworowsubtableleft{K -- FP:  0.68, FN:  0.51}{N -- FP:  0.37, FN:  0.42} &                           \tworowsubtableleft{K -- \cellcolor{red!25}FT:  0.84, MT:  0.28}{N -- \cellcolor{red!25}FT:  0.33, MT:  0.20} &     \tworowsubtableleft{K -- \cellcolor{red!25}Unsafe:  0.60}{N -- \cellcolor{red!25}Unsafe:  0.57} \\ \midrule
\ref{av:cam-lidar-v2} \tworowsubtablecenter{Cam-LiDAR Fusion}{ at Detection (v2)} & \ref{att:frust-trans} Frustum Translate &  \tworowsubtableleft{K -- FP:  0.68, FN:  0.52}{N -- FP:  0.37, FN:  0.42} &                           \tworowsubtableleft{K -- \cellcolor{red!25}FT:  0.39, \cellcolor{red!25}MT:  0.53}{N -- FT:  0.18, MT:  0.31} &     \tworowsubtableleft{K -- \cellcolor{red!25}Unsafe:  0.35}{N -- \cellcolor{red!25}Unsafe:  0.41} \\ \midrule
               \ref{av:ttt} \tworowsubtablecenter{Cam-LiDAR Fusion}{ at Tracking} & \ref{att:frust-trans} Frustum Translate &  \tworowsubtableleft{K -- FP:  0.69, FN:  0.52}{N -- FP:  0.37, FN:  0.42} &                       \tworowsubtableleft{K -- FT:  -0.01, \cellcolor{red!25}MT:  0.41}{N -- \cellcolor{green!25}FT:  -0.01, MT:  0.35} & \tworowsubtableleft{K -- \cellcolor{green!25}Unsafe:  0.00}{N -- \cellcolor{green!25}Unsafe:  0.01} \\ \midrule
\bottomrule
\end{tabular}
    \end{adjustbox}
    \label{tab:results_attack_collection}
\end{table*}

\end{document}